\documentclass[11pt]{llncs}
\usepackage{geometry}
\usepackage{Xiao-llncs}

\newcommand{\Adv}{\ensuremath{\mathcal{A}}\xspace}

\newcommand{\algo}[1]{\ensuremath{\mathsf{#1}}\xspace}

\newcommand{\Bad}{\ensuremath{\mathsf{Bad}}\xspace}

\newcommand{\bits}{\ensuremath{\{0,1\}}\xspace}

\newcommand{\CCY}{\mathsf{CCY}}

\newcommand{\cind}{\ensuremath{\stackrel{\text{c}}{\approx}}\xspace}

\newcommand{\com}{\ensuremath{\mathsf{com}}\xspace}
\newcommand{\Com}{\ensuremath{\mathsf{Com}}\xspace}

\newcommand{\decom}{\ensuremath{\mathsf{decom}}\xspace}

\renewcommand{\epsilon}{\varepsilon}

\newcommand{\EXEC}{\ensuremath{\mathsf{EXEC}}\xspace}
\newcommand{\ExtCom}{\ensuremath{\mathsf{ExtCom}}\xspace}
\newcommand{\Ext}{\ensuremath{\mathsf{Ext}}\xspace}

\newcommand{\Func}{\ensuremath{\mathcal{F}}\xspace}

\newcommand{\Good}{\ensuremath{\mathsf{Good}}\xspace}

\newcommand{\idind}{\ensuremath{\stackrel{\text{i.d.}}{=\mathrel{\mkern-3mu}=}}\xspace}
\newcommand{\IDEAL}{\ensuremath{\mathsf{IDEAL}}\xspace}
\newcommand{\IgnoreInvalidTrans}{\ensuremath{\mathsf{IgnoreInvalidTrans}}\xspace}

\newcommand{\init}{\ensuremath{\mathsf{init}}\xspace}

\newcommand{\Lang}{\ensuremath{\mathcal{L}}\xspace}

\newcommand{\Naturals}{\ensuremath{\mathbb{N}}\xspace}

\newcommand{\negl}{\ensuremath{\mathsf{negl}}\xspace}
\newcommand{\no}{\mathrm{no}}

\newcommand{\NP}{\ensuremath{\mathbf{NP}}\xspace}

\newcommand{\Output}{\ensuremath{\mathsf{Out}}\xspace}

\renewcommand{\paragraph}{\para}

\newcommand{\pick}{\ensuremath{\leftarrow}\xspace}
\newcommand{\poly}{\ensuremath{\mathsf{poly}}\xspace}
\newcommand{\PPT}{\ensuremath{\mathrm{PPT}}\xspace}

\newcommand{\Prot}{\ensuremath{\mathrm{\Pi}}\xspace}

\newcommand{\REAL}{\ensuremath{\mathsf{REAL}}\xspace}
\newcommand{\Recon}{\ensuremath{\mathsf{Recon}}\xspace}

\newcommand{\Relation}{\ensuremath{\mathcal{R}}\xspace}

\newcommand{\SimExt}{\ensuremath{\mathcal{SE}}\xspace}
\newcommand{\SecPar}{\ensuremath{\lambda}\xspace}

\newcommand{\Set}[1]{\ensuremath{\{#1\}}\xspace}

\newcommand{\sfpar}{\ensuremath{\mathsf{par}}\xspace}
\newcommand{\Share}{\ensuremath{\mathsf{Share}}\xspace}

\newcommand{\Sim}{\ensuremath{\mathcal{S}}\xspace}

\newcommand{\simless}{\ensuremath{\mathsf{Sim}\text{-}\mathsf{less}}\xspace}
\newcommand{\sind}{\ensuremath{\stackrel{\text{s}}{\approx}}\xspace}
\newcommand{\ST}{\ensuremath{\mathsf{ST}}\xspace}

\newcommand{\Trans}{\ensuremath{\tau}\xspace}
\renewcommand{\tilde}{\widetilde}

\newcommand{\vcom}{\overline{\mathsf{com}}}
\newcommand{\vdecom}{\overline{\mathsf{decom}}}

\newcommand{\Vect}[1]{\ensuremath{\mathrm{{\bf #1}}}\xspace}
\newcommand{\Verify}{\ensuremath{\mathsf{Verify}}\xspace}
\newcommand{\View}{\ensuremath{\mathsf{View}}\xspace}
\newcommand{\view}{\ensuremath{\mathsf{v}}\xspace}

\newcommand{\VSS}{\ensuremath{\mathsf{VSS}}\xspace}
\newcommand{\weak}{\ensuremath{\mathsf{weak}}\xspace}
\newcommand{\wExtCom}{\ensuremath{\mathsf{wExtCom}}\xspace}

\newcommand{\revise}[1]{#1}
\newcommand{\xor}{\ensuremath{\oplus}\xspace}

\newcommand{\yes}{\mathrm{yes}}

\newcommand{\sample}{\gets}

\newcommand{\ra}{\rightarrow}

\newcommand{\secpar}{\lambda}

\newcommand{\out}{\mathsf{out}}
\newcommand{\bit}{\{0,1\}}

\newcommand{\hil}{\mathcal{H}}

\newcommand{\defeq}{:=}

\newcommand{\redunderline}[1]{\textcolor{red}{\underline{\textcolor{black}{#1}}}}

\newcommand{\A}{\mathcal{A}}
\newcommand{\B}{\mathcal{B}}

\newcommand{\ext}{\mathsf{Ext}}

\newcommand{\calX}{\mathcal{X}}
\newcommand{\calY}{\mathcal{Y}}

\newcommand{\ot}{\otimes}
\newcommand{\fail}{\mathsf{Fail}}

\newcommand{\Acc}{\mathsf{Acc}}

\newcommand{\inp}{\mathsf{inp}}

\newcommand{\QMA}{\mathbf{QMA}}
\newcommand{\BQP}{\mathbf{BQP}}

\newcommand*{\ipro}[2]{\langle #1|#2\rangle}
\newcommand{\TD}{\mathsf{TD}}

\newcommand{\reginp}{\mathsf{Inp}}

\newcommand*{\regX}{\mathbf{X}}
\newcommand*{\regY}{\mathbf{Y}}
\newcommand*{\regZ}{\mathbf{Z}}
\newcommand{\regW}{\mathbf{W}}

\newcommand*{\regM}{\mathbf{M}}
\newcommand*{\regout}{\mathbf{Out}}
\newcommand*{\reganc}{\mathbf{Anc}}

\newcommand*{\regA}{\mathbf{A}}

\newcommand*{\regB}{\mathbf{B}}
\newcommand{\regother}{\mathbf{Other}}
\newcommand{\regV}{\mathbf{V}}

\newcommand*{\regD}{\mathbf{D}}

\newcommand{\test}{\mathsf{test}}
\newcommand{\siml}{\mathsf{Sim}}
\newcommand{\abort}{\mathsf{a}}
\newcommand{\nonabort}{\mathsf{na}}
\newcommand{\comb}{\mathsf{comb}}

\newcommand{\compind}{\cind}
\newcommand{\statind}{\sind}

\newcommand{\val}{\mathsf{val}}

\newcommand{\execution}[2]{\langle #1, #2 \rangle}

\newcommand{\pcombsuc}{p_{\comb}^{\mathsf{suc}}(1^\secpar,1^{\epsilon^{-1}},\{\Pi_i\}_{i\in C},\A,\ket{\psi_\init})}

\newcommand{\Exp}{\mathsf{Exp}}

\newcommand{\OUT}{\mathsf{OUT}}
\newcommand{\sfQ}{\mathsf{Q}}
\newcommand{\sfR}{\mathsf{R}}
\newcommand{\final}{\mathsf{final}}

\newcommand{\amp}{\mathsf{amp}}
\newcommand{\Amp}{\mathsf{Amp}}

\newenvironment{boxfig}[2]{\begin{figure}[#1]\fbox{\begin{minipage}{0.97\linewidth}
                        \vspace{0.2em}
                        \makebox[0.025\linewidth]{}
                        \begin{minipage}{0.95\linewidth}
            {{
                        #2 }}
                        \end{minipage}
                        \vspace{0.2em}
                        \end{minipage}}}{\end{figure}}

\newtheorem{myclaim}[theorem]{Claim}

\usepackage[para,online,flushleft]{threeparttable}

\makeatletter
\def\@fnsymbol#1{\ensuremath{\ifcase#1\or *\or \dagger\or \ddagger\or
   \mathsection\or \mathparagraph\or \|\or **\or \dagger\dagger
   \or \ddagger\ddagger \else\@ctrerr\fi}}
\makeatother

\begin{document}


\title{Post-Quantum Simulatable Extraction with Minimal Assumptions: Black-Box and Constant-Round}

\author{
Nai-Hui Chia\inst{1}
\and 
 Kai-Min Chung\inst{2}\thanks{Kai-Min Chung is supported in part by Ministry of Science and Technology, Taiwan, under Grant no. MOST 109-2223-E-001-001-MY3, the 2021 Academia Sinica Investigator Award (AS-IA-110-M02), and the Air Force Office of Scientific Research under award number FA2386-20-1-4066.}
\and 
Xiao Liang\inst{3}\thanks{Xiao Liang is supported in part by Omkant Pandey's DARPA SIEVE Award HR00112020026 and NSF grants 1907908 and 2028920. Any opinions, findings and conclusions or recommendations expressed in this material are those of the author(s) and do not necessarily reflect the views of the United States Government, DARPA, or NSF.} 
\and 
Takashi Yamakawa\inst{4}
}

\institute{
Indiana University Bloomington, IN, USA\\ 
\email{naichia@iu.edu}
 \and
  Academia Sinica, Taipei, Taiwan\\
 \email{kmchung@iis.sinica.edu.tw}
\and
 Stony Brook University, NY, USA\\ \email{liang1@cs.stonybrook.edu}
 \and
 NTT Social Informatics Laboratories, Tokyo, Japan\\
 \email{takashi.yamakawa.ga@hco.ntt.co.jp}
}
\let\oldaddcontentsline\addcontentsline
\def\addcontentsline#1#2#3{}
\maketitle
\def\addcontentsline#1#2#3{\oldaddcontentsline{#1}{#2}{#3}}

\phantomsection
\addcontentsline{toc}{section}{Abstract}
\begin{abstract}
From the minimal assumption of post-quantum semi-honest oblivious transfers, we build the first  {\em $\epsilon$-simulatable} two-party computation (2PC) against quantum polynomial-time (QPT) adversaries that is both constant-round and black-box (for both the construction and security reduction).  A recent work by Chia, Chung, Liu, and  Yamakawa (FOCS'21) shows that post-quantum 2PC with standard simulation-based security is impossible in constant rounds, unless either $\NP \subseteq \BQP$ or relying on non-black-box simulation. The $\epsilon$-simulatability we target is a relaxation of the standard simulation-based security that allows for an arbitrarily small noticeable simulation error $\epsilon$.
Moreover, when quantum communication is allowed, we can further weaken the assumption to post-quantum secure one-way functions (PQ-OWFs), while maintaining the constant-round and black-box property.\\[-0.5em]

\hspace{3ex}Our techniques also yield the following set of {\em constant-round and black-box} two-party protocols secure against QPT adversaries, only assuming black-box access to PQ-OWFs:
\begin{itemize}
\item
extractable commitments for which the extractor is also an $\epsilon$-simulator; 
\item
$\epsilon$-zero-knowledge commit-and-prove whose commit stage is extractable with $\epsilon$-simulation;
\item $\epsilon$-simulatable coin-flipping;
\item
$\epsilon$-zero-knowledge arguments of knowledge for $\NP$ for which the knowledge extractor is also an $\epsilon$-simulator;
\item
$\epsilon$-zero-knowledge arguments for $\QMA$. 
\end{itemize}
\hspace{3ex}At the heart of the above results is a black-box extraction lemma showing how to efficiently extract secrets from QPT adversaries while disturbing their quantum state in a controllable manner, i.e., achieving $\epsilon$-simulatability of the post-extraction state of the adversary.

\keywords{Simulation \and Extraction \and Post-Quantum}
\end{abstract}

\tableofcontents
\addcontentsline{toc}{section}{Table of Contents}
\clearpage

	\pagenumbering{arabic}
	
\section{Introduction}\label{sec:intro}
Extractability is an important concept in cryptography. A typical example is extractable commitments,  
which enable an extractor to extract a committed message from a malicious committer. 
Extractable commitments have played a central role in several major cryptographic tasks, including (but not limited to) secure two-party and multi-party computation (e.g., \cite{TCC:CDMW09,TCC:PasWee09,STOC:Goyal11,TCC:GLPPS15}), zero-knowledge (ZK) protocols (e.g., \cite{TCC:Rosen04,JC:Lindell13}), concurrent zero-knowledge protocols (e.g., \cite{FOCS:PraRosSah02,TCC:MOSV06}), non-malleable commitments (e.g., \cite{FOCS:GLOV12,C:Kiyoshima14}) etc. 
Recently, two concurrent works by Grilo, Lin, Song, and Vaikuntanathan \cite{EC:GLSV21} and Bartusek, Coladangelo, Khurana, and Ma \cite{C:BCKM21b} (based on earlier works  \cite{C:CreKil88,C:BBCS91,C:DFLSS09,C:BouFeh10}) demonstrate new applications of extractable commitments in quantum cryptography.
They show that quantumly secure extractable commitments are sufficient for constructing maliciously secure quantum oblivious transfers (OTs), which can be compiled into general-purpose quantum MPC~\cite{C:IshPraSah08,EC:DGJMS20}.\footnote{They actually rely on extractable and \emph{equivocal} commitments. However, since equivocality can be added easily, extractable commitments are the essential building block.} 

As noted in \cite{EC:GLSV21}, it is surprisingly non-trivial to construct quantumly secure extractable commitments. The reason is that quantum extractability requires an extractor to extract the committed message \emph{while simulating the committer's post-execution state.}
However, known rewinding-based classical extraction techniques are not directly applicable as it is unclear if they could provide any simulation guarantee when used against quantum adversaries. 
To address this issue, recent works \cite{EC:GLSV21,C:BCKM21b} propose new polynomial-round \emph{quantum} constructions of
quantumly secure extractable commitments 
from post-quantum one-way functions (PQ-OWFs), which are functions efficiently computable in the classical sense but one-way against quantum polynomial-time (QPT) adversaries.  
Relying on assumptions stronger than PQ-OWFs,  \emph{classical} constructions of quantumly secure extractable commitments (which we call \emph{post-quantum} extractable commitments) are known \cite{STOC:BitShm20,TCC:AnaLap20,BLS21,C:HalSmiSon11,AFRICACRYPT:LunNie11}.
However, those constructions require (at least) the existence of OTs. 

Moreover, all existing post-quantum extractable commitments make {\em non-black-box} use of their building-block primitives. This is not ideal as {\em black-box} constructions are often preferred over non-black-box ones. A black-box construction only depends
on the input/output behavior of its building-block cryptographic primitive(s). In particular, such a construction is independent of the specific implementation or code of the building-block primitive. Black-box constructions enjoy certain advantages. For example, they remain valid even if the building-block primitive/oracle is based on a {\em physical} object such as a noisy channel or tamper-proof hardware \cite{wyner1975wire,FOCS:CreKil88,TCC:GLMMR04}. Also, since the efficiency of black-box constructions does not depend on the implementation details of the primitive, their efficiency can be theoretically independent of the code of lower-level primitives. Indeed, it has been an important theme to obtain black-box constructions for major cryptographic objects, e.g., \cite{STOC:Kilian88,C:DamIsh05,STOC:IKLP06,STOC:IKOS07,TCC:Haitner08,C:IshPraSah08,TCC:PasWee09,STOC:Goyal11,FOCS:GLOV12,C:LinPas12,C:Kiyoshima14,STOC:GOSV14,TCC:GGMP16,C:HazVen16,EC:GarKiyPan18,TCC:KhuOstSri18,ICALP:ChaLiaPan20,SCN:GLPV20,EC:GKLW21,C:LiaPan21,C:ChiChuYam21}. 

In the classical setting, it is well-known that constant-round extractable commitments can be obtained assuming only black-box access to OWFs \cite{
TCC:PasWee09, 
SIAM:DolDwoNao00,FOCS:PraRosSah02,TCC:Rosen04}.\footnote{The term ``black-box'' here refers to both black-box construction and black-box extraction.}
Therefore, it is natural to ask the following analog question  in the quantum setting: Is it possible to construct constant-round post-quantum extractable commitments assuming only black-box access to PQ-OWFs? We remark that this question is open even if we do not require the scheme to be constant-round or black-box.



\paragraph{The Black-Box Extraction Barrier.}
We observe that the recent lower bound on black-box post-quantum ZK \cite{FOCS:CCLY21} suggests a negative answer to the above question. 
Namely, if we have constant-round post-quantum extractable commitment with black-box extraction, then we can construct constant-round post-quantum ZK arguments for $\NP$ with black-box simulation based on standard techniques (see \Cref{sec:extcom_to_zk} for details). 
However, \cite{FOCS:CCLY21} showed that such a ZK argument cannot exist unless $\NP\subseteq \BQP$, which seems unlikely.\footnote{A concurrent work by Lombardi, Ma, and Spooner \cite{LMS21} showed that the impossibility of \cite{FOCS:CCLY21} can be avoided if we consider a stronger computational model for simulators. We provide more discussion in \Cref{sec:concurrent}.}

\paragraph{$\epsilon$-Simulation Security.}
On the other hand, another recent work \cite{C:ChiChuYam21} showed that we can bypass the impossibility result by relaxing the requirement of ZK to the so-called $\epsilon$-ZK~\cite{DNS04,STOC:BitKalPan18,EC:FleGoyJai18}. 
The standard ZK property requires a simulator to simulate the verifier's view in a way that no distinguisher can distinguish it from the real one with non-negligible advantage. 
In contrast, the 
$\epsilon$-ZK property only requires the existence of a simulator such that for any noticeable $\epsilon(\secpar)$, the simulated view can be distinguished from the real one with advantage at most $\epsilon$. 
As explained in \cite{C:ChiChuYam21}, $\epsilon$-ZK is still useful in several applications of ZK. 
The results in \cite{C:ChiChuYam21} suggest the possibility of post-quantum extractable commitments if we relax the simulation requirement on the extractor to a similar $\epsilon$-close\footnote{Throughout this paper, ``$\epsilon$-close'' means that the adversary's distinguishing advantage is at most $\epsilon$.} version. We will refer to this weakened notion as \emph{extractability with $\epsilon$-simulation}.\footnote{In the main body, we call it \emph{strong} extractability with $\epsilon$-simulation since we also define a weaker variant of that.} 
It seems natural to hope that the techniques in \cite{C:ChiChuYam21} could be used in the context of extractable commitments.
Indeed, by plugging the ZK argument from \cite{C:ChiChuYam21} into the OT-based construction~\cite{STOC:BitShm20,BLS21,C:HalSmiSon11,AFRICACRYPT:LunNie11}, we can obtain a non-black-box construction of constant-round post-quantum extractable commitments with $\epsilon$-simulation, assuming constant-round post-quantum OTs.
However, if we focus on \emph{black-box constructions} from the \emph{minimal assumption} of PQ-OWFs, it is unclear if the techniques in \cite{C:ChiChuYam21} would help.  
Therefore, we ask the following question:
\begin{quote}
{\bf Question 1:} \emph{Is it possible to  have  constant-round  post-quantum extractable commitments with $\epsilon$-simulation, assuming only black-box access to PQ-OWFs?
}
\end{quote}

In the more general context of 2PC and MPC, the implication of \cite{FOCS:CCLY21} is that to obtain constant-round constructions with post-quantum security, we have to either 
\begin{enumerate}
    \item rely on non-black-box simulation, {\em or}
    \item aim for a relaxed security notion (e.g., $\epsilon$-close simulation security).
\end{enumerate}
The first approach was taken 
in 
\cite{EC:ABGKM21} (based on \cite{STOC:BitShm20}), leading to a constant-round post-quantum MPC protocol with non-black-box simulation.  
On the other hand, the second approach has not been explored in the existing literature of post-qauntum 2PC or MPC (except for the special case of ZK as in \cite{C:ChiChuYam21}). 
It is possible to construct constant-round post-quantum 2PC with $\epsilon$-close simulation by combining constant-round post-quantum semi-honest OTs and the constant-round post-quantum $\epsilon$-ZK in \cite{C:ChiChuYam21}. However, the naive approach will lead to a non-black-box construction.   
In contrast, 
in the classical setting, constant-round \emph{black-box constructions} of 2PC \cite{TCC:PasWee09} and MPC \cite{TCC:CDMW09,STOC:Goyal11} are known from the minimal assumption of constant-round semi-honest OT. 
The above discussion suggests that one has to relax the security requirement when  considering the post-quantum counterparts of these tasks. 
We will refer to 2PC and MPC with $\epsilon$-close simulation as $\epsilon$-2PC and $\epsilon$-MPC respectively. 
Then, an interesting question is:
\begin{quote}\label{quote:Q1}
{\bf Question 2:} \emph{Do there exist constant-round black-box post-quantum $\epsilon$-2PC and $\epsilon$-MPC, assuming only constant-round semi-honest OTs secure against QPT adversaries?
}
\end{quote}

\subsection{Our Results}
\begin{table}[t]
\setlength\tabcolsep{0.5eM}
\begin{center}
\begin{minipage}[c]{\textwidth} \scriptsize
\begin{center}
\begin{threeparttable}
\caption{Comparison of Quantumly Secure Extractable Commitment. }
\label{tbl:comparison}
\begin{tabular}{lllllll}
\toprule
\multicolumn{1}{c}{Reference} 
&\multicolumn{1}{c}{\#Round}
&\multicolumn{1}{c}{Cla. Const.}
&\multicolumn{1}{c}{BB Const.}
&\multicolumn{1}{c}{BB Ext.}
&\multicolumn{1}{c}{Siml. Err.} &\multicolumn{1}{c}{Assumption}\\
\midrule
\cite{EC:GLSV21}&$\poly(\secpar)$ &  &&\checkmark& $\negl$ & OWF \\\hline
\cite{C:BCKM21b}&$\poly(\secpar)$ & &\checkmark &\checkmark& $\negl$ & OWF \\\hline
\cite{STOC:BitShm20}&$O(1)$ &\checkmark& & & $\negl$ & QFHE+QLWE \\\hline
folklore\footnote{As noted in \cite{BLS21}, the construction is implicit in  \cite{STOC:BitShm20,C:HalSmiSon11,AFRICACRYPT:LunNie11}.}&$\poly(\secpar)$ &\checkmark& &\checkmark & $\negl$ & OT \\\hline
folklore+\cite{C:ChiChuYam21}&$O(1)$ &\checkmark& &\checkmark & $\epsilon$ &$O(1)$-round OT \\\hline
Ours&$O(1)$&\checkmark & \checkmark &\checkmark& $\epsilon$ & OWF \\
\bottomrule\\
\end{tabular}
The ``Cla. Const.", ``BB Const.", and ``BB Ext." columns indicate if the scheme relies on 
classical constructions, 
black-box constructions, and extraction, respectively. 
In the ``Siml. Err." column, $\negl$ and $\epsilon$ mean that the construction achieves the standard quantum extractability and quantum extractability with $\epsilon$-simulation, respectively.
In ``Assumption" column, QFHE and QLWE means quantum fully homomorphic encryption and the quantum hardness of learning with errors, respectively. 
\end{threeparttable}
\end{center}
\end{minipage}
\end{center}
\end{table}

We answer {\bf Question 1} affirmatively and address {\bf Question 2} partially, showing a positive answer only for the two-party case. 
We first construct  constant-round black-box post-quantum extractable commitments with $\epsilon$-simulation from PQ-OWFs. 
See \Cref{tbl:comparison} for comparisons among quantumly secure extractable commitments. 
Such commitments imply new constant-round and black-box protocols for general-purpose 2PC secure against QPT adversaries. In particular, we get
\begin{itemize}
    \item post-quantum $\epsilon$-2PC from semi-honest OTs, and
    \item post-quantum $\epsilon$-2PC from PQ-OWFs, assuming that quantum communication is possible. (Henceforth, we will use OWFs to denote PQ-OWFs.)
\end{itemize}

As an intermediate tool to achieve the above results, we construct a constant-round post-quantum $\epsilon$-ZK commit-and-prove, assuming only black-box access to OWFs. 
Black-box zero-knowledge commit-and-prove \cite{STOC:IKOS07,FOCS:GLOV12,STOC:GOSV14,TCC:HazVen18,TCC:KhuOstSri18,C:Kiyoshima20} is a well-studied primitive in classical cryptography. It enables a prover to commit to some message and later to prove in zero-knowledge that the committed message satisfies a given predicate in a {\em black-box} manner. In addition to being secure in the post-quantum setting, our construction enjoys the extra property that the commit stage is extractable (albeit with only $\epsilon$-simulation of the adversary's post-extraction state). Such a constant-round $\epsilon$-simulatable ExtCom-and-Prove protocol implies the following set of two-party protocols:
\begin{itemize}
    \item constant-round black-box post-quantum coin-flipping with $\epsilon$-simulation,
    \item constant-round black-box post-quantum $\epsilon$-ZK arguments of knowledge for $\NP$ with $\epsilon$-simulating knowledge extractor, and
    \item constant-round black-box $\epsilon$-ZK arguments for $\QMA$. 
\end{itemize}
In the following, we provide more discussion about them.

\paragraph{Coin-Flipping.}
 Coin-flipping is a two-party protocol used to generate a uniformly random string that cannot be biased by either of parties (w.r.t.\ the standard simulation-based security). In the classical setting, constant-round black-box constructions from OWFs are known \cite{TCC:PasWee09}. On the other hand, known post-quantum constructions are based on stronger assumptions (like QLWE) than OWFs, and require either 
polynomial rounds  \cite{AFRICACRYPT:LunNie11} or 
non-black-box simulation \cite{EC:ABGKM21}.   
Our construction can be understood as the post-quantum counterpart of the classical construction by Pass and Wee \cite{TCC:PasWee09}, albeit with $\epsilon$-simulation. 

\paragraph{Arguments of Knowledge with Simulating Extractor.}
Arguments of knowledge intuitively require an extractor to extract a witness from any efficient malicious prover whenever it passes the verification. 
In the classical setting, constant-round black-box constructions from OWFs are known \cite{TCC:PasWee09}. 
In the post-quantum setting, there are two existing notions of arguments of knowledge depending on whether we require the extractor to simulate the prover's post-execution state or not.  
For the ``without-simulation'' version, Unruh \cite{EC:Unruh12} gave a polynomial-round black-box construction from OWFs.\footnote{Though Unruh originally assumes \emph{injective} OWFs, \cite{C:ChiChuYam21} pointed out that any OWF suffices.}  
For the ``with-simulation'' version, all existing constructions require both polynomial rounds {\em and} assumptions stronger than OWFs (like QLWE) \cite{C:HalSmiSon11,AFRICACRYPT:LunNie11,ananth2021concurrent}.\footnote{Though not claimed explicitly, it seems also possible to obtain constant-round construction with non-black-box simulation from QLWE and QFHE based on \cite{STOC:BitShm20}.} 
Our construction improves both the round complexity and the required assumption, at the cost of weakening ZK and extractability to their $\epsilon$-simulation variants.
On the other hand, we note that the construction in \cite{ananth2021concurrent} achieves \emph{proofs of knowledge}, while ours only achieves \emph{arguments of knowledge}. 
We also note that even without knowledge extractability, our construction improves the construction in \cite[Section 6]{C:ChiChuYam21}, which is a \emph{non-black-box construction} of constant-round $\epsilon$-ZK arguments for $\NP$ from OWFs.  

\paragraph{ZK Arguments for $\QMA$.}
$\QMA$ is a quantum analog of $\NP$. 
Known constructions of ZK proofs or arguments for $\QMA$ rely on either polynomial-round communication \cite{BJSW20,FOCS:BroGri20,BraYue} or non-black-box simulation \cite{STOC:BitShm20}. 
If we relax the ZK requirement to $\epsilon$-ZK, constant-round black-box $\epsilon$-ZK {\em proofs} were already constructed in \cite{C:ChiChuYam21}; but that construction needs to assume collapsing hash functions, which are stronger than OWFs. 
Our construction improves the assumption to the existence of OWFs at the cost of weakening the soundness to the computational one (i.e., an argument system).  

\subsection{Discussion}

\paragraph{Minimality of Assumptions.}
We can show that OWFs and semi-honest OTs are the minimal assumptions for post-quantum extractable commitments with $\epsilon$-simulation and $\epsilon$-2PC, respectively. 
The former is straightforward (since any computationally-hiding and statistically-binding commitments without extractability already imply OWFs), but the latter needs more explanations. 
First, we note that $\epsilon$-2PC trivially implies $\epsilon$-simulatable semi-honest OTs because the latter is a special case of the former. Next, we remark that $\epsilon$-simulatable semi-honest security implies the standard semi-honest indistinguishability-based security (with negligible distinguishing advantage). 
This is a special case of a folklore that $\epsilon$-simulation security suffices for indistinguishability-based applications. The reason is that we can set the simulation error $\epsilon$ after an adversary's distinguishing advantage $\delta$ is fixed so that $\epsilon \ll \delta$, because the simulator only appears in the security proof for the indistinguishability-based security.
Finally, we remark that semi-honest indistinguishability-based security implies semi-honest simulation-based security. In summary, $\epsilon$-2PC implies standard semi-honest OTs. 

We also remark that, as observed in \cite{C:BCKM21b,EC:GLSV21}, it is unclear if OWFs are necessary for \emph{quantum constructions} of 2PC/MPC, and it may be possible to construct them based on a weaker assumption.

\paragraph{Other Potential Applications.}
A recent work by Bitansky, Lin, and Shmueli \cite{BLS21} gave a generic construction of post-quantum non-malleable commitments from post-quantum extractable commitments. 
We believe that our $\epsilon$-simulatable extractable commitments can be used in their construction as a building block. 
It is reasonable to expect that we can prove the standard non-malleability as defined in \cite{BLS21} even though we start from $\epsilon$-simulatable extractability; This is because non-malleability is an indistinguishability-based security and $\epsilon$-simulation usually suffices for indistinguishability-based applications as mentioned in the previous paragraph. 
If the above idea works, we will get $\log^{\star}(\secpar)$-round post-quantum non-malleable commitments solely from OWFs. 
\cite{BLS21} obtains their $\log^{\star}(\secpar)$-round protocol based on much stronger assumptions of QFHE and QLWE; they also presents
a polynomial-round classical (resp.\ quantum) protocol based on OTs (resp.\ OWFs).
While our intuition above is plausible, a formal proof of it is out of scope of this work, and thus is left for future work.

\begin{takashienv}{open problems, hidden}
\paragraph{Open Problems.}
We leave the following open problems.
\begin{itemize}
    \item Can we extend our results on 2PC to MPC? We note that constant-round black-box MPC is known in the classical setting \cite{STOC:Goyal11}. 
    \item Can we extend our result on the quantum construction of 2PC to support quantum functionalities? 
    We note that polynomial-round construction is possible by combining our result and \cite{EC:DGJMS20}. 
    \item Can we construct a (not necessarily constant-round or black-box) post-quantum extractable commitment (rather than the $\epsilon$-simulation version) from OWF?
    Such a construction is known from OT, but not known soley from OWF. 
    We remark that the impossibility of \cite{FOCS:CCLY21} does not apply to super-constant-round constructions. 
\end{itemize}
\end{takashienv}

\subsection{Concurrent Work}\label{sec:concurrent}
A concurrent work by Lombardi, Ma, and Spooner \cite{LMS21} observed that the impossibility of~\cite{FOCS:CCLY21} implicitly assumed a computational model for simulators, which they call measured-runtime expected quantum polynomial time ($\mathsf{EQPT}_m$), and   
showed that the impossibility can be circumvented if we consider a stronger  model  called coherent-runtime expected quantum polynomial time ($\mathsf{EQPT}_c$).
Roughly speaking, the difference between $\mathsf{EQPT}_m$ and $\mathsf{EQPT}_c$ is that the latter allows the simulator to coherently run multiple computations with different runtimes so that they can interfere with each other.  
(See~\cite{LMS21} for more details.) 
Then, they construct constant-round post-quantum zero-knowledge proofs (or arguments) and extractable commitments with $\mathsf{EQPT}_c$ simulators and extractors. 
Though ZK with $\mathsf{EQPT}_c$ simulators is weaker than the standard ZK with polynomial-time simulators, they show that it implies $\epsilon$-ZK. 
Therefore, their notion of ZK  with $\mathsf{EQPT}_c$ simulation  lies between the standard ZK and $\epsilon$-ZK. 
Their formalization of $\mathsf{EQPT}_c$ simulation is very interesting as this enables us to reduce the simulation error to be negligible, which makes the state of affairs be similar to the classical case.

On the other hand, we believe that the gap between  ZK with $\mathsf{EQPT}_c$ simulation and $\epsilon$-ZK is not too large from the perspective of (theoretical) applications: 
The definition of $\mathsf{EQPT}_c$ models a simulator as a quantum circuit of a superpolynomial sizes with a certain property, and in particular, if we need a polynomial-time simulator, we must truncate the simulation after running polynomially many steps.  
However, in that case, there occurs a noticeable simulation error, which is similar to $\epsilon$-ZK. 
For this reason, we do not find any application for which ZK with $\mathsf{EQPT}_c$ simulation suffices but $\epsilon$-ZK does not. 
We make a similar observation on extractability with $\mathsf{EQPT}_c$ extraction and extractability with $\epsilon$-close simulation as well. 

Below, we give a comparison of results of~\cite{LMS21} and our work.
Among many others, \cite{LMS21} constructed constant-round post-quantum extractable commitments with $\mathsf{EQPT}_c$ extraction assuming super-polynomially secure OWFs or polynomially secure collapse-binding commitments. 
Though they achieve a stronger notion of extractability than ours as explained above, they rely on non-black-box constructions and stronger assumptions than the polynomial hardness of OWFs. 
On the other hand, the main focus of this work is black-box constructions from the minimal assumption of the polynomial hardness of OWFs (or OTs for 2PC).  
There are similar advantages and disadvantages for all constructions given in~\cite{LMS21}. 
Thus, the results of~\cite{LMS21} are incomparable to ours. 

A natural question that arises from the above comparison is if we can obtain protocols that take advantages of both works. That is, can we construct constant-round black-box post-quantum extractable commitments (resp. 2PC) with $\mathsf{EQPT}_c$ extraction (resp. simulation) and negligible simulation errors assuming only the minimal assumption of polynomially secure OWFs (resp. OTs)?   
This might be achieved by combining the techniques of this paper and~\cite{LMS21}.
We leave it as an interesting future work. 








\section{Technical Overview}
We give technical overview for our results on extractable commitments, ExtCom-and-Prove, and 2PC. 
The other applications claimed before follows from our ExtCom-and-Prove protocol via rather standard techniques. Therefore, we refer the reader to \Cref{sec:e-extcom-n-prove} for corresponding constructions.

\subsection{Extractable Commitment with $\epsilon$-Simulation}\label{sec:overview_extcom}


The main technical tool for constructing  extractable commitments with $\epsilon$-simulation is a generalization of the recent extract-and-simualate technique of \cite{C:ChiChuYam21}. 

\paragraph{Extract-and-Simulation Lemma in \cite{C:ChiChuYam21}.}
We briefly recall the extract-and-simulate lemma shown in \cite[Lemma 4.2]{C:ChiChuYam21}.\footnote{In \cite{C:ChiChuYam21}, the lemma was called ``extraction lemma". Here, we add ``simulation" to emphasize that the extractor not only extracts but also simulates the adversary's state.}
At a high level, that lemma can be interpreted as follows.\footnote{There are two versions of their lemma: the statistically-binding case and the strong collapse-binding case. The abstraction given here is a generalization of the statistically-binding case.}
Let $\A$ be a quantum algorithm with an initial state $\rho$.
Suppose that $\A$ outputs some unique classical string $s^*$  or otherwise outputs a failure symbol $\fail$. 
Then, there exists a simulation-extractor $\SimExt$ such that for any noticeable function $\epsilon$ (on the security parameter), the following two experiments are $\epsilon$-close:

\smallskip
\begin{tabular}{l|l}
    \begin{minipage}[t]{0.40\textwidth}
\underline{$\Exp_{\mathsf{real}}$}\\
~\vspace{1.4mm}\\
\text{Run} $\A(\rho)$,\\
\emph{If} $\A$ outputs $\fail$,\\
$~~~~$\emph{Output} $\fail$\\
\emph{Else output} $\A$\text{'s~final~state}.
    \end{minipage} 
        &~
    \begin{minipage}[t]{0.55\textwidth}
\underline{$\Exp_{\mathsf{ext}}$}\\
$(s_\ext,\rho_\ext)\gets \SimExt^{\A(\rho)}(1^{\epsilon^{-1}})$\\
\text{Run} $\A(\rho_\ext)$,\\
\emph{If} $\A$ outputs $\fail$ ~$\lor$~$s_\ext\neq s^*$,\\
$~~~~$\emph{Output} $\fail$\\
\emph{Else output} $\A$\text{'s~final~state}.
    \end{minipage}
    \end{tabular}



\paragraph{Generalizing the Lemma.}
Note that their lemma will enable us to extract $s^*$ from $\A$ {\em only if $\Adv$ reveals the value $s^*$ at the end}. As shown in \cite{C:ChiChuYam21}, 
this already suffices for the constant-round ZK proof by Goldreich and Kahan \cite{JC:GolKah96}, where the verifier first commits to the challenge and opens it (i.e., ``reveals it at the end'') later. 
However, this does not seem to help obtain extractable commitments, because the committed message is not revealed {\em at the end the commit stage} (i.e., before decommitment happens); but the definition of extractable commitments does require extraction before decommitment happens.


To deal with this issue, we generalize the \cite{C:ChiChuYam21} lemma as follows.
Let 
$\A$ be a quantum algorithm that on an initial state $\rho$, outputs a classical symbol $\mathsf{Succ}$ or $\fail$.
Moreover, suppose that 
there are a unique classical string $s^*$ and 
a ``simulation-less extractor" $\ext_{\simless}$ that outputs $s^*$, or otherwise $\fail$. Also, suppose that 
\begin{equation}\label[Inequality]{eq:Unruh-guarantee}
    \Pr[\ext_{\simless}^{\A(\rho)}=s^*]\geq \left(\Pr[\A(\rho)=\mathsf{Succ}]\right)^c-\negl(\secpar)
\end{equation}
    for some constant $c$. 
Our generalized lemma says that the $\epsilon$-closeness between  $\Exp_{\mathsf{real}}$ and $\Exp_{\mathsf{ext}}$ holds in this setting as well.   
   
One can think of $\A$ as a joint execution of a malicious committer and honest receiver where it outputs $\mathsf{Succ}$ if and only if the receiver accepts. 
In this setting, one can understand the above lemma as a lifting lemma from ``simulation-less extractor" to ``$\epsilon$-simulation extractor" in the setting where the extracted string is unique.  
In the main body, we present the lemma in a more specific form (\Cref{lem:extract_and_simulate}), where it is integrated with 
Watrous' rewinding lemma \cite{SIAM:Watrous09} and
Unruh's rewinding lemma \cite{EC:Unruh12}, because that is more convenient for our purpose. 
We will overview the  intuition behind the above generalized lemma toward the end of this subsection. 



\paragraph{Weakly Extractable Commitment.}
Next, we explain how to construct post-quantum extractable commitments using our extract-and-simulate lemma. 
We go through the following two steps: 
\begin{enumerate}
    \item \label[Step]{step:weak_ext}
    Construct a commitment scheme $\wExtCom$ that satisfies a weak version of post-quantum extractability with $\epsilon$-simulation. 
    \item \label[Step]{step:strong_ext}
    Upgrade $\wExtCom$ into a scheme $\ExtCom$ with full-fledged post-quantum extractability with $\epsilon$-simulation (which we call \emph{strong} extractability with $\epsilon$-simulation to distinguish it from the weak one).
\end{enumerate}

We first explain \Cref{step:weak_ext}, the construction of $\wExtCom$. 
Actually, our construction of $\wExtCom$ is exactly the same as the classical extractable commitments from OWFs given in \cite{TCC:PasWee09}, which are in turn based on earlier works \cite{SIAM:DolDwoNao00,FOCS:PraRosSah02,TCC:Rosen04}. 
Let $\Com$ be a computationally-hiding and statistically-binding commitment scheme (say, Naor's commitment \cite{JC:Naor91}).
Then, the commitment scheme $\wExtCom$ works as follows.
\begin{description}
    \item[Commit Stage:]~
    \begin{enumerate}
    \item \label[Step]{step:overview_commit}
    To commit to a message $m$, the committer $C$ generates $k=\omega(\log \secpar)$ pairs of $2$-out-of-$2$ additive secret shares $\{(v_i^0,v_i^1)\}_{i=1}^{k}$, i.e., they are uniformly chosen conditioned on that $v_i^0\oplus v_i^1=m$ for each $i\in [k]$.  
    Then, $C$ commits independently to each $v_{i}^{b}$ ($b \in \bits$) in parallel by using $\Com$. We denote these commitments by $\{(\com_i^0,\com_i^1)\}_{i=1}^{k}$. 
    \item
    \label[Step]{step:overview_challenge}
    $R$ randomly chooses $\Vect{c}=(c_1,...,c_k)\pick \bit^k$ and sends it to $C$.
    \item 
      \label[Step]{step:overview_response}
    $C$ decommits $\{\com_{i}^{c_i}\}_{i=1}^{k}$ to  $\{v_{i}^{c_i}\}_{i=1}^{k}$, and $R$ checks that the openings are valid.
    \end{enumerate}
     \item[Decommit Stage:]~
    \begin{enumerate}
    \item $C$ sends $m$ and opens all the remaining commitments; $R$ checks that all openings are valid {\em and} $v_i^0\oplus v_i^1=m$ for all $i\in [k]$.
    \end{enumerate}
\end{description}

Suppose that a malicious committer $C^*$ generates commitments $\{(\com_i^0,\com_i^1)\}_{i=1}^{k}$ in \Cref{step:overview_commit}, and let $\rho$ be its internal state at this point. 
Then, we consider $\A(\rho)$ that works as follows:
\begin{itemize}
    \item Choose  $\Vect{c}=(c_1,...,c_k)\gets \bit^k$ at random.
    \item 
    Send $\Vect{c}$ to $C^*$ and
    simulate \Cref{step:overview_response} of $C^*$ in the commit stage to get $\{v_{i}^{c_i}\}_{i=1}^{k}$ and the corresponding decommitment information.
    \item 
    If all the openings are valid, output $\mathsf{Succ}$; otherwise output $\fail$. 
\end{itemize}

To use our extract-and-simulate lemma, we need to construct a simulation-less extractor $\ext_{\simless}$ satisfying \Cref{eq:Unruh-guarantee}. 
A natural idea is to use Unruh's rewinding lemma~\cite{EC:Unruh12}. 
His lemma directly implies that if $\A$ returns $\mathsf{Succ}$ with probability $\delta$, then we can obtain valid $\{v_{i}^{c_i}\}_{i=1}^{k}$ and $\{v_{i}^{c'_i}\}_{i=1}^{k}$ for two uniformly random challenges, $\Vect{c} = (c_1, \ldots, c_k)$ and $\Vect{c}' = (c'_1, \ldots, c'_k)$, with probability at least $\delta^3$. 
In that case, unless $\Vect{c}=\Vect{c}'$ (which happens with negligible probability), we can ``extract"  $ m = v_{i}^{0}\oplus v_{i}^{1}$ from position $i \in [k]$ that satisfies $c_i\neq c'_i$. 
However, such an ``extractor" does not satisfy the assumption for our generalized extract-and-simulate lemma in general, because $v_{i}^{0}\oplus v_{i}^{1}$ may be different for each $i \in [k]$. 

Therefore, to satisfy this requirement, we have to introduce an additional assumption that $\{(\com_i^0,\com_i^1)\}_{i=1}^{k}$ is {\em consistent}, i.e., if we denote the corresponding committed messages as $\{(v_i^0,v_i^1)\}_{i=1}^{k}$, then there exists a unique $m$ such that $v_i^0\oplus v_i^1=m$ for all $i\in [k]$.\footnote{The corresponding message is well-defined (except for negligible probability) since we assume that $\Com$ is statistically binding.}  
With this assumption, we can apply our generalized extract-and-simulate lemma. 
It enables us to extract the committed message {\em and} simultaneously $\epsilon$-simulate $C^*$'s state, conditioned on that the receiver accepts in the commit stage. 
The case where the receiver rejects can be easily handled using Watrous' rewinding lemma \cite{SIAM:Watrous09} as we will explain later.
As a result, we get an $\epsilon$-simulating extractor that works well conditioned on that the commitments generated in \Cref{step:overview_commit} are consistent. 
We will refer to such a weak notion of simulation-extractability as {\em weak extractability with $\epsilon$-simulation} (see \Cref{def:epsilon-sim-ext-com:weak} for the formal definition).

Moreover, since Unruh's rewinding lemma naturally gives a simulation-less extractor in the parallel setting (where $C^*$ interacts with many copies of $R$ in parallel), we can prove the parallel version of the weak extractability with $\epsilon$-simulation similarly.
More generally, we prove that $\wExtCom$ satisfies a further generalized notion of extractability which we call the {\em special parallel weak extractability with $\epsilon$-simulation} (see \Cref{def:epsilon-sim-ext-com:parallel_weak} for the formal definition). 
Roughly speaking, it requires an $\epsilon$-simulating extractor to work in $n$-parallel execution as long as the commitments in some subset of $[n]$ are consistent and the committed messages in those sessions determine a unique value. 
This parallel extractability will play an important role  in the weak-to-strong compiler which we discuss next.





\paragraph{Weak-to-Strong Compiler.}
The reason why we cannot directly prove that 
$\wExtCom$ satisfies the strong extractability with $\epsilon$-simulation is related to an issue that is often referred to as \emph{over-extraction} in the classical literature (e.g., \cite{FOCS:GLOV12,EC:GGJS12,C:Kiyoshima14}). 
Over-extraction means that an extractor may extract some non-$\bot$ message from an invalid commitment, instead of detecting the invalidness of the commitment.  
In particular, there does not exist a unique ``committed message" when the commitment is ill-formed in $\wExtCom$, and extraction of such a non-unique message may collapse the committer's state.
To deal with this issue, we have to add some mechanism which enables a receiver (and thus the extractor) to detect invalidness of the commitment.  

One possible approach is to revisit the techniques developed in the classical setting, performing necessary surgery to make the proof work against QPT adversaries. However, as demonstrated by the above cited works, existing techniques in the classical setting are already delicate; even if it would work eventually, such a non-black-box treatment would further complicate the proof undesirably.
Therefore, we present an alternative approach that deviates from existing ones in the classical setting; as we will show later, this new approach turns to be quantum-friendly.


Roughly speaking, our construction $\ExtCom$ works as follows: 
\begin{description}
    \item[Commit Stage:]~
    \begin{enumerate}
    \item \label[Step]{step:overview_strong_commit}
    The committer $C$ generates shares $\{\view_i\}_{i=1}^{n}$ of a {\em verifiable secret sharing} ($\VSS$) scheme (see \Cref{def:VSS}) of the message to be committed to, and then commits to each $\view_i$ using $\wExtCom$ separately in parallel.
    \item \label[Step]{step:overview_strong_commit:coin-flipping}
    $C$ and the receiver $R$ jointly run a ``one-side simulatable" coin-flipping protocol based on $\wExtCom$ to generate a random subset $T$ of $[n]$ of a certain size.\footnote{We remark that it is a non-trivial task to construct constant-round two-party coin-flipping from OWFs in the quantum setting, achieving the (even $\epsilon$-)simulation-based security {\em against both parties}.
    Indeed, that will be one application of the strongly extractable commitment with $\epsilon$-simulation, which we are now constructing. 
    However, this is not a circular reasoning. Here, we need simulation-based security only against a malicious receiver. For such a one-side simulatable coin-flipping, the weakly extractable commitment $\wExtCom$ (with $\epsilon$-simulation) suffices.}
    Specifically, they do the following:
    \begin{enumerate}
        \item $R$ commits to a random string $r_1$ by $\wExtCom$.
        \item $C$ sends a random string $r_2$ in the clear.
        \item $R$ opens $r_1$. Then, both parties derive the subset $T$ from $r_1\oplus r_2$.
    \end{enumerate}
    \item $C$ opens the commitments corresponding to the subset $T$, and $R$ checks their validness and consistency. 
    \end{enumerate}
     \item[Decommit Stage:]~
    \begin{enumerate}
    \item $C$ opens all the commitments. $R$ checks those openings are valid. If they are valid, $R$ runs the reconstruction algorithm of $\VSS$ to recover the committed message.
    \end{enumerate}
\end{description}
Using a similar argument as that for the soundness of the {\em MPC-in-the-head paradigm} \cite{STOC:IKOS08,STOC:Goyal11},\footnote{To avoid disturbing the current discussion, we will provide more details of this argument in \Cref{sec:overview:e-extcom-n-prove}, where it is used again to establish the security of our ExtCom-and-Prove.} we can show that if a malicious committer passes the verification in the commit stage, then: 
\begin{enumerate}
\item 
Most of the commitments of $\wExtCom$ generated in \Cref{step:overview_strong_commit} are valid {\em as a commitment}; {\bf and}\item
The committed shares in those valid commitments determines a {\em unique} message that can be recovered by the reconstruction algorithm of $\VSS$. 
\end{enumerate} 
Then, we can apply the special parallel weak extractability with $\epsilon$-simulation of $\wExtCom$ to show the strong extractability  with $\epsilon$-simulation of $\ExtCom$. 
We remark that essentially the same proof can also be used to show that the \emph{parallel} execution of $\ExtCom$ is still strongly extractable with $\epsilon$-simulation. We refer to this as the parallel-strong extractability with $\epsilon$-simulation; it will play a critical role in our construction of ExtCom-and-Prove (see \Cref{sec:overview:e-extcom-n-prove}).

\paragraph{Dealing with Rejection in Commit Stage.}
So far, we have only focused on the case where the receiver accepts in the commit stage.  
However, the definition of (both weak and strong) extractability requires that the final state should be simulated even in the case where the receiver rejects in the commit stage.
In this case, of course, the extractor does not need to extract anything, and thus the simulation is straightforward.
A non-trivial issue, however, is that the extractor does not know if the receiver rejects in advance. 
This issue can be solved by a technique introduced in \cite{STOC:BitShm20}. The idea is to just guess if the receiver accepts, and runs the corresponding extractor assuming that the guess is correct. This gives an intermediate extractor that succeeds with probability almost $1/2$ and its output correctly simulates the desired distribution conditioned on that it does not abort. 
Such an extractor can be compiled into a full-fledged extractor that does not abort by Watrous' rewinding lemma \cite{SIAM:Watrous09}.

\paragraph{Proof Idea for the Generalized Extract-and-Simulate Lemma.}
Finally, we briefly explain the idea for the proof of our generalized extract-and-simulate lemma. The basic idea is similar to the original extract-and-simulate lemma in \cite{C:ChiChuYam21}---Use Jordan's lemma to decompose the adversary's internal state into ``good" and ``bad" subspaces, and amplify the extraction probability in the good subspace while effectively ignoring the bad-subspace components.
However, the crucial difference is that in \cite{C:ChiChuYam21}, they define those subspaces with respect to the success probability of $\A$ whereas we define them with respect to the success probability of $\ext_{\simless}$. 
That is, for a noticeable $\delta$, 
we apply Jordan's lemma to define a subspaces $S_{<\delta}$ and $S_{\geq \delta}$ such that 
\begin{enumerate}
    \item \label{item:overview_ext_prob}
    When $\ext_{\simless}$'s input is in $S_{<\delta}$ (resp.\ $S_{\geq \delta}$), it succeeds in extracting $s^*$ with probability $<\delta$ (resp.\ $\geq \delta$).   
    \item \label{item:overview_amplification} Given a state in $S_{\geq \delta}$, we can extract $s^*$ with overwhelming probability within $O(\delta^{-1})$ steps.
\item \label{item:overview_no_interference}
The above procedure does not cause any interference between $S_{<\delta}$ and $S_{\geq \delta}$.
\end{enumerate}

We define $\SimExt$ to be an algorithm that runs the procedure in \Cref{item:overview_amplification} and outputs $s$ (which is supposed to be $s^*$ in the case of success) and the post-execution state of $\A$. 
First, we consider simpler cases where the initial state of the experiments is a pure state $\ket{\psi}$ that is in either $S_{\geq \delta}$ or $S_{< \delta}$.

\begin{description}
\item[Case of] {\bf $\ket{\psi}\in S_{\geq \delta}$:}
In this case, \Cref{item:overview_amplification} implies that $\SimExt$ outputs $s^*$ with overwhelming probability. In general, such an almost-deterministic quantum procedure can be done (almost) without affecting the state (e.g., see the {\em Almost-as-Good-as-New Lemma} in~\cite[Lemma 2.2]{Aar05}). 
Therefore, $\Exp_{\mathsf{real}}$ and $\Exp_{\mathsf{ext}}$ are negligibly indistinguishable in this case. 
~\\
\item[Case of] {\bf $\ket{\psi}\in S_{< \delta}$:}
For any state $\ket{\psi_{<\delta}}\in S_{<\delta}$, \Cref{item:overview_ext_prob} implies
$$
    \Pr[\ext_{\simless}^{\A(\ket{\psi_{<\delta}})}=s^*]\leq \delta.
    $$
On the other hand, our assumption (i.e., \Cref{eq:Unruh-guarantee}) implies
$$
    \Pr[\ext_{\simless}^{\A(\ket{\psi_{<\delta}})}=s^*]\geq \left(\Pr[\A(\ket{\psi_{<\delta}})=\mathsf{Succ}]\right)^c-\negl(\secpar)
    $$
    for some constant $c$. 
By combining them, we have 
$$
   \Pr[\A(\ket{\psi_{<\delta}})=\mathsf{Succ}]\leq \left(\delta+\negl(\secpar)\right)^{1/c}.
    $$
We note that the second output of $\SimExt$ in $\Exp_{\mathsf{ext}}$ is in $S_{<\delta}$ if the initial state is in $S_{<\delta}$ by \Cref{item:overview_no_interference}.
Therefore, if we run $\Exp_{\mathsf{real}}$ or $\Exp_{\mathsf{ext}}$ with an initial state in $S_{<\delta}$, it outputs $\fail$ with probability $>1-\left(\delta+\negl(\secpar)\right)^{1/c}$.\takashi{Mar.31 2022: I fixed a simple typo here. This was $\ge\left(\delta+\negl(\secpar)\right)^{1/c}$ before, which was obviously wrong.}
Recall that when an experiment outputs $\fail$, no information about the internal state of $\Adv$ is revealed. 
Thus, the distinguishing advantage between those experiments can be bounded by $O(\delta^{1/c})$. 
\end{description}

In general, the initial state is a superposition of $S_{<\delta}$ component and $S_{\geq \delta}$ component. Thanks to \Cref{item:overview_no_interference}, we can reduce the general case to the above two cases. When doing that, there occurs an additional loss of the $4$-th power of $\delta$ due to a technical reason (that appears in \Cref{lem:state-close}).
Still, we can bound the distinguishing advantage between the two experiments by $O(\delta^{1/(4c)})$. This can be made to be an arbitrarily small noticeable function because $\delta$ is an arbitrarily small noticeable function.
This suffices for establishing the $\epsilon$-closeness of those experiments.

\subsection{Black-Box $\epsilon$-Simulatable ExtCom-and-Prove}\label{sec:overview:e-extcom-n-prove}
Black-box zero-knowledge commit-and-prove allows a committer to commit to some message $m$ (the Commit Stage), and later prove in zero-knowledge that the committed $m$ satisfies some predicate $\phi$ (the Prove Stage). What makes this primitive non-trivial is the requirement of black-box use of cryptographic building blocks; otherwise, this task can be fulfilled easily by giving a standard commitment to $m$ first, and then running any zero-knowledge system over the commitment in a non-black-box manner.

\para{MPC-in-the-Head.} In the classical setting, black-box zero-knowledge commit-and-prove has been constructed following the so-called ``MPC-in-the-head'' paradigm \cite{STOC:IKOS07,FOCS:GLOV12}. To commit to $m$, the committer will imagine $n(\secpar)$ virtual parties ``in his head'', who jointly execute a {\em $(n,t)$-verifiable secret sharing} (VSS) scheme to share the message $m$. Roughly speaking, such a VSS scheme ensures that if only $\le t$ parties are corrupted, then all the honest parties will learn their shares properly and can always recover the $m$ by exchanging their shares if they want; however, the $\le t$ number of corrupted parties learns no information about $m$. Denote the views of the $n$ virtual parties during the VSS sharing stage execution as $\Set{\view_i}_{i\in[n]}$. To finish the Commit Stage, the committer commits to these $n$ views in parallel, using independent instances of a statistically-binding commitment $\Com$ for each view separately. 

To prove the committed $m$ satisfies a predicate $\phi$, the committer continues by asking the $n$ virtual parties to execute an $(n,t)$-secure MPC with $\Set{\view_i}_{i\in [n]}$ as their respective input; this MPC is executed for a functionality that collects all the views from each party, runs the VSS reconstruction algorithm to recover the value $m$, and finally outputs $\phi(m)$ to each party. Denote the views of the $n$ parties during this MPC execution as $\Set{\view'_i}_{i\in[n]}$. The committer commits to $\Set{\view'_i}_{i\in[n]}$ in parallel, using independent instances of a statistically-binding commitment for each $\view'_i$ separately. Next, the committer and receiver will execute a coin-flipping protocol to determine a random size-$t$ subset $T\subseteq [n]$; the committer decommits to $\Set{\view_i}_{i \in T}$ and $\Set{\view'_i}_{i \in T}$, and the receiver accepts if and only if these views are {\em consistent} w.r.t.\ the VSS and MPC execution. Roughly speaking, this means that for each pair $i, j \in T$, $\view_i$ and $\view_j$ (resp.\ $\view'_i$ and $\view'_j$) contain consistent incoming/outgoing messages, and they are the messages computed following the honest parties' algorithm w.r.t.\ the VSS (resp.\ the MPC) protocol.

To see why this construction is zero-knowledge, observe that a simulator $\Sim$ can pre-decide a size-$t$ set $\tilde{T}\subseteq[n]$ and commit to ``fake'' consistent views for both the VSS and MPC execution for parties in set $\tilde{T}$, {\em without knowing the actual value $m$}. This can be done because both the VSS and MPC reveals no information if only $t$ views are leaked.  For the views in $[n] \setminus \tilde{T}$, $\Sim$ simply commits to all-0 strings of proper length. Then, $\Sim$ will bias the coin-flipping result to $\tilde{T}$ using the simulator for the coin-flipping protocol against the malicious receiver. In this way, only the faked views in the set $\tilde{T}$ need to be revealed, and they will pass the receiver's consistency check as they are faked to be consistent.  

Soundness can be proven as follows. Due to the security property of the VSS and MPC, corrupting $\le t$ parties during the execution will not change the value $\phi(m)$ learned by the $\ge n-t$ honest parties. Therefore, to lie about $\phi(m)$, there must be $> t$ number of virtual parties being corrupted. However, this will necessarily yield many inconsistent pairs of views. Since the coin-flipping step determines a size-$t$ (pseudo-)random subset $T \subseteq [n]$ for consistency check, the inconsistent views will be caught by the receiver, except  with negligible probability (by setting $n$ and $t$ properly). We remark that the actual proof for soundness requires a more involved argument to formalize the above intuition (See \Cref{lemma:e-ExtCom-n-Prove:soundness} for details). 

From the above discussion, it should be clear that the coin-flipping must achieve simulation-based security against corrupted receivers (for zero-knowledge), while it only needs to be indistinguishability-based (IND) secure against corrupted committer (for soundness).

\para{Our Construction.} Our construction follows the above paradigm with the following modifications. To make the commitment stage extractable, we ask the committer to use the $\epsilon$-simulatable (strongly) extractable commitment $\algo{ExtCom}$ we constructed in \Cref{sec:overview_extcom}, in place of $\algo{Com}$. As mentioned in \Cref{sec:overview_extcom}, $\algo{ExtCom}$ maintains its $\epsilon$-simulatable extractability even when used in parallel; therefore, if suffices for the committer's parallel commitments to the views. For a malicious QPT committer, the parallel-strong extractability of $\ExtCom$ allows us to extract the committed value while performing a $\epsilon$-simulation for the committer's post-extraction state. Thus, we obtain the $\epsilon$-simulatable extractability of the Commit Stage.

To obtain soundness and $\epsilon$-ZK against QPT adversaries, we only need to build a coin-flipping that is {\em  $\epsilon$-simulatable against QPT malicious receiver, and IND-secure against QPT malicious committer}, to replace the classical coin-flipping protocol used to determine $T$. Once we have such a coin-flipping, soundness and $\epsilon$-ZK can be established as in the above classical setting.

We construct such a coin-flipping using the $\algo{wExtCom}$ constructed in \Cref{sec:overview_extcom} (note that we do not need the strongly extractable $\algo{ExtCom}$). This protocol is the same as \Cref{step:overview_strong_commit:coin-flipping} of our $\algo{wExtCom}$-to-$\algo{ExtCom}$ compiler: 
\begin{enumerate}
\item
The receiver uses $\algo{wExtCom}$ to commit to a random string $r_1$; 
\item
The committer samples and sends a random string $r_2$; 
\item 
The receiver decommits to $r_1$. The coin-flipping result is determined as $r\coloneqq r_1 \xor r_2$, which determines the size-$t$ subset $T$. 
\end{enumerate}
The pseudo-randomness of $r$ against the corrupted committer follows from the computationally-hiding property of $\algo{wExtCom}$. This allows us to prove soundness. To simulate for a corrupted receiver (i.e., to bias the coin-fliping result to a random $\tilde{r}$ that $\Sim$ obtained from the ideal-world functionality), $\Sim$ simply runs the $\epsilon$-simulation extractor to extract the string $r^*_1$ committed in $\algo{wExtCom}$ by the malicious receiver, and set $r_2 \coloneqq \tilde{r}\xor r^*_1$. Note that the extractor is also a $\epsilon$-simulator for the post-extraction state; thus, we obtain $\epsilon$-zero-knowledge.

\subsection{Black-Box $\epsilon$-Simulatable 2PC}

At a high level, our construction consists of the following 3 steps:
\begin{enumerate}
    \item \label[Step]{item:overview:e-2PC:blueprint:1}
    There exists a known constant-round black-box compiler {\em in the post-quantum setting}, which converts bounded-parallel string-OTs between two parties to general-purpose 2PC \cite{C:IshPraSah08}.
    \item \label[Step]{item:overview:e-2PC:blueprint:2}
    There also exists a constant-round black-box compiler in the post-quantum setting, which converts semi-honest bit-OTs to bounded-parallel string-OTs between two parties, in the $\Func^t_\textsc{so-com}$-hybrid model \cite{TCC:CDMW09,EC:GLSV21}. $\Func^t_\textsc{so-com}$ is a two-party ideal functionality formalized recently in \cite{EC:GLSV21}: it allows a committer to commit to an a-priori fixed polynomial number $t(\secpar)$ of messages in parallel, and later decommit to a subset of these commitments named by the receiver (thus, ``\textsc{so}'' stands for ``selectively opening''). 
\item \label[Step]{item:overview:e-2PC:blueprint:3}
   Therefore, all we need to do is to construct a constant-round black-box $\epsilon$-simulatable protocol that implements $\Func^t_\textsc{so-com}$, and then rely on the ``non-concurrent composition'' lemma in the stand-alone setting (developed in \cite{JC:Canetti00}) to obtain the bounded-parallel two-party string-OTs (and eventually 2PC via \Cref{item:overview:e-2PC:blueprint:1}). In the following, we show how to obtain such a realization of $\Func^t_\textsc{so-com}$ using the ExtCom-and-Prove from \Cref{sec:overview:e-extcom-n-prove}.
\end{enumerate}
\begin{remark}
\Cref{item:overview:e-2PC:blueprint:1,item:overview:e-2PC:blueprint:2} have been used to construct black-box and constant-round {\em stand-alone secure} 2PC (actually, even MPC) protocols in the classical setting \cite{TCC:PasWee09,TCC:CDMW09,STOC:Goyal11}. However, to our best knowledge, these two steps have not been explicitly and rigorously formalized {\em in the stand-alone setting}. There are some subtleties regarding the composition of protocols that require extra caution. One reason for the lack of a rigorous treatment is that the main focus of \cite{C:IshPraSah08,TCC:CDMW09} is the universally-composable (UC) model \cite{FOCS:Canetti01}. The recent work \cite{EC:GLSV21} made it explicit by formally defining and constructing bounded-parallel two-party string-OTs and the $F^t_\textsc{so-com}$ functionality. We provide further clarification of these subtleties in \Cref{sec:e-2pc:classical-recipe}.
\end{remark}

To construct an $\epsilon$-simulatable protocol for $\Func^t_\textsc{so-com}$, we ask the committer to commit to his $t$ messages $(m_1, \ldots, m_t)$ using the Commit Stage of the ExtCom-and-Prove. To decommit to the subset of messages determined by the receiver's challenge set $I \subseteq [n]$, the committer first sends $\Set{m_i}_{i\in I}$ to the receiver; then, both parties execute the Prove Stage, where the committer proves the following predicate:
\begin{equation}
\phi_{I, \Set{m_i}_{i\in I}}(x) = 
\begin{cases}
1 & \text{if}~ (x = m'_1 \| \ldots\| m'_t) \wedge (\forall i\in I, m'_i = m_i)\\
0 & \text{otherwise}
\end{cases}.
\end{equation} 
The proof of security is straightforward: if the committer is corrupted, simulation can be done via the $\epsilon$-simulatable extractability of the Commit Stage; if the receiver is corrupted, simulation can be done via the $\epsilon$-ZK property of the ExtCom-and-Prove.

Here is one caveat: in \Cref{item:overview:e-2PC:blueprint:3}, we actually need to rely on a variant of the non-concurrent composition lemma by Canetti \cite{JC:Canetti00}, which should:
\begin{enumerate}
    \item hold in the post-quantum setting. That is, it should hold for classical protocols secure against QPT adversaries; {\bf and}
    \item
    allow composition of {\em $\epsilon$-simulatable} protocols. Roughly speaking, it requires that if $\pi^\Func$ is a post-quantum $\epsilon$-simulatable protocol for an ideal functionality $\mathcal{G}$ in the $\Func$-hybrid model and $\phi$ is a post-quantum $\epsilon$-simulatable protocol for the ideal functionality $\Func$ in the plain model, then replacing the $\Func$ calls  in $\pi^\Func$ with $\phi$ yields a post-quantum $\epsilon$-simulatable protocol for $\mathcal{G}$ in the plain model,  as long as the execution of $\phi$ is not interleaved with other parts of $\pi^{(\cdot)}$.
\end{enumerate}
Fortunately, these requirements can be obtained by generalizing the proof of the non-concurrent composition lemma in \cite{JC:Canetti00} to the $\epsilon$-simulatable security against QPT adversaries. We present a formal treatment of it in \Cref{sec:e-non-concurrent-composition}.

\section{Preliminaries}
\paragraph{Basic Notations.}
We denote by $\secpar$ the security parameter throughout the paper.
For a positive integer $n\in\mathbb{N}$, $[n]$ denotes the set $\{1,2,...,n\}$.
For a finite set $\calX$, $x\sample \calX$ means that $x$ is uniformly chosen from $\calX$.

A function $f:\mathbb{N}\ra [0,1]$ is said to be \emph{negligible} if for all polynomial $p$ and sufficiently large $\secpar \in \mathbb{N}$, we have $f(\secpar)< 1/p(\secpar)$; it is said to be \emph{overwhelming} if $1-f$ is negligible, and said to be \emph{noticeable} if there is a polynomial $p$ such that $f(\secpar)\geq  1/p(\secpar)$ for sufficiently large $\secpar\in \mathbb{N}$.
We denote by $\poly$ an unspecified polynomial and by $\negl$ an unspecified negligible function.

We use PPT and QPT to mean (classical) probabilistic polynomial time and quantum polynomial time, respectively.
For a classical probabilistic or quantum algorithm $\A$, $y\sample \A(x)$ means that $\A$ is run on input $x$ and outputs $y$.
An adversary (or malicious party) is modeled as a non-uniform QPT algorithm $\A$ (with quantum advice) that is specified by a sequence of polynomial-size quantum circuits and quantum advices $\{\A_\secpar, \rho_\secpar\}_{\secpar\in\mathbb{N}}$.
In an execution with the security parameter $\secpar$, $\A$ runs $\A_{\secpar}$ taking $\rho_\secpar$ as the advice. 
We often omit the index  $\secpar$ and just write $\A(\rho)$ to mean a non-uniform QPT algorithm specified by  $\{\A_\secpar, \rho_\secpar\}_{\secpar\in \mathbb{N}}$
for simplicity. 


We use the bold font (like $\regX$) to denote quantum registers. 
For a quantum state $\rho$, $\|\rho\|_{tr}$ denotes the trace norm of $\rho$.

\subsection{Quantum Computation}
\paragraph{Interactive Quantum Machines and Oracle-Aided Quantum Machines.}
We rely on the definition of interactive quantum machines and oracle-aided quantum machines that are given oracle access to an interactive quantum machine, following \cite{EC:Unruh12}.
Roughly, an interactive quantum machine $\A$ is formalized by a unitary over registers $\regM$ for receiving and sending messages,  and $\regA$ for maintaining $\A$'s internal state.
For two interactive quantum machines $\A$ and $\B$ that share the same message register $\regM$, an interaction between $\A$ and $\B$ proceeds by alternating invocations of $\A$ and $\B$ while exchanging messages over  $\regM$.

An oracle-aided quantum machine $\mathcal{S}$ given oracle access to an interactive quantum machine $\A$ with an initial internal state $\rho$ (denoted by $\mathcal{S}^{\A(\rho)}$) is allowed to apply the unitary part of $\A$ (the unitary obtained by deferring all measurements by $\A$ and omitting these measurements)  and its inverse in a black-box manner. $\mathcal{S}$ is only allowed to act on $\A$'s internal register $\regA$ through oracle access.
We refer to \cite{EC:Unruh12} for formal definitions of interactive quantum machines and black-box access to them.

\paragraph{Indistinguishability of Quantum States.}
We define computational and statistical indistinguishability of quantum states similarly to \cite{STOC:BitShm20,C:ChiChuYam21}.

We may consider random variables over bit strings or over quantum states. 
This will be clear from the context. 
For ensembles of random variables $\mathcal{X}=\{X_i\}_{\secpar\in \mathbb{N},i\in I_\secpar}$ and $\mathcal{Y}=\{Y_i\}_{\secpar\in \mathbb{N},i\in I_\secpar}$  over the same set of indices $I=\bigcup_{\secpar\in\mathbb{N}}I_\secpar$ and a function $\delta$,       
we use $\mathcal{X}\compind_{\delta}\mathcal{Y}$ to mean that for any non-uniform QPT algorithm $\{\A_\secpar, \rho_{\secpar}\}_{\secpar\in\mathbb{N}}$, there exists a negligible function $\negl(\cdot)$ such that for all $\secpar\in\mathbb{N}$, $i\in I_\secpar$, we have
\[
|\Pr[\A_\secpar(X_i,\rho_\secpar)]-\Pr[\A_\secpar(Y_i,\rho_\secpar)]|\leq \delta(\secpar) + \negl(\secpar).
\]
In particular, when the above holds for $\delta=0$, we say that $\calX$ and $\calY$ are computationally indistinguishable, and simply write $\calX\compind \calY$. Unless stated differently, throughout this paper, computational indistinguishability is always w.r.t.\ non-uniform QPT adversaries.

Similarly, we use $\calX\statind_{\delta}\calY$ to mean that for any unbounded time  algorithm $\A$, there exists a negligible function $\negl(\cdot)$ such that for all $\secpar\in\mathbb{N}$, $i\in I_\secpar$, we have 
\[
|\Pr[\A(X_i)]-\Pr[\A(Y_i)]|\leq \delta(\secpar) + \negl(\secpar).\footnote{In other words, $\calX\statind_{\delta}\calY$ means that there exists a negligible function $\negl(\cdot)$ such that the trace distance between $\rho_{X_i}$ and $\rho_{Y_i}$ is at most $\delta(\secpar) + \negl(\secpar)$ for all $\secpar\in \mathbb{N}$ and $i\in I_\secpar$ where $\rho_{X_i}$ and $\rho_{Y_i}$ denote the density matrices corresponding to $X_{i}$ and $Y_{i}$.}
\]
In particular, when the above hold for $\delta=0$, we say that $\calX$ and $\calY$ are statistically indistinguishable, and simply write $\calX\statind \calY$.
Moreover, 
we write $\calX \equiv \calY$ to mean
that $X_i$ and $Y_i$ are distributed identically for all $i\in I$

When we consider an ensemble $\calX$ that is only indexed by $\secpar$, (i.e., $I_\secpar=\{\secpar\}$), we write $\calX=\{X_\secpar\}_\secpar$ for simplicity.

\subsection{Technical Lemmas}

\para{Serfling's Inequality.}
The following {\em two-sided} version of Serfling's inequality is taken from \cite{C:BouFeh10}. 
\begin{lemma}[Serfling's Inequality {\cite{serfling1974probability,C:BouFeh10}}]\label{lem:Serfling}
Let $\vb{b} \in \bits^n$ be a bit string with $\mu\cdot n$ non-zero bits (i.e., a $\mu$-fraction is 1's). Let the random variables
$(Y_1,Y_2,\ldots,Y_k)$ be obtained by sampling $k$ random entries from $\vb{b}$ {\em without replacement}. Let $\bar{Y} \coloneqq \frac{1}{k}\sum_i^k Y_i$. Then, for any $\delta >0$, it holds that
$$\Pr[|\bar{Y} - \mu| >\delta]\le 2\exp(\frac{-2\delta^2kn}{n-k+1}).$$ 
\end{lemma}

\para{Watrous' Rewinding Lemma.} The following is Watrous' rewinding lemma \cite{SIAM:Watrous09} in the form of  \cite[Lemma 2.1]{STOC:BitShm20}.
\begin{lemma}[Watrous' Rewinding Lemma \cite{SIAM:Watrous09}]\label{lem:quantum_rewinding}
There is a quantum algorithm $\sfR$ that gets as input the following:
\begin{itemize}
\item A quantum circuit $\sfQ$ that takes $n$-input qubits in register $\reginp$ and outputs a classical bit $b$ (in a register outside $\reginp$)  and an $m$-qubit output.  
\item An $n$-qubit state $\rho$ in register $\reginp$.
\item A number $T\in \mathbb{N}$ in unary.
\end{itemize}

$\sfR(1^T,\sfQ,\rho)$ executes in time $T\cdot|\sfQ|$  and outputs a distribution over $m$-qubit states  $D_{\rho}\defeq \sfR(1^T,\sfQ,\rho)$  with the following guarantees.

For an $n$-qubit state $\rho$, denote by $\sfQ_{\rho}$ the conditional distribution of the output distribution $\sfQ(\rho)$,
conditioned on $b = 0$, and denote by $p(\rho)$ the probability that $b = 0$. If there exist $p_0, q \in (0,1)$, $\gamma \in (0,\frac{1}{2})$
such that:
\begin{itemize}
    \item  Amplification executes for enough time: $T\geq \frac{\log (1/\gamma)}{4p_0(1-p_0)}$,
    \item  There is some minimal probability that $b = 0$: For every $n$-qubit state $\rho$, $p_0\leq p(\rho)$,
    \item  $p(\rho)$ is input-independent, up to $\gamma$ distance: For every $n$-qubit state $\rho$, $|p(\rho)-q|<\gamma$, and
    \item  $q$ is closer to $\frac{1}{2}$: $p_0(1-p_0)\leq q(1-q)$,
\end{itemize}
then for every $n$-qubit state $\rho$,
\begin{align*}
    \TD(\sfQ_{\rho},D_{\rho})\leq 4\sqrt{\gamma}\frac{\log(1/\gamma)}{p_0(1-p_0)}.
\end{align*}
\end{lemma}

\para{Unruh's Rewinding Lemma.} The following lemma is proven in \cite{EC:Unruh12}. 
\begin{lemma}[Unruh's Rewinding Lemma {\cite[Lemma 7]{EC:Unruh12}}]\label{lem:Unruh_rewinding}
Let $C$ be a finite set. 
Let $\{\Pi_{i}\}_{i\in C}$ be orthogonal projectors on a Hilbert space $\hil$.
Let $\ket{\psi}\in \hil$ be a unit vector. 
Let $\alpha\defeq \sum_{i\in C}\frac{1}{|C|}\|\Pi_i\ket{\psi}\|^2$ and $\beta\defeq \sum_{i,j\in C}\frac{1}{|C|^2}\|\Pi_i\Pi_j\ket{\psi}\|^2$ Then we have $\beta\geq \alpha^3$. 
\end{lemma}  

\xiao{Serfling's Inequality, and the whole sections \Cref{add-prelim:VSS} and \Cref{add-prelim:mpc_in_the_head} can go to the appedix. We can make a section called ``Additional Preliminaries'' to collect them together.}

\subsection{Verifiable Secret Sharing Schemes}\label{add-prelim:VSS}
We present in \Cref{def:VSS} the definition of verifiable secret sharing (VSS) schemes \cite{FOCS:CGMA85}. We remark that \cite{STOC:BenGolWig88,EC:CDDHR99} implemented $(n+1, \lfloor n/3 \rfloor)$-perfectly secure VSS schemes. These constructions suffice for all the applications in the current paper.
\begin{definition}[Verifiable Secret Sharing]\label{def:VSS}
An $(n + 1, t)$-perfectly secure VSS scheme $\Prot_\VSS$ consists of a pair of protocols $(\VSS_\Share, \VSS_\Recon)$ that implement respectively the sharing and reconstruction phases as follows.
\begin{itemize}
\item {\bf Sharing Phase $\VSS_\Share$:} 
Player $P_{n+1}$ (referred to as dealer) runs on input a secret $s$ and randomness $r_{n+1}$, while any other player $P_i$ $(i \in [n])$ runs on input a randomness $r_i$. During this phase players can send (both
private and broadcast) messages in multiple rounds.
\item {\bf Reconstruction Phase $\VSS_\Recon$:}
Each shareholder sends its view $v_i$ $(i \in [n])$ of the Sharing Phase to each other player, and on input the views of all players (that can include bad or empty views) each player outputs a reconstruction of the secret $s$.
\end{itemize}
All computations performed by honest players are efficient. The computationally unbounded adversary can corrupt up to t players that can deviate from the above procedures. The following security properties
hold.
\begin{enumerate}
\item {\bf Perfectly Verifiable-Committing:} \label[Property]{item:def:VSS:vc}
if the dealer is dishonest, then one of the following two cases happen (i.e., with probability 1): 
\begin{enumerate}
\item \label[Case]{item:def:VSS:vc:case:1}
During the Sharing Phase, honest players disqualify the dealer, therefore they output a special value $\bot$ and will refuse to play the reconstruction phase; 
\item \label[Case]{item:def:VSS:vc:case:2}
During the Sharing Phase, honest players do not disqualify the dealer. Therefore such a phase determines a unique value $s^*$
that belongs to the set of possible legal values that does not include $\bot$, which will be reconstructed by the honest players during the
reconstruction phase.
\end{enumerate} 
\item \label[Property]{item:def:VSS:secrecy}
{\bf Secrecy:} 
if the dealer is honest, then the adversary obtains no information about the shared secret before running the protocol $\Recon$. More accurately, there exists a PPT oracle machine $\Sim^{(\cdot)}$ such that for any message $m$, and every (potentially inefficient) adversary $\Adv$ corrupting a set $T$ of parties with $|T|\le t$ during the Sharing Phase $\VSS_\Share(m)$ (denote $\Adv$'s view in this execution as $\View_{\Adv, T}(1^\secpar, m)$), the following holds: 
$\Set{\View_{\Adv, T}(1^\secpar, m)} \idind \Set{\Sim^\Adv(1^\secpar, T)}$.
 


\item {\bf Correctness:}
 if the dealer is honest throughout the protocols, then each honest player will output the shared secret $s$ at the end of protocol $\Recon$.
\end{enumerate}
\end{definition}

\subsection{Information-Theoretic MPC and the MPC-in-the-Head Paradigm}
\label{add-prelim:mpc_in_the_head}
We first recall {\em information-theoretically secure} MPC and relevant notions that will be employed in the MPC-in-the-head paradigm shown later.

\para{Information-Theoretic MPC.} We now define MPC in the information-theoretic setting (i.e., secure against unbounded adversaries). 


\begin{definition}[Perfectly/Statistically-Secure MPC]\label{def:MPC}
Let $f:(\Set{0,1}^*)^n \mapsto (\Set{0,1}^*)^n$ be an n-ary functionality, and let $\Prot$ be a protocol. We say that $\Prot$ {\em $(n,t)$-perfectly (resp., statistically) securely} computes $f$ if for every static, malicious, and (possibly-inefficient) probabilistic adversary $\Adv$ in the real model, there exists a probabilistic adversary $\Sim$ of comparable complexity (i.e., with running time polynomial in that of $\Adv$) in the ideal model, such that for every $I \subset [n]$ of cardinality at most $t$, every $\vec{x} = (x_1,\dots,x_n) \in (\Set{0,1}^*)^n$ (where $|x_1|=\dots=|x_n|$), and every $z \in \Set{0,1}^*$, it holds that: 
$$\Set{\REAL_{\Prot,\Adv(z),I}(\vec{x})} \idind \Set{\IDEAL_{f,\Sim(z),I}(\vec{x})} ~~~~\big(\text{resp.,}~\Set{\REAL_{\Prot,\Adv(z),I}(\vec{x})} \sind \Set{\IDEAL_{f,\Sim(z),I}(\vec{x})}\big).$$   
\end{definition}
Recall that the MPC protocol from \cite{STOC:BenGolWig88} achieves $(n,t)$-perfect security (against static and malicious adversaries) with $t$ being a constant fraction of $n$.
\begin{theorem}[\cite{STOC:BenGolWig88}]\label{thm:BGW88}
Consider a synchronous network  with  pairwise  private  channels. Then,  for every $n$-ary functionality $f$, there exists a protocol that $(n,t)$-perfectly securely computes $f$ in the presence of a static malicious adversary for any $t < n/3$.
\end{theorem}

\para{Consistency, Privacy, and Robustness.} We now define some {notation} related to MPC protocols. Their roles will become clear when we discuss the MPC-in-the-head technique later. 
\begin{definition}[View Consistency]
\label{def:view-consistency}
A view $\View_i$ of an honest player $P_i$ during an MPC computation $\Prot$ contains input and randomness used in the computation, and all messages received from and sent to the communication tapes. A pair of views $(\View_i,\View_j)$ is {\em consistent} with each other if 
\begin{enumerate}
  \item Both corresponding players $P_i$ and $P_j$ individually computed each outgoing message honestly by using the random tapes, inputs and incoming messages specified in $\View_i$ and $\View_j$ respectively, and:
  \item All output messages of $P_i$ to $P_j$ appearing in $\View_i$ are consistent with incoming messages of $P_j$ received from $P_i$ appearing in $\View_j$ (and vice versa).  
\end{enumerate}   
\end{definition}
\begin{remark}[View Consistency of VSS]\label{rmk:VSS:view-consistency}
Although \Cref{def:view-consistency} defines view consistency for MPC protocols, we will also refer to the view consistency for the execution of verifiable secret sharing schemes (\Cref{def:VSS}). The views $(\view_i, \view_j)$ of players $i$ and $j$ (excluding the dealer) during the execution of $\VSS_\Share$ is said to be consistent if any only if $(\view_i, \view_j)$ satisfies the two requirements in \Cref{def:view-consistency}.
\end{remark}

We further define the notions of correctness, privacy, and robustness for multi-party protocols. 

\begin{definition}[Semi-Honest Computational Privacy]\label{def:t-privacy}
Let $1\leq t<n$, let $\Prot$ be an MPC protocol, and let $\Adv$ be any {\em static, PPT, and semi-honest} adversary. We say that $\Prot$ realizes a function $f:(\Set{0,1}^*)^n \mapsto (\Set{0,1}^*)^n$ with {\em semi-honest $(n,t)$-computational privacy} if there is a PPT simulator $\Sim$ such that for any inputs $x,w_1,\dots,w_n$, every subset $T \subset [n]$ $(|T| \leq t)$ of players corrupted by $\Adv$, and every $D$ with circuit size at most $\mathsf{poly(\SecPar)}$, it holds that 
{\fontsize{10.5pt}{0pt}\selectfont
\begin{equation}
\big|\Pr[D(\View_T(x,w_1,\dots,w_n)) =1]- \Pr[D(\Sim(T,x,\Set{w_i}_{i \in T},f_T(x,w_1,\dots,w_n)))=1]\big| \leq \negl(\SecPar),
\end{equation}}
where $\View_T(x,w_1,\dots,w_n)$ is the joint view of all players.
\end{definition}

\begin{definition}[Statistical/Perfect Correctness]\label{def:MPC-correctness}
Let $\Prot$ be an MPC protocol. We say that $\Prot$ realizes a deterministic n-party functionality $f(x,w_1,\dots,w_n)$ with {\em perfect (resp., statistical)} correctness if for all inputs $x,w_1,\dots,w_n$, the probability that the output of some party is different from the output of some party is different from the actual output of $f$ is 0 (resp., negligible in $k$), where the probability is over the independent choices of the random inputs $r_1,\dots,r_n$ of these parties.   
\end{definition}

\begin{definition}[Perfect/Statistical Robustness] \label{def:t-robustness}
Assume the same setting as the previous definition. We say that $\Prot$ realizes $f$ with {\em $(n,t)$-perfect (resp., statistical) robustness} if in addition to being perfectly (resp., statistical) correct in the presence of a semi-honest adversary as above, it enjoys the following {\em robustness} property against any computationally unbounded malicious adversary corrupting a set $T$ of at most $t$ parties, and for any inputs $(x,w_1,\dots,w_n)$: if there is no $(w_1',\dots,w_n')$ such that $f(x,w_1',\dots,w_n')=1$, then the probability that some uncorrupted player outputs 1 in an execution of $\Prot$ in which the inputs of the honest parties are consistent with $(x,w_1,\dots,w_n)$ is 0 (resp., negligible in \SecPar).    
\end{definition}

\para{MPC-in-the-Head.} MPC-in-the-head is a technique developed for constructing  black-box ZK protocols from MPC protocols \cite{STOC:IKOS07}. Intuitively, the MPC-in-the-head idea works as follows.  Let $\Func_\textsc{zk}$ be the zero-knowledge functionality for an \NP language. Assume there are $n$ parties holding a witness in a secret-sharing form. $\Func_\textsc{zk}$ takes as public input $x$ and one share from each party, and outputs 1 iff the secret reconstructed from the shares is a valid witness.
To build a ZK protocol, the prover runs in his head an execution of MPC w.r.t.\ $\Func_\textsc{zk}$ among $n$ imaginary parties, each one participating in the protocol with a share of the witness. Then, it commits to the view of each party separately. The verifier obtains $t$ randomly chosen views, checks that such views are ``consistent'' (see \Cref{def:view-consistency}), and accepts if the output of every party is 1. The idea is that, by selecting the $t$ views at random, $V$ will catch inconsistent views if the prover cheats.

We emphasize that, in this paradigm, a malicious prover decides the randomness of each virtual party, including those not checked by the verifier (corresponding to honest parties in the MPC execution). Therefore, MPC protocols with standard computational security may fail to protect against such attacks. We need to ensure that the adversary cannot force a wrong output even if it additionally controls the honest parties' random tapes. The $(n,\lfloor n/3 \rfloor)$-perfectly secure MPC protocol in \Cref{thm:BGW88} suffices for this purpose (see also \Cref{rmk:exact-mpc-requriements:mpc-in-the-head}).

One can extend this technique further (as in \cite{FOCS:GLOV12}), to prove a general predicate $\phi$ about an arbitrary value $\alpha$.  Namely, one can consider the functionality $\Func_\phi$ in which party $i$ participates with input a VSS share $[\alpha]_i$. $\Func_\phi$ collects all such shares, and outputs 1 iff $\phi(\VSS_\Recon([\alpha]_1,\ldots,[\alpha]_n)) = 1$.

\begin{remark}[Exact Security Requirements on the Underlying MPC.] 
\label{rmk:exact-mpc-requriements:mpc-in-the-head}
To be more accurate, any MPC protocol that achieves {\em semi-honest $(n,t)$-computational privacy (as per \Cref{def:t-privacy}) and $(n,t)$-perfect robustness (as per \Cref{def:t-robustness})} will suffice for the MPC-in-the-head application.\footnote{It is also worth noting that the $(n,t)$-perfect robustness could be replaced with {\em adaptive $(n,t)$-statistical robustness}. See \cite[Section 4.2]{STOC:IKOS07} for more details.} These two requirements are satisfied by any $(n,t)$-perfectly secure MPC (and, in particular, the one from \Cref{thm:BGW88}).
\end{remark}

\subsection{Post-Quantum Extractable Commitment}
We give a definition of post-quantum (strongly) extractable commitments with $\epsilon$-simulation. We will omit the security parameter from the input to parties when it is clear from the context.
\begin{definition}[Post-Quantum Commitment]\label{def:commitment}
A {\em post-quantum commitment scheme}  $\Prot$ is a classical interactive protocol between interactive \PPT machines $C$ and $R$. Let $m\in \bits^{\ell(\secpar)}$ (where $\ell(\cdot)$ is some polynomial) is a message that $C$ wants to commit to. The protocol consists of the following stages:
\begin{itemize}
\item
{\bf Commit Stage:} $C(m)$ and $R$ interact with each other to generate a transcript (which is also called a commitment) denoted by $\com$,\footnote{That is, we regard the whole transcript as a commitment.} 
$C$'s state $\ST_{C}$, and 
$R$'s output $b_{\mathrm{com}}\in\bit$ indicating acceptance $(i.e., b_{\mathrm{com}}=1)$ 
or rejection $(i.e., b_{\mathrm{com}}=0)$.
 We denote this execution by $(\com,\ST_{C},b_{\mathrm{com}}) \gets \langle C(m), R \rangle(1^\secpar)$.
 When $C$ is honest, $\ST_C$ is classical, but 
 when we consider a malicious quantum committer $C^*(\rho)$, we allow it to generate any quantum state $\ST_{C^*}$. 
 Similarly, a malicious quantum receiver $R^*(\rho)$ can output any quantum state, which we denote by $\OUT_{R^*}$ instead of $b_{\mathrm{com}}$. 
\item
{\bf Decommit Stage:}
$C$ generates a decommitment $\decom$ from $\ST_C$.
We denote this procedure by $\decom \gets C(\ST_C)$.\footnote{We could define $\ST_C$ to be $\decom$ itself w.l.o.g. However, we define them separately because this is more convenient when we define ExtCom-and-Prove in \Cref{def:com-n-prove}, which is an extension of post-quantum extractable commitments.} 
Then it sends a message $m$ and decommitment $\decom$ to $R$, 
and $R$ outputs a bit
$b_{\mathrm{dec}}\in\bit$ indicating acceptance $(i.e., b_{\mathrm{dec}}=1)$ 
or rejection $(i.e., b_{\mathrm{dec}}=0)$.
We assume that $R$'s verification procedure is deterministic and denote it
by $\Verify(\com,m,\decom)$.\footnote{Note that $\Verify$ is well-defined since our syntax does not allow $R$ to keep a state from the commit stage.}
W.l.o.g., we assume that $R$ always rejects (i.e., $\Verify(\com,\cdot,\cdot) = 0$) whenever $b_\mathrm{com} = 0$. (Note that w.l.o.g., $\com$ can include $b_\mathrm{com}$ because we can always modify the protocol to ask $R$ to send $b_\mathrm{com}$ as the last round message.)
\end{itemize}
The scheme satisfies the following correctness requirement:
\begin{enumerate}
\item
{\bf Correctness.} For any $m \in \bits^{\ell(\secpar)}$, it holds that
\begin{equation*}
\Pr[b_{\mathrm{com}}=b_{\mathrm{dec}}=1 : 
\begin{array}{l}
(\com, \ST_{C}, b_{\mathrm{com}}) \gets \langle C(m),R \rangle(1^\secpar) \\
\decom \gets C(\ST_C)\\
b_{\mathrm{dec}}\gets \Verify(\com,m,\decom)
\end{array}
] = 1.
\end{equation*}
\xiao{make sure we have correctness 1; otherwise, change it to $1-\negl(\secpar)$.}
\takashi{I believe all constructions have correctness $1$.}
\end{enumerate}
\end{definition}

\begin{definition}[Computationally Hiding]\label{def:comp_hiding}
A post-quantum commitment $\Prot$ is {\em computationally hiding} if for any $m_0, m_1 \in \bits^{\ell(\secpar)}$ and any 
non-uniform QPT receiver $R^*(\rho)$, the following holds:
\begin{equation*}
\{\OUT_{R^*}:(\com, \ST_{C},\OUT_{R^*})\gets \langle C(m_0),R^*(\rho) \rangle(1^\secpar) \}_{\secpar} \cind \{\OUT_{R^*}:(\com, \ST_{C},\OUT_{R^*})\langle C(m_1),R^*(\rho) \rangle(1^\secpar) \}_{\secpar}.
\end{equation*}
\end{definition}

\begin{definition}[Statistically Binding]\label{def:stat-binding} 
A post-quantum commitment $\Prot$ is {\em statistically binding} if for any unbounded-time comitter $C^*$, the following holds: 
\begin{align*}
    \Pr[
    \begin{array}{l}
    \exists~m,m',\decom,\decom',~m\neq m' \\
    \land~ \Verify(\com,m,\decom)=\Verify(\com,m',\decom')=1
    \end{array}
    :(\com,\ST_{C^*},b_{\mathrm{com}}) \gets \langle C^*, R \rangle(1^\secpar)]=\negl(\secpar).
\end{align*}
\naihui{$b_{com} = 1$?}
\takashi{I added an additional assumption that $\Verify$ always outputs $0$ when $b_{com}$ is $0$ in the decommit stage in Definition 1.}
\end{definition}

\begin{definition}[Committed Values]\label{def:val}
For a post-quantum commitment $\Prot$, 
we define the value function as follows:
\begin{equation*}
    \val_{\Prot}(\com)\defeq 
    \begin{cases}
    m&\text{~if~}\exists\text{~unique~}m\text{~s.t.~}\exists~\decom, \Verify(\com,m,\decom)=1\\
    \bot &\text{otherwise}
    \end{cases}.
\end{equation*}
We say that $\com$ is valid if $\val_{\Prot}(\com)\neq \bot$ and invalid if $\val_{\Prot}(\com)=\bot$.
\end{definition}

Then we give the definition of the strong extractability with $\epsilon$-simulation. 
The definition is similar to that of post-quantum extractable commitments in \cite{STOC:BitShm20,BLS21} except that we allow an (arbitrarily small) noticeable approximation error similarly to post-quantum $\epsilon$-zero-knowledge \cite{C:ChiChuYam21}. 
We note that we call it the \emph{strong} extractability since we also define a weaker version of extractability in \Cref{def:epsilon-sim-ext-com:weak} in \Cref{sec:weak_ext_construction}. 

\begin{definition}[Strong Extractability with $\epsilon$-Simulation]\label{def:epsilon-sim-ext-com:strong}
A commitment scheme $\Prot$ is {\em  strongly extractable with $\epsilon$-simulation} if there exists a QPT algorithm $\SimExt$ (called the $\epsilon$-simulation strong-extractor) such that for any noticeable $\epsilon(\secpar)$ and any non-uniform QPT $C^*(\rho)$, 
\begin{equation*}
\big\{ \SimExt^{C^*(\rho)}(1^\secpar,1^{\epsilon^{-1}}) \big\}_\secpar
\cind_\epsilon
\big\{(\val_\Prot(\com), \ST_{C^*}):(\com,\ST_{C^*},b_{\mathrm{com}}) \gets \langle C^*(\rho), R \rangle(1^\secpar)\big\}_\secpar.  
\end{equation*}
\end{definition}

We also define the parallel version.
\begin{definition}[Parallel-Strong Extractability with $\epsilon$-Simulation]\label{def:epsilon-sim-ext-com:parallel_strong}
A commitment scheme $\Prot$ is {\em  parallelly strongly extractable with $\epsilon$-simulation} if 
for any integer $n=\poly(\secpar)$, 
there exists a QPT algorithm $\SimExt_\sfpar$ (called the $\epsilon$-simulation parallel-strong-extractor) such that for any noticeable $\epsilon(\secpar)$ and any non-uniform QPT $C^*(\rho)$, 
\begin{align*}
&\big\{ \SimExt_\sfpar^{C^*(\rho)}(1^\secpar,1^{\epsilon^{-1}}) \big\}_\secpar\\
\cind_\epsilon
&\big\{(\Lambda_{\{b_{\mathrm{com},j}\}_{j=1}^{n}}(\{\val(\com_j)\}_{j=1}^{n}), \ST_{C^*}):(\{\com_j\}_{j=1}^{n},\ST_{C^*},\{b_{\mathrm{com},j}\}_{j=1}^{n})\sample\execution{C^*(\rho)}{R^n}(1^\SecPar)\big\}_\secpar
\end{align*}
where 
$(\{\com_j\}_{j=1}^{n},\ST_{C^*},\{b_{\mathrm{com},j}\}_{j=1}^{n})\sample\execution{C^*(\rho)}{R^n}(1^\SecPar)$
means that $C^*(\rho)$ interacts with $n$ copies of the honest receiver $R$ in parallel and the execution results in transcripts  $\{\com_j\}_{j=1}^{n}$, the final state $\ST_{C^*}$, and outputs $\{b_{\mathrm{com},j}\}_{j=1}^{n}$ of each copy of $R$ and 
$$\Lambda_{\{b_{\mathrm{com},j}\}_{j=1}^{n}}(\{\val(\com_j)\}_{j=1}^{n}) \coloneqq 
\begin{cases}
\{\val_\Prot(\com_j)\}_{j=1}^{n} & \text{if}~\forall~j\in[n]~ b_{\mathrm{com},j} = 1 \\
\bot & \text{otherwise}
\end{cases}.
$$
\end{definition}
\begin{remark}
We remark that the above definition only requires the extractor to extract the committed values when $R$ accepts in all the parallel sessions. In particular, when $R$ accepts in some sessions but not in others, the extractor does not need to extract the committed values at all. 
An alternative stronger (and probably more natural) definition would require the extractor to extract $\val_\Prot(\com_j)$ for all $j\in [n]$ such that $R$ accepts in the $j$-th session. 
But we define it in the above way since it suffices for our purpose
and we do not know if our construction satisfies the stronger one.
\end{remark}

\section{Extract-and-Simulate Lemma}\label{sec:extract_and_simulate}
We prove a lemma that can be seen as an $\epsilon$-simulation variant of Unruh's rewinding lemma (\Cref{lem:Unruh_rewinding}) in typical applications.
This lemma is the technical core of all the results in this paper.

\subsection{Statement of Extract-and-Simulate Lemma}
Our lemma is stated as follows.
\begin{lemma}[Extract-and-Simulate Lemma]
\label{lem:extract_and_simulate}
Let $C$ be a finite set.
Let $\{\Pi_{i}\}_{i\in C}$ be orthogonal projectors on a Hilbert space $\hil$ such that the measurement $\{\Pi_i,I-\Pi_i\}$ can be efficiently implemented. 
Let $\ket{\psi_\init}\in \hil$ be a unit vector.

Suppose that there are a subset $S\in C^2$ and a QPT algorithm $\A=(\A_0,\A_1)$ that satisfies the following:
\begin{enumerate}
    \item 
    \label{item:extract_and_simulate_overwhelming}
    $S$ consists of an overwhelming fraction of $C^2$, i.e., $\frac{|S|}{|C|^2}=1-\negl(\secpar)$.
    \item \label{item:extract_and_simulate_s_i}
    For all $i\in C$, there exists a classical string $s_i$ such that 
    $$\Pr[\A_0\left(i,\frac{\Pi_i\ket{\psi_\init}}{\|\Pi_i\ket{\psi_\init}\|}\right)=s_i]=1.
    $$
    \item 
    \label{item:extract_and_simulate_s_star}
    There exists a classical string $s^*$ such that 
    for any $(i,j)\in S$, 
      $$\Pr[\A_1\left(i,j,s_i,s_j\right)=s^*]=1.
    $$
\end{enumerate}

Let 
$\Exp(\secpar,\{\Pi_i\}_{i\in C},\ket{\psi_\init})$ be an experiment that works as follows:
\begin{itemize}
    \item Choose $i\sample C$.
    \item Apply the measurement $\{\Pi_{i},I-\Pi_{i}\}$ on $\ket{\psi_\init}$.
    \begin{itemize}
        \item If the state is projected onto $\Pi_i$, the experiment outputs $i$, the classical string $s^*$, and the resulting state $\frac{\Pi_i\ket{\psi_\init}}{\left\|\Pi_i\ket{\psi_\init}\right\|}$.\footnote{We stress that we do not assume that the experiment is efficient. Especially, it may be computationally hard to find $s^*$ from $\frac{\Pi_i\ket{\psi_\init}}{\left\|\Pi_i\ket{\psi_\init}\right\|}$.}
        \item If the state is projected onto $I-\Pi_i$, the experiment outputs $i$, $\bot$, and the resulting state $\frac{(I-\Pi_i)\ket{\psi_\init}}{\left|(I-\Pi_i)\ket{\psi_\init}\right|}$. 
    \end{itemize}
\end{itemize}

Then, there is a QPT algorithm $\SimExt$ such that for any noticeable $\epsilon$, 
$$
\{\SimExt(1^\secpar,1^{\epsilon^{-1}},\{\Pi_i\}_{i\in C},\A,\ket{\psi_\init})\}_{\secpar}
\statind_{\epsilon} 
\{\Exp(\secpar,\{\Pi_i\}_{i\in C},\ket{\psi_\init})\}_{\secpar}.
$$
\end{lemma} 

\subsection{Proof of the Extract-and-Simulate Lemma}
We prove the extract-and-simulate lemma (\Cref{lem:extract_and_simulate}). 
Throughout this subsection, we use the notations defined in \Cref{lem:extract_and_simulate}. 

\smallskip
\noindent\textbf{Structure of the Proof.}
The high-level structure of our proof is similar to those of the ($\epsilon$-)zero-knowledge properties of protocols in \cite{STOC:BitShm20,C:ChiChuYam21}. 
We first construct extractors $\ext_{\siml,\abort}$ and $\ext_{\siml,\nonabort}$ that work in the ``aborting case" and ``non-aborting case", respectively where we say that the experiment $\Exp(\secpar,\{\Pi_i\}_{i\in C},\ket{\psi_\init})$ aborts if its second output is $\bot$.  
Then we consider a combined simulator $\ext_{\siml,\comb}$ that randomly guesses if the experiment aborts, runs either of $\ext_{\siml,\abort}$ or  $\ext_{\siml,\nonabort}$ that corresponds to the guessed case, and returns a failure symbol $\fail$ if the guess turns out to be wrong.
Then, $\ext_\comb$ correctly works conditioned on that the output is not $\fail$, and it returns $\fail$ with probability almost $1/2$. 
By applying Watrous' rewinding lemma ( \Cref{lem:quantum_rewinding}) to $\ext_{\siml,\comb}$, we can convert it to a full-fledged simulator.\\

Let $\Exp_\abort(\secpar,\{\Pi_i\}_{i\in C},\ket{\psi_\init})$ and $\Exp_\nonabort(\secpar,\{\Pi_i\}_{i\in C},\ket{\psi_\init})$ be the same as $\Exp(\secpar,\{\Pi_i\}_{i\in C},\ket{\psi_\init})$  except that they output a failure symbol $\fail$ in aborting and non-aborting case, respectively.
That is, they work as follows where differences from $\Exp(\secpar,\{\Pi_i\}_{i\in C},\ket{\psi_\init})$ are marked by red underlines:

\smallskip
\noindent
$\Exp_\abort(\secpar,\{\Pi_i\}_{i\in C},\ket{\psi_\init})$:
\begin{itemize}
    \item Choose $i\sample C$.
    \item Apply the measurement $\{\Pi_{i},I-\Pi_{i}\}$ on $\ket{\psi_\init}$.
    \begin{itemize}
        \item \redunderline{If the state is projected onto $\Pi_i$, the experiment outputs $\fail$.} 
        \item If the state is projected onto $I-\Pi_i$, the experiment outputs $i$, $\bot$, and the resulting state $\frac{(I-\Pi_i)\ket{\psi_\init}}{\left\|(I-\Pi_i)\ket{\psi_\init}\right\|}$.  
    \end{itemize}
\end{itemize}

\smallskip
\noindent
$\Exp_\nonabort(\secpar,\{\Pi_i\}_{i\in C},\ket{\psi_\init})$:
\begin{itemize}
    \item Choose $i\sample C$.
    \item Apply the measurement $\{\Pi_{i},I-\Pi_{i}\}$ on $\ket{\psi_\init}$.
    \begin{itemize}
        \item If the state is projected onto $\Pi_i$, the experiment outputs 
        $i$, the classical string $s^*$, and 
        the resulting state $\frac{\Pi_i\ket{\psi_\init}}{\left\|\Pi_i\ket{\psi_\init}\right\|}$.
        \item \redunderline{If the state is projected onto $I-\Pi_i$, the experiment outputs  $\fail$.}  
    \end{itemize}
\end{itemize}

We give simulation extractors for each of these experiments.
\begin{lemma}[Extract-and-Simulate for the Aborting Case]\label{lem:aborting_case}
There is a QPT algorithm $\ext_{\siml,\abort}$ such that  for any noticeable $\epsilon$, 
$$
\{\ext_{\siml,\abort}(1^\secpar,1^{\epsilon^{-1}},\{\Pi_i\}_{i\in C},\A,\ket{\psi_\init})\}_{\secpar}
\equiv 
\{\Exp_\abort(\secpar,\{\Pi_i\}_{i\in C},\ket{\psi_\init})\}_{\secpar}.\footnote{$\ext_{\siml,\abort}$ does not need to take $1^{\epsilon^{-1}}$ or $\A$ as part of its input, but we include them in input for notational convenience.}
$$ 
\end{lemma}
\begin{proof}[Proof of \Cref{lem:aborting_case}]
Since $\Exp_\abort$ can be run efficiently (because it never outputs $s^*$), $\ext_{\siml,\abort}$ just needs to run $\Exp_\abort$.
\end{proof}

\begin{lemma}[Extract-and-Simulate for the Non-aborting Case]\label{lem:non-aborting_case}
There is a QPT algorithm $\ext_{\siml,\nonabort}$ such that for any noticeable $\epsilon$, 
$$
\{\ext_{\siml,\nonabort}(1^\secpar,1^{\epsilon^{-1}},\{\Pi_i\}_{i\in C},\A,\ket{\psi_\init})\}_{\secpar}
\statind_{\epsilon} 
\{\Exp_\nonabort(\secpar,\{\Pi_i\}_{i\in C},\ket{\psi_\init})\}_{\secpar}.
$$ 
\end{lemma}

Since the proof of \Cref{lem:non-aborting_case} is the most non-trivial technical part, we defer it to \Cref{sec:proof_non-aborting_case} after some preparations in \Cref{sec:preparation_for_non-aborting_case}.

Given \Cref{lem:aborting_case} and \Cref{lem:non-aborting_case}, the rest of the proof of \Cref{lem:extract_and_simulate} is very similar to the corresponding part of the $\epsilon$-zero-knowledge property of the protocols in \cite{C:ChiChuYam21}.
We give the full proof for completeness.

Let $\ext_{\siml,\comb}$ be an algorithm that works as follows:

\smallskip
\noindent
$\ext_{\siml,\comb}(1^\secpar,1^{\epsilon^{-1}},\{\Pi_i\}_{i\in C},\A,\ket{\psi_\init})$:
\begin{enumerate}
    \item Set $\epsilon'\defeq\frac{\epsilon^2}{3600\log^4(\secpar)}$.
    \item Choose $\mathsf{mode}\sample \{\abort,\nonabort\}$.
    \item Run and output $\ext_{\siml,\mathsf{mode}}(1^\secpar,1^{{\epsilon'}^{-1}},\{\Pi_i\}_{i\in C},\A,\ket{\psi_\init})$.
\end{enumerate}

\begin{lemma}[$\ext_{\siml,\comb}$ Simulates $\Exp$ with Probability almost $1/2$]\label{lem:comb}
Let $\pcombsuc$ be the probability that $\ext_{\siml,\comb}(1^\secpar,1^{\epsilon^{-1}},\{\Pi_i\}_{i\in C},\A,\ket{\psi_\init})$ does not return $\fail$, and let 
$$D_{\mathsf{ext},\comb}(1^\secpar, 1^{\epsilon^{-1}}, \{\Pi_i\}_{i\in C}, \A, \ket{\psi_\init})$$
 be a conditional  distribution of $\ext_{\siml,\comb}(1^\secpar,1^{\epsilon^{-1}},\{\Pi_i\}_{i\in C},\A, \ket{\psi_\init})$, conditioned on that it does not return $\fail$.
Then we have 
\begin{align}
    \left|\pcombsuc-1/2\right|\leq \epsilon'/2+\negl(\secpar). \label{eq:pcombsuc}
\end{align}
Moreover, we have 
\begin{align}
\{D_{\mathsf{ext},\comb}(1^\secpar,1^{\epsilon^{-1}},\{\Pi_i\}_{i\in C},\A, \ket{\psi_\init})\}_{\secpar}
\statind_{4\epsilon'} 
\{\Exp(\secpar,\{\Pi_i\}_{i\in C},\ket{\psi_\init})\}_{\secpar}. \label{eq:simlcomb}
\end{align}
\end{lemma}
\begin{proof}(sketch.)
The intuition behind this proof is as follows. 
By Lemma \ref{lem:aborting_case} and \ref{lem:non-aborting_case}, $\ext_{\siml,\abort}$ and $\ext_{\siml,\nonabort}$ almost simulate $\Exp$ conditioned on that $\Exp$ aborts and does not abort, respectively.
Therefore, if we randomly guess  if $\Exp$ aborts and runs either of $\ext_{\siml,\abort}$ or $\ext_{\siml,\nonabort}$ that successfully works for the guessed case, the output distribution is close to  the real output distribution of $\Exp$ conditioned on that the guess is correct, which happens with probability almost $1/2$.

A formal proof can be obtained based on the above intuition and is exactly the same as the proof of \cite[Lemma 5.5]{C:ChiChuYam21}
except for  notational adaptations.  
\end{proof}

Then, we convert $\ext_{\siml,\comb}$ into a full-fledged simulator that does not return $\fail$ by using Watrous' rewinding lemma (Lemma \ref{lem:quantum_rewinding}).
Namely, we let $\sfQ$ be a quantum algorithm that takes $\ket{\psi_\init}$ as input and outputs $\ext_{\siml,\comb}(1^\secpar,1^{\epsilon^{-1}},\{\Pi_i\}_{i\in C},\A,\ket{\psi_\init})$ where $b\defeq 0$ if and only if  it does not return $\fail$, $p_0\defeq \frac{1}{4}$, $q\defeq \frac{1}{2}$, $\gamma\defeq \epsilon'$, and $T\defeq 2\log (1/\epsilon')$.
Then it is easy to check that the conditions for Lemma \ref{lem:quantum_rewinding} is satisfied by Eq. \ref{eq:pcombsuc} in Lemma \ref{lem:comb} (for sufficiently large $\secpar$).
Then by using Lemma \ref{lem:quantum_rewinding}, we can see that $\sfR(1^T,\sfQ,\ket{\psi_\init})$ runs in time $T\cdot |\sfQ|=\poly(\secpar)$ and its output (seen as a mixed state) has a trace distance bounded by $4\sqrt{\gamma}\frac{\log(1/\gamma)}{p_0(1-p_0)}$ from  $D_{\mathsf{ext},\comb}(1^\secpar,1^{\epsilon^{-1}},\{\Pi_i\}_{i\in C},\A,\ket{\psi_\init})$.
Since we have $\gamma=\epsilon'=\frac{\epsilon^2}{3600\log^4(\secpar)}=1/\poly(\secpar)$, we have $4\sqrt{\gamma}\frac{\log(1/\gamma)}{p_0(1-p_0)}< 30\sqrt{\gamma} \log^2 (\secpar)=\frac{\epsilon}{2}$ for sufficiently large $\secpar$
 where we used $\log(1/\gamma)=\log(\poly(\secpar))=o(\log^2(\secpar))$.
Thus, by combining the above and Eq. \ref{eq:simlcomb} in Lemma \ref{lem:comb}, if we define $\ext_\siml(1^\secpar,1^{\epsilon^{-1}},\{\Pi_i\}_{i\in C},\A,\ket{\psi_\init})\defeq \sfR(1^T,\sfQ,\ket{\psi_\init})$, then we have 
\begin{align*}
\{\ext_\siml(1^\secpar,1^{\epsilon^{-1}},\{\Pi_i\}_{i\in C},\A,\ket{\psi_\init})\}_{\secpar}
\statind_{\frac{\epsilon}{2}+4\epsilon'} 
  \{\Exp(\secpar,\{\Pi_i\}_{i\in C},\ket{\psi_\init})\}_{\secpar}.
\end{align*}
We can conclude the proof of \Cref{lem:extract_and_simulate} by noting that  we have
    $\frac{\epsilon}{2}+4\epsilon'< \epsilon$
since we have $\epsilon'= \frac{\epsilon^2}{3600\log^4(\secpar)} < \frac{\epsilon}{8}$.

\subsection{Preparation for Proof of \Cref{lem:non-aborting_case}}\label{sec:preparation_for_non-aborting_case}
For proving \Cref{lem:non-aborting_case}, we prepare the following three lemmas.

The first is a simple technical lemma  that is a variant of \cite[Lemma 3.1]{C:ChiChuYam21}. 
\revise{\begin{lemma}[Variant of {\cite[Lemma 3.1]{C:ChiChuYam21}}]\label{lem:state-close}
Let $\ket{\phi_b}=\ket{\phi_{b,0}}+\ket{\phi_{b,1}}$ be a normalized quantum state in a Hilbert space $\hil$.  \takashi{I removed the assumption that $\bra{\phi_{b,0}}\ket{\phi_{b,1}}=0$ because this was not used.}
Let $F$ be a quantum algorithm  that takes a state in $\hil$ as input and outputs a quantum state (not necessarily in $\hil$) or a classical failure symbol $\fail$. 
Suppose that we have  
$$\Pr[F\left(\frac{\ket{\phi_{b,0}}\bra{\phi_{b,0}}}{\|\ket{\phi_{b,0}}\|^2}\right)=\fail]\geq 1-\gamma$$
for $b\in \bit$
and  
$\|\ket{\phi_{1,1}}-\ket{\phi_{1,0}}\|\le \delta$. \takashi{Note that this is Euclidean distance instead of trace distance unlike the previous version.}   
Then for any distinguisher $D$, it holds that\footnote{In the previous version of this paper, we claimed that the bound was $4\gamma^{1/4} + \|\ket{\phi_{0,1}}\bra{\phi_{0,1}}-\ket{\phi_{1,1}}\bra{\phi_{1,1}}\|_{tr}$, but there was a flaw in the proof in the case of $\ket{\phi_{0,1}}\neq \ket{\phi_{1,1}}$.} 
\begin{align*}
    |\Pr[D(F(\ket{\phi_0}\bra{\phi_0}))=1]
    -\Pr[D(F(\ket{\phi_1}\bra{\phi_1}))=1]|\leq (12\gamma^{1/2}+2\delta)^{1/2}. 
\end{align*}
\end{lemma}}
We give the proof in \Cref{sec:proof_state-close}.

\revise{
The second lemma is a variant of the gentle measurement lemma shown in \cite{C:ChiChuYam21}. 
\begin{lemma}[{\cite[Lemma 3.2]{C:ChiChuYam21}}]\label{lem:gentle_measurement} \takashi{The previous version also used this lemma in the case of $\nu=\negl$ by just saying "the gentle measurement lemma".}
    Let $\ket{\psi}_\regX$ be a (not necessarily normalized) state over register $\regX$ and $U$ be a unitary over registers $(\regX,\regY,\regZ)$.  
    Suppose that a measurement of register $\regZ$ of $U\ket{\psi}_\regX\ket{0}_{\regY,\regZ}$ results in a deterministic value except for probability $\nu$, i.e., 
    there is $z^*$ such that 
    \begin{align*}
        \|(I-\ket{z^*}\bra{z^*})_{\regZ} U\ket{\psi}_\regX\ket{0}_{\regY,\regZ}\|^2\le \nu. 
    \end{align*}
    If we let $R:= (\ket{0}\bra{0})_{\regY,\regZ}U^\dagger (\ket{z^*}\bra{z^*})_{\regZ} U$, 
then we have 
\begin{align*}
    \|
    \ket{\psi}_\regX\ket{0}_{\regY,\regZ}-
   R \ket{\psi}_\regX\ket{0}_{\regY,\regZ}\|\le \sqrt{\nu}.
\end{align*}
\end{lemma}
}

The third is a variant of \cite[Lemma 3.3]{C:ChiChuYam21}. 
\takashi{In the previous version, this lemma was in the appendix, and we introduced an intermediate lemma that is a consequence of this lemma. However, I found that the proof became a little bit ambiguous because of this. Thus, I decided to directly use this lemma in the main body.
}
\begin{lemma}[A variant of {\cite[Lemma 3.3]{C:ChiChuYam21}}]\label{lem:amplification}
Let $\Pi$ be a projection over a Hilbert space $\hil_\regX \ot \hil_\regY$. 
For any noticeable function $\delta=\delta(\lambda)$, there exists an orthogonal decomposition $(S_{<\delta}, S_{\geq \delta})$ of $\hil_\regX \ot \hil_\regY$ that satisfies the following:
\begin{enumerate}
\item(\textbf{$S_{<\delta}$ and $S_{\geq \delta}$ are invariant under $\Pi$ and $(\ket{0}\bra{0})_{\regY}$.}) \label{item:amplification_invariance_projection}
For any $\ket{\psi}_{\regX,\regY}\in S_{<\delta}$, we have 
\begin{align*}
\Pi \ket{\psi}_{\regX,\regY}\in S_{<\delta},~~~~~ (I_{\regX}\otimes(\ket{0}\bra{0})_{\regY})\ket{\psi}_{\regX,\regY}\in S_{<\delta}.
\end{align*}
Similarly, 
 for any $\ket{\psi}_{\regX,\regY}\in S_{\geq \delta}$, we have 
\begin{align*}
\Pi \ket{\psi}_{\regX,\regY}\in S_{\geq \delta},~~~~~ (I_{\regX}\otimes(\ket{0}\bra{0})_{\regY})\ket{\psi}_{\regX,\regY}\in S_{\geq \delta}.
\end{align*}
\item(\textbf{$\Pi$ succeeds with probability $<\delta$ and $\geq \delta$ in $S_{<\delta}$ and $S_{\geq \delta}$.})  \label{item:amplification_success_probability}
For any quantum state $\ket{\phi}_{\regX}\in \hil_{\regX}$ s.t.  $\ket{\phi}_{\regX}\ket{0}_{\regY}\in S_{<\delta}$ we have 
\begin{align*}
\|\Pi\ket{\phi}_{\regX}\ket{0}_{\regY}\|^2< \delta. 
\end{align*}
Similarly, for any quantum state $\ket{\phi}_{\regX}\in \hil_{\regX}$ s.t.  $\ket{\phi}_{\regX}\ket{0}_{\regY}\in S_{\geq \delta}$ we have 
\begin{align*}
\|\Pi\ket{\phi}_{\regX}\ket{0}_{\regY}\|^2\geq \delta. 
\end{align*}
 \item(\textbf{Unitary for amplification.})
For any $T\in \mathbb{N}$, there exists a unitary  $U_{\amp,T}$ over $\hil_\regX \ot \hil_\regY \ot \hil_\regB \ot \hil_\reganc$ where $\regB$ is a register to store a qubit and $\reganc$ is a register to store ancillary qubits with the following properties:
 \label{item:amplification}
 \begin{enumerate}
 \item(\textbf{Mapped onto $\revise{\Pi(I_{\regX}\otimes(\ket{0}\bra{0})_{\regY})}$ when $\regB$ contains $1$.})
For any quantum state $\ket{\psi}_{\regX,\regY}\in \hil_{\regX}\otimes \hil_{\regY}$, we can write
 \[
 \ket{1}\bra{1}_{\regB} U_{\amp,T}\ket{\psi}_{\regX,\regY}\ket{0}_{\regB,\reganc}=\sum_{anc}\ket{\psi'_{anc}}_{\regX,\regY}\ket{1}_{\regB}\ket{anc}_{\reganc}
 \]
 by using sub-normalized states $\ket{\psi'_{anc}}_{\regX,\regY}$ that are in the span of $\revise{\Pi(I_{\regX}\otimes(\ket{0}\bra{0})_{\regY})}$.
 \label{item:amplification_map_to_pi}
\item(\textbf{Amplification of success probability in $S_{\geq \delta}$.})  \label{item:amplification_amplification}
For any noticeable function $\nu=\nu(\secpar)$, there is $T=\poly(\secpar)$ such that  
for any quantum state $\ket{\phi}_{\regX}\in \hil_{\regX}$ s.t.  $\ket{\phi}_{\regX}\ket{0}_{\regY}\in S_{\geq \delta}$, we have 
 \[
 \|\ket{1}\bra{1}_\regB U_{\amp,T}\ket{\phi}_{\regX}\ket{0}_{\regY}\ket{0}_{\regB,\reganc}\|^2
\geq 1-\nu. 
 \]
 \item(\textbf{$S_{<\delta}$ and $S_{\geq \delta}$ are invariant under $U_{\amp,T}$}). 
 For any  quantum state $\ket{\psi_{<\delta}}_{\regX,\regY}\in  S_{<\delta}$ and any $b,anc$,
we can write 
  \[
U_{\amp,T}\ket{\psi_{< \delta}}_{\regX,\regY}\ket{b,anc}_{\regB,\reganc}=\sum_{b',anc'}\ket{\psi'_{<\delta,b',anc'}}_{\regX,\regY}\ket{b',anc'}_{\regB,\reganc}
  \]
  by using sub-normalized states $\ket{\psi'_{<\delta, b',anc'}}_{\regX,\regY} \in S_{<\delta}$.
  
  Similarly, 
 for any  quantum state $\ket{\psi_{\geq \delta}}_{\regX,\regY}\in  S_{\geq \delta}$ and any $b,anc$,
we can write 
  \[
U_{\amp,T}\ket{\psi_{\geq \delta}}_{\regX,\regY}\ket{b, anc}_{\regB,\reganc}=\sum_{b', anc'}\ket{\psi'_{\geq \delta, b', anc'}}_{\regX,\regY}\ket{b', anc'}_{\regB,\reganc}
  \]
  by using sub-normalized states $\ket{\psi'_{\geq \delta, b',anc'}}_{\regX,\regY} \in S_{\geq \delta}$. 
  \label{item:amplification_invariance}
 \end{enumerate}
 \item(\textbf{Efficient Implementation of $U_{\amp,T}$}.) 
 There exists a QPT algorithm $\Amp$ (whose description is independent of $\Pi$) that takes as input  $1^T$, a description of quantum circuit that perform a measurement $(\Pi, I_{\regX,\regY}-\Pi)$, and a state $\ket{\psi}_{\regX,\regY,\regB,\reganc}$, and outputs $U_{\amp,T}\ket{\psi}_{\regX,\regY,\regB,\reganc}$.
 Moreover, $\Amp$ uses the measurement circuit for only implementing an oracle that apply unitary to write a measurement result in a designated register in $\reganc$, and it acts on $\regX$ only through the oracle access.
 \label{item:amplification_efficiency}
\end{enumerate}
\end{lemma}

\revise{The only difference from \cite[Lemma 3.3]{C:ChiChuYam21} is \Cref{item:amplification_map_to_pi} where 
we require that an output of $U_{\amp,T}$ to be in the span of $\Pi(I_{\regX}\otimes(\ket{0}\bra{0})_{\regY})$ when $\regB$ contains $1$ whereas  \cite[Lemma 3.3]{C:ChiChuYam21} only requires it to be in the span of $\Pi$.\footnote{In the previous version of this paper, we only required the state to be in the span of $\Pi$ similarly to \cite[Lemma 3.3]{C:ChiChuYam21}. However, we found that we needed the above stronger requirement due to a technical reason. In particular,this is used in the proof of \Cref{claim:always_extract_s_star}.}
We can prove \Cref{lem:amplification} by using Jordan's lemma in a very similar manner to that for the proof of  \cite[Lemma 3.3]{C:ChiChuYam21}. Thus, we defer the proof to \Cref{sec:proof_lem_amplification}.}

\subsection{Proof of \Cref{lem:non-aborting_case}}
\label{sec:proof_non-aborting_case}
By using \Cref{lem:Unruh_rewinding,lem:amplification,lem:state-close,lem:gentle_measurement}, 
we prove \Cref{lem:non-aborting_case}, which completes the proof of \Cref{lem:extract_and_simulate}. 
\begin{proof}[Proof of \Cref{lem:non-aborting_case}]
First, we define a ``simulation-less extractor" $\ext_{\simless}$ that works as follows:

\smallskip
\noindent$\ext_{\simless}(1^\secpar,\{\Pi_i\}_{i\in C},\A,\ket{\psi})$: 
\begin{enumerate}
    \item 
     \label{step:choose_i_and_j}
    Uniformly choose $(i,j)\sample C^2$ and immediately output $\fail$ if $(i,j)\notin S$. 
    \item 
    \label{step:projections}
    For $k\in \{i,j\}$ in the order of $k=j$, $k=i$,\footnote{Whichever order is fine for our purpose. We specify it just for completeness of the description of the algorithm.} do the following:
    \begin{enumerate}
    \item Perform the measurement $\{\Pi_k,I-\Pi_k\}$, and immediately output $\fail$ if the state is projected onto $I-\Pi_k$. Otherwise, go to the next step with the residual state $\ket{\psi'_k}$. \footnote{Note that $\ket{\psi'_j}=\frac{\Pi_j\ket{\psi}}{|\Pi_j\ket{\psi}|}$ and $\ket{\psi'_i}=\frac{\Pi_i\Pi_j\ket{\psi}}{|\Pi_i\Pi_j\ket{\psi}|}$}
    \item Run $s_i \leftarrow \A_0(k,\ket{\psi'_k})$ in a non-destructive way, i.e., in a way such that the state $\ket{\psi'_k}$ is preserved. This is possible since the output of $\A_0$ is deterministic as required in \Cref{item:extract_and_simulate_s_i} of \Cref{lem:extract_and_simulate}.  
    \end{enumerate}
    \item Run $s\leftarrow \A_1(i,j,s_i,s_j)$ and output $s$. 
\end{enumerate}

Then, by combining \Cref{lem:Unruh_rewinding} and \Cref{item:extract_and_simulate_s_star} of \Cref{lem:extract_and_simulate}, we can show the following claim.
\begin{myclaim}\label{cla:simless}
For any $\ket{\psi}\in \hil$, $\ext_{\simless}(1^\secpar,\{\Pi_i\}_{i\in C},\A,\ket{\psi})$ outputs $s^*$ whenever it does not output $\fail$ and it holds that  
$$
\Pr[\ext_{\simless}(1^\secpar,\{\Pi_i\}_{i\in C},\A,\ket{\psi})=s^*]\geq \left(\sum_{i\in C}\frac{1}{|C|}\|\Pi_i\ket{\psi}\|^2\right)^3- \left(1-\frac{|S|}{|C|^2}\right).
$$
\end{myclaim}
\begin{proof}[Proof of \Cref{cla:simless}]
The former part is clear from \Cref{item:extract_and_simulate_s_star} of \Cref{lem:extract_and_simulate}. 
For the latter part, we only have to lower bound the probability that $\ext_{\simless}$ does not return $\fail$. 
If we remove the condition for outputting $\fail$ in 
the first step of $\ext_{\simless}$, it is clear that the probability to not output $\fail$ is
$\sum_{i,j\in C^2}\frac{1}{|C|^2}\|\Pi_i\Pi_j\ket{\psi}\|^2$, which is lower bounded by 
$\left(\sum_{i\in C}\frac{1}{|C|}\|\Pi_i\ket{\psi}\|^2\right)^3$ by \Cref{lem:Unruh_rewinding}.  
Moreover, it is easy to see that the probability to output $\fail$ in the first step of $\ext_{\simless}$ is $\frac{|S|}{|C|^2}$. By union bound, the latter part of the claim follows.
\end{proof}


\revise{
Our next step is to apply \Cref{lem:amplification} with respect to a projection corresponding to the success of $\ext_{\simless}$. 
Let $U_{\simless}$ be the unitary that represents $\ext_{\simless}(1^\secpar,\{\Pi_i\}_{i\in C},\A,\cdot)$.  
More precisely, we define $U_{\simless}$ over registers the input register $\reginp$, working register $\regW$, and  output register $\regout$ so that $\ext_{\simless}(1^\secpar,\{\Pi_i\}_{i\in C},\A,\cdot))$ can be described as follows:

\smallskip
\noindent\textbf{$\ext_{\simless}(1^\secpar,\{\Pi_i\}_{i\in C},\A,\cdot))$}:
It takes a quantum state $\ket{\psi}$ in the register $\reginp$ and initializes registers $\regW$ and $\regout$ to be $\ket{0}_{\regW,\regout}$. 
Then it applies the unitary $U_{\simless}$, measures the register $\regout$ in the standard basis to obtain $s$,  and outputs $s$. 
\smallskip

We define a projection $\Pi$ over $(\reginp,\regW,\regout)$ as
\begin{align}\label{eq:def_pi}
\Pi \defeq U_{\simless}^\dagger \left(\sum_{s\neq \fail}\ket{s}\bra{s}\right)_{\regout}  U_{\simless}.
\end{align}
Then the following claim immediately follows from the former half of \Cref{cla:simless}.
\begin{myclaim}\label{cla:measure_s_star}
Given any state in the span of $\Pi(I_{\reginp}\otimes(\ket{0}\bra{0})_{\regW,\regout})$, if we apply $U_{\simless}$ and then measure register $\regout$, then the measurement outcome is always $s^*$
\end{myclaim}  
We apply \Cref{lem:amplification} for the above $\Pi$ where $\hil_\regX:=\hil_\reginp$, $\hil_\regY:= \hil_\regW \ot \hil_\regout$,  $\delta:= \left(\frac{\epsilon}{5}\right)^{12}-\left(1-\frac{|S|}{|C|^2}\right)$, and $T=\poly(\secpar)$ is chosen in such a way that \Cref{item:amplification_amplification} of \Cref{lem:amplification} holds for $\nu:=\frac{\epsilon^4}{16}$.  
Then we have a decomposition $(S_{<\delta}, S_{\geq \delta})$ of $\hil_\regX\ot \hil_\regY$ and a unitary $U_{\amp,T}$ over $\hil_\regX\ot \hil_\regY \ot \hil_\regB \ot \hil_\reganc$ that satisfy the requirements in \Cref{lem:amplification}. 
We denote by $\regother$ to mean the registers $\regW$, $\regout$, $\regB$, and $\reganc$ for brevity.
We construct the extractor $\ext_{\siml,\nonabort}$ for \Cref{lem:non-aborting_case} as follows:

\smallskip
\noindent
$\ext_{\siml,\nonabort}(1^\secpar,1^{\epsilon^{-1}},\{\Pi_i\}_{i\in C},\A,\ket{\psi_\init})$:
\begin{enumerate}
\item Set $\ket{\psi_\init}$ in register $\reginp$ and initlialize register $\regother$ to be $\ket{0}$. 
\item Apply $U_{\amp,T}$ by using the algorithm $\Amp$ in \Cref{item:amplification_efficiency} of \Cref{lem:amplification}. 
\item \label{step:measure_b}
Measure register $\regB$ and let $b$ be the outcome. 
If $b=0$, output $\fail$ and immediately halt. Otherwise, proceed to the next step. 
\item \label{step:extract_s} Apply $U_{\simless}$, measure register $\regout$ to obtain an outcome $s_{\ext}$, and apply $U_{\simless}^\dagger$.
\item Apply $U_{\amp,T}^\dagger$ by using the algorithm $\Amp$ in \Cref{item:amplification_efficiency} of \Cref{lem:amplification}. 
\item Measure register $\regother$. If the outcome is not the all $0$'s string, output $\fail$ and immediately halt. Otherwise, let $\ket{\psi_{\mathsf{mid}}}$ be the state in register $\reginp$ at this point, and proceed to the next step.
\item Choose $i\sample C$.
\item Apply the measurement $\{\Pi_{i},I-\Pi_{i}\}$ on $\ket{\psi_{\mathsf{mid}}}$.
    \begin{itemize}
        \item If the state is projected onto $\Pi_i$, output $i$, the classical string $s_\ext$ and the resulting state $\frac{\Pi_i\ket{\psi_{\mathsf{mid}}}}{\left|\Pi_i\ket{\psi_{\mathsf{mid}}}\right|}$.
        \item If the state is projected onto $I-\Pi_i$, output  $\fail$.  
    \end{itemize}
\end{enumerate}
\smallskip 

We can easily see the following claim: 
\begin{myclaim}\label{claim:always_extract_s_star}
Whenever Step \ref{step:extract_s} of $\ext_{\siml,\nonabort}$ is invoked, $s_{\ext}$ obtained in the step is always equal to $s^*$. Moreover, the step does not change the state in registers $\reginp$ and $\regother$, that is, the states before and after the step are identical.  
\end{myclaim}
\begin{proof}[Proof of \Cref{claim:always_extract_s_star}]
Whenever  Step \ref{step:extract_s} is invoked, the bit $b$ obtained in Step \ref{step:measure_b} is equal to $1$. In this case, by \Cref{item:amplification_map_to_pi} of \Cref{lem:amplification}, the state in registers $\reginp$, $\regW$, and $\regout$ is in the span of $\Pi(I_{\reginp}\otimes(\ket{0}\bra{0})_{\regW,\regout})$. Then, \Cref{cla:measure_s_star} implies that $s_\ext$ is always equal to $s^*$. Then the measurement of $\regout$ does not collapse the state and thus  the step does not change the state.
\end{proof}

The rest of the proof is similar to that of \cite[Claim 4.5]{C:ChiChuYam21}. 
Let $R$ be an operator defined as follows:
\begin{align*}
R:=(\ket{0}\bra{0})_{\regother}U_{\amp,T}^{\dagger}(\ket{1}\bra{1})_{\regB}U_{\amp,T}.
\end{align*}
Let $\Pi_{<\delta}$ and $\Pi_{\ge \delta}$ be projections onto $S_{<\delta}$ and $S_{\ge \delta}$, respectively.  
To apply Lemma \ref{lem:state-close}, 
we define states $\ket{\phi_0}=\ket{\phi_{0,0}}+\ket{\phi_{0,1}}$ and $\ket{\phi_1}=\ket{\phi_{1,0}}+\ket{\phi_{1,1}}$ over $(\regD,\reginp,\regother)$ 
where $\regD$ is an additional one-qubit register as follows:   
\begin{align*}
&\ket{\phi_{0}}:= \ket{1}_{\regD}\ket{\psi_\init}_{\reginp}\ket{0}_{\regother},\\
&\ket{\phi_{0,0}}:= \ket{1}_{\regD}\Pi_{< \delta}\ket{\psi_\init}_{\reginp}\ket{0}_{\regother},\\
&\ket{\phi_{0,1}}:= \ket{1}_{\regD}\Pi_{\ge \delta}\ket{\psi_\init}_{\reginp}\ket{0}_{\regother},\\
&\ket{\phi_{1}}:=\ket{1}_{\regD} R \ket{\psi_\init}_{\reginp}\ket{0}_{\regother}+\alpha \ket{0}_{\regD}\ket{0}_{\reginp}\ket{0}_{\regother},\\
&\ket{\phi_{1,0}}:=\ket{1}_{\regD} R \Pi_{< \delta}  \ket{\psi_\init}_{\reginp}\ket{0}_{\regother}+\alpha \ket{0}_{\regD}\ket{0}_{\reginp}\ket{0}_{\regother},\\
&\ket{\phi_{1,1}}:= \ket{1}_{\regD}R \Pi_{\ge \delta} \ket{\psi_\init}_{\reginp}\ket{0}_{\regother}
\end{align*}
for $\alpha:=\sqrt{1-\|R \ket{\psi_\init}_{\reginp}\ket{0}_{\regother}\|^2}$ (so that $\ket{\phi_{1}}$ is a normalized state). 
Let $F$ be a quantum algorithm that works as follows: 
\begin{description}
\item $F\left(\ket{\phi}_{\regD,\reginp,\regother}\right)$:
It measures $\regD$, and outputs $\fail$ if the outcome is $0$. 
Otherwise,  it samples $i\sample C$ and applies the measurement $\{\Pi_{i},I-\Pi_{i}\}$ on register $\reginp$. If the state is projected onto $\Pi_i$, output $i$, the measurement outcome $s$ of the second register, and the resulting state in the third register.
Otherwise, it outputs  $\fail$.  
\end{description} 

It is easy to see that 
\begin{align*}
    \Exp_\nonabort(\secpar,\{\Pi_i\}_{i\in C},\ket{\psi_\init})\equiv F(\ket{\phi_0}\bra{\phi_0}). 
\end{align*}

Moreover, by the definition of $\ext_{\siml,\nonabort}$ and \Cref{claim:always_extract_s_star}, we can see that 
\begin{align*}
   \ext_{\siml,\nonabort}(1^\secpar,1^{\epsilon^{-1}},\{\Pi_i\}_{i\in C},\A,\ket{\psi_\init})\equiv F(\ket{\phi_1}\bra{\phi_1}). 
\end{align*}
Thus, it suffices to prove that the distinguishing advantage between $F(\ket{\phi_0}\bra{\phi_0})$ and $F(\ket{\phi_1}\bra{\phi_1})$ is at most $\epsilon$.  
To apply \Cref{lem:state-close}, 
we prove the following claim. 
\begin{myclaim}\label{claim:condition_check}
The following hold:
\begin{enumerate}
 \item $\Pr[F\left(\frac{\ket{\phi_{b,0}}\bra{\phi_{b,0}}}{\|\ket{\phi_{b,0}}\|^2}\right)=\fail]\geq 1-\left(\frac{\epsilon}{5}\right)^4$ 
for $b\in \bit$.
 \item $\|\ket{\phi_{1,1}}-\ket{\phi_{0,1}}\|\le \left(\frac{\epsilon}{2}\right)^2$.
\end{enumerate}
\end{myclaim}
\begin{proof}[Proof of Claim \ref{claim:condition_check}]~ \\
\paragraph{First item.}
We can write $\Pi_{< \delta}\ket{\psi_\init}_{\reginp}\ket{0}_{\regother}=\ket{\psi_{< \delta}}_{\reginp}\ket{0}_{\regother}$. 
Then we have 
\begin{align*}
    \Pr[F\left(\frac{\ket{\phi_{0,0}}\bra{\phi_{0,0}}}{\|\ket{\phi_{0,0}}\|^2}\right)\neq \fail]&= \sum_{i\in C}\frac{1}{|C|}\left\|\Pi_i \frac{\ket{\psi_{<\delta}}}{\|\ket{\psi_{<\delta}}\|}\right\|^2\\
    &\leq \left(\Pr[\ext_{\simless}\left(1^\secpar,\{\Pi_i\}_{i\in C},\A,\frac{\ket{\psi_{<\delta}}}{\|\ket{\psi_{<\delta }}\|}\right)=s^*]+\left(1-\frac{|S|}{|C|^2}\right)\right)^{1/3}\\
    &\leq \left(\delta+\left(1-\frac{|S|}{|C|^2}\right)\right)^{1/3}\\
    &= \left(\frac{\epsilon}{5}\right)^4     
\end{align*}
where the first inequality follows from \Cref{cla:simless}, the second inequality follows from
$\ket{\psi_{< \delta}}_{\reginp}\ket{0}_{\regother}\in S_{<\delta}$ and 
\Cref{item:amplification_success_probability} of \Cref{lem:amplification}, and the final equality follows from $\delta= \left(\frac{\epsilon}{5}\right)^{12}-\left(1-\frac{|S|}{|C|^2}\right)$. 
This completes the proof of the first item for the case of $b=0$. 
The case of $b=1$ can be proven similarly noting that $R \Pi_{< \delta}  \ket{\psi_\init}_{\reginp}\ket{0}_{\regother}\in S_{<\delta}$ by  \Cref{item:amplification_invariance_projection,item:amplification_invariance} of Lemma \ref{lem:amplification}. 

\paragraph{Second Item.} 
By 
\Cref{item:amplification_amplification} of \Cref{lem:amplification}, we have 
$$
\|(\ket{1}\bra{1})_{\regB}U_{\amp,T}\Pi_{\ge t}\ket{\psi_\init}_{\reginp}\ket{0}_{\regother}\|^2\le \nu.
$$
Thus, \Cref{lem:gentle_measurement} implies
\begin{align*}
   \|\Pi_{\ge t}\ket{\psi_\init}_{\reginp}\ket{0}_{\regother}-R\Pi_{\ge t}\ket{\psi_\init}_{\reginp}\ket{0}_{\regother}\|\le \nu^{1/2}. 
\end{align*}
Since $\nu=\frac{\epsilon^4}{16}$, this immediately implies the second item of the claim.
\end{proof}
By \Cref{lem:state-close} and \Cref{claim:condition_check} 
the distinguishing advantage between $F(\ket{\phi_0}\bra{\phi_0})$ and $F(\ket{\phi_1}\bra{\phi_1})$ is at most $\left(12\left(\frac{\epsilon}{5}\right)^2+2\left(\frac{\epsilon}{2}\right)^2\right)^{1/2}<\epsilon$. This completes the proof of \Cref{lem:non-aborting_case}.
}
\end{proof}


\section{Black-Box $\epsilon$-Simulation-Extractable Commitments in Constant Rounds}
\label{sec:sim-ext-com}
In this section, we construct a post-quantum commitment scheme that satisfies the (parallel) strong extractability with $\epsilon$-simulation.
Namely, we prove the following lemma.
\begin{lemma}\label{lem:extcom}
Assume the existence of post-quantum secure OWFs. Then, there exists a constant-round construction of 
post-quantum commitment that satisfies computational hiding (\Cref{def:comp_hiding}), statistical binding (\Cref{def:stat-binding}), and 
(parallel) strongly extractable commitment with $\epsilon$-simulation  (\Cref{def:epsilon-sim-ext-com:strong,def:epsilon-sim-ext-com:parallel_strong}). Moreover, this construction makes only black-box use of the assumed OWF.
\end{lemma}

Toward proving that, we first construct a scheme that satisfies a weaker notion of $\epsilon$-simulatable extractability in \Cref{sec:weak_ext_construction}.
In \Cref{sec:handling-over-extraction}, we present a compiler that converts the weak scheme in \Cref{sec:weak_ext_construction} into one that satisfies the (parallel) strong extractability with $\epsilon$-simulation.

\subsection{Weakly Extractable Commitment}\label{sec:weak_ext_construction}

We construct a commitment scheme that satisfies weak notions of extractability defined in \Cref{def:epsilon-sim-ext-com:weak,def:epsilon-sim-ext-com:parallel_weak} based on OWFs. 
The description of the scheme is given in \Cref{figure:wext-com}, where $\Com$ is a statistically-binding and computationally-hiding commitment scheme (e.g., Naor's commnitment).   
We remark that the scheme is identical to the \emph{classical} extractable commitment in \cite{TCC:PasWee09}, which in turn is based on earlier works~\cite{SIAM:DolDwoNao00,FOCS:PraRosSah02,TCC:Rosen04}.  
\begin{ProtocolBox}[
label={figure:wext-com},
]{Extractable Commitment Scheme $\wExtCom$}
\begingroup\fontsize{10pt}{0pt}\selectfont
The extractable commitment scheme, based on any commitment scheme $\Com$, works in the following way.\\
~\\
{\bf Input:}
\begin{itemize}[topsep=0.2em,itemsep=0.2em]
\item
both the  committer $C$ and the receiver $R$ get security parameter $1^\SecPar$ as the common input.
\item
$C$ gets a string $m \in \bits^{\ell(\SecPar)}$ as his private input, where $\ell(\cdot)$ is a polynomial 
\end{itemize}
~\\
{\bf Commitmment Phase:}
\begin{enumerate}[topsep=0.2em,itemsep=0.2em]
\item \label[Step]{item:extcom:com}
The committer $C$ commits using \Com to $k = \SecPar$ pairs of strings $\Set{(v^0_i, v^1_i)}_{i=1}^{k}$ where $(v^0_i ,v^1_i) = (\eta_i, m \xor \eta_i)$ and $\eta_i$ are random strings in $\Set{0,1}^{\ell}$ for $1 \le i \le k$.\footnotemark
We denote those commitments by $\vcom=\{\com_i^0,\com_i^1\}_{i=1}^{k}$.
\item \label[Step]{item:extcom:challenge}
Upon receiving a challenge $\Vect{c} = (c_1, \ldots, c_k)$ from the receiver $R$, $S$ opens the commitments to $\Vect{v}\defeq (v^{c_1}_1, \ldots, v^{c_k}_k)$ with the corresponding decommitment $\vdecom\defeq (\decom^{c_1}_1, \ldots, \decom^{c_k}_k)$.
\item \label[Step]{item:extcom:open}
$R$ checks that the openings are valid.
\end{enumerate}
~\\
{\bf Decommitment Phase:}
\begin{itemize}[topsep=0.2em,itemsep=0.2em]
\item 
$C$ sends $\sigma$ and opens the commitments to all $k$ pairs of strings.
$R$ checks that all the openings are valid, and also that $m = v^0_1 \xor v^1_1 = \cdots = v^0_k \xor v^1_k$.
\end{itemize}
\endgroup
\end{ProtocolBox}
\footnotetext{Actually, the scheme will be secure as long as we use \Com to commit $k = \omega(\log\SecPar)$ pairs of strings.}
\para{Proof of Security.} The correctness and the statistically-binding property of $\wExtCom$ follows straightforwardly from that of $\Com$. The computationally-hiding property of $\wExtCom$ can be reduced to that of $\Com$ by standard arguments.
\begin{lemma}[Computational Hiding]\label{lem:wExtCom_comp_hiding}
$\wExtCom$ is computationally hiding.
\end{lemma}
\begin{proof}[Proof (sketch)]
Since this can be proven similarly to the classical counterpart in \cite{TCC:PasWee09}, we only give a proof sketch. For messages $m_0,m_1$ and $j\in [k+1]$, we consider a hybrid $\mathsf{Hyb}_j$ where $(v_i^0,v_i^1)$ are 2-out-of-2 secret shares of $m_0$ for 
$i\geq j$
and those of $m_1$ for $i\leq j-1$.
What we should show is that 
$\mathsf{Hyb}_0$ and $\mathsf{Hyb}_{k+1}$ are computationally indistinguishable from the view of a malicious receiver.
By a standard hybrid argument, it suffices to prove the computational indistinguishability 
between $\mathsf{Hyb}_j$ and $\mathsf{Hyb}_{j+1}$.
This can be reduced to the computational hiding property of $\Com$ by guessing $c_{j}$ and embedding the instance of computational hiding of $\Com$ into $\com_j^{1-c_j}$. The reduction works as long as the guess is correct, which occurs with probability $1/2$.
\end{proof}

We prove that $\wExtCom$ satisfies a weak version of the extractability which we call the \emph{weak extractability with $\epsilon$-simulation}.
Intuitively, it requires the simulation-extractor to perform extraction {\em and} $\epsilon$-simulation properly, \emph{as long as the commitment is valid}. 
A formal definition is given below.
\begin{definition}[Weak Extractability with $\epsilon$-Simulation]\label{def:epsilon-sim-ext-com:weak}
A commitment scheme $\Prot$ is {\em weakly extractable with $\epsilon$-simulation} if there exists a QPT algorithm $\SimExt_\weak$ (called the $\epsilon$-simulation weak-extractor) such that for any noticeable $\epsilon(\secpar)$ and any non-uniform QPT $C^*(\rho)$, 
\begin{align*}
&\big\{ \Gamma_\com(m_\ext,\widetilde{\ST}_{C^*}):(\com,m_\ext,\widetilde{\ST}_{C^*})\gets\SimExt^{C^*(\rho)}(1^\secpar,1^{\epsilon^{-1}}) \big\}_\secpar\\
\cind_\epsilon
&
\big\{ \Gamma_\com(\val_\Prot(\com),\ST_{C^*}):(\com,\ST_{C^*},b_{\mathrm{com}}) \gets \langle C^*(\rho), R \rangle(1^\secpar) \big\}_\secpar
\end{align*}
where $\Gamma_\com(m, \ST_{C^*}) \coloneqq 
\begin{cases}
(m, \ST_{C^*}) & \text{if}~ \val_\Prot(\com) \ne \bot \\
\bot & \text{otherwise}
\end{cases}$.
\end{definition}
\begin{lemma}[Weak Extractability with $\epsilon$-Simulation]\label{lem:weak_ext}
$\wExtCom$ is weakly extractable with $\epsilon$-simulation (as per \Cref{def:epsilon-sim-ext-com:weak}).
\end{lemma}


Before proving \Cref{lem:weak_ext}, we prepare several definitions.

\begin{definition}[Validness of $\vcom$]\label{def:valid_vcom}
For a sequence $\vcom=\{\com_{i}^{0},\com_{i}^{1}\}_{i=1}^{k}$ of commitments of the scheme $\Com$, we say that $\vcom$ is valid if there exists $m\in \bit^\ell$ such that 
$\val_\Com(\com_{i}^{b})\neq \bot$ for all $i\in[k]$ and $b\in \bit$ and
$\val_\Com(\com_{i}^{0})\oplus \val_\Com(\com_{i}^{1})=m$ for all $i\in[k]$ where $\val_\Com(\com_{i}^{b})$ is the value function as defined in \Cref{def:val}.
We denote by $\val_\Com(\vcom)$ to mean such $m$ if $\vcom$ is valid and otherwise  $\bot$. 
\end{definition}
\begin{definition}[Accepting Opening of $\vcom$]\label{def:accepting_opening_vcom}
For a sequence $\vcom=\{\com_{i}^{0},\com_{i}^{1}\}_{i=1}^{k}$ of commitments of the commitment scheme $\Com$ and $\Vect{c}=(c_1,...,c_k)\in \bit^k$, we say that $(\Vect{v}=(v_1,...,v_k),\vdecom=(\decom_1,...,\decom_k))$ is an accepting opening of $\vcom$ w.r.t. $\Vect{c}$ if $\Verify_\Com(\com_i^{c_i},v_i,\decom_i)=1$ for all $i\in[k]$. 
\end{definition}

Then we prove \Cref{lem:weak_ext}.
\begin{proof}[Proof of \Cref{lem:weak_ext}]
For simplicity, we assume that $\Com$ satisfies perfect binding. It is straightforward to extend the proof to the statistically binding case by excluding the bad case where any commitment of $\Com$ is not bounded to a unique message, which happens with a negligible probability.

Remark that the weak extractability with $\epsilon$-simulation only requires the extractor to correctly extract and simulate if the commitment generated in the commit stage is valid in the sense of \Cref{def:val}. 
When the commitment is valid, $\vcom$ generated in \Cref{item:extcom:com} is also valid in the sense of \Cref{def:valid_vcom} (because otherwise a committer cannot pass the verification in the decommitment stage).
Therefore, it suffices to prove that the extractor works for any fixed valid $\vcom$.

Let $C^*(\rho)$ be a non-uniform QPT malicious committer.
For $\Vect{c}\in\bit^k$, let $U_{\Vect{c}}$ be the unitary corresponding to the action of $C^*$ in \Cref{item:extcom:challenge}. That is, for the state $\rho'$ before \Cref{item:extcom:challenge}, it applies $U_{\Vect{c}}$ to get $U_{\Vect{c}}\rho' U_{\Vect{c}}^\dagger$ and measures designated registers $\regV$ and $\regD$ to get the message $\Vect{v}$ and opening information $\vdecom$ in \Cref{item:extcom:challenge}.
Let $\Pi^{\test}_{\Vect{c}}$ be the projection that maps onto states that contain an accepting opening $\Vect{v}$ and $\vdecom$ of $\vcom$ w.r.t. $\Vect{c}$ (as defined in \Cref{def:accepting_opening_vcom}) in $\regV\otimes \regD$. 
For $\Vect{c}\in\bit^k$, we define $\Pi_{\Vect{c}}\defeq U_{\Vect{c}}^{\dagger}
\Pi^{\test}_{\Vect{c}}
U_{\Vect{c}}$.

We apply \Cref{lem:extract_and_simulate} for $\{\Pi_{\Vect{c}}\}_{\Vect{c}\in\bit^k}$ with the following correspondence.
\begin{itemize}
    \item $\hil$ is the internal space of $C^*$.
    \item The initial state is $\rho'$.\footnote{Though we assume that the initial state $\ket{\psi_\init}$ is a pure state in \Cref{lem:extract_and_simulate}, the lemma holds for any mixed state since a mixed state can be seen as a probability distribution over pure states.}
    \item $C=\bit^k$.
    \item $S=\{((c_1,...,c_k),(c'_1,...,c'_k)):\exists i\in[k]\text{~s.t.~}c_i\neq c'_i\}$
    \item $\A_0$ applies $U_{\Vect{c}}$ on its input, measures $\regV$ to get $\Vect{v}$, applies $U_{\Vect{c}}^\dagger$, and outputs $\Vect{v}$.
    \item $\A_1$ is given as input $(\Vect{c},\Vect{c'})\in S$, $\Vect{v}_\Vect{c}=(v_1^{c_1},...,v_k^{c_k})$, and $\Vect{v}_\Vect{c'}=(v_1^{c'_1},...,v_k^{c'_k})$. $\A_1$ outputs $v_i^{c_i}\oplus v_i^{c'_i}$ for the smallest $i\in[k]$ such that $c_i\neq c'_i$. Note that such $i$ exists since we assume $(\Vect{c},\Vect{c'})\in S$.
\end{itemize}

If $\vcom$ is valid, we can see that the assumptions for \Cref{lem:extract_and_simulate} are satisfied as follows:
\begin{enumerate}
    \item By the definition of $S$, it is easy to see that $\frac{|S|}{|C|^2}=1-2^{-k}=1-\negl(\secpar)$.
    \item 
    For any $\Vect{c}$, if $\A_0$ takes a state in the span of $\Pi_\Vect{c}$ as input, it outputs $s_{\Vect{c}}\defeq (\val_\Com(\com_{1}^{c_1}),...,\val_\Com(\com_{k}^{c_k}))$ with probability $1$ 
    by the definition of $\Pi_{\Vect{c}}$ and the perfect binding property of $\Com$. 
    \item For any $(\Vect{c},\Vect{c'})\in S$, if $\A_1$ takes as input the $s_{\Vect{c}}$ and $s_{\Vect{c'}}$ defined as follows:
$$
\begin{cases}
s_{\Vect{c}}= (\val_\Com(\com_{1}^{c_1}), \ldots,\val_\Com(\com_{k}^{c_k}))\\
s_{\Vect{c'}}= (\val_\Com(\com_{1}^{c'_1}), \ldots,\val_\Com(\com_{k}^{c'_k}))
\end{cases};
$$
then, it outputs $s^*\defeq \val_{\Com}(\vcom)$ as defined in \Cref{def:valid_vcom} since we assume that $\vcom$ is valid. 
\end{enumerate}
Let $\widetilde{\SimExt}$ be the $\epsilon$-simulation extractor of \Cref{lem:extract_and_simulate} in the above setting. 
Then \Cref{lem:extract_and_simulate}  gives us the following:
$$
\{\widetilde{\SimExt}(1^\secpar,1^{\epsilon^{-1}},\{\Pi_{\Vect{c}}\}_{\Vect{c}\in\bit^k},\A,\rho')\}_{\secpar}
\statind_{\epsilon} 
\{\Exp(\secpar,\{\Pi_{\Vect{c}}\}_{\Vect{c}\in\bit^k},\rho')\}_{\secpar}
$$
where $\Exp(\secpar,\{\Pi_{\Vect{c}}\}_{\Vect{c}\in\bit^k},\rho')$ is as defined in \Cref{lem:extract_and_simulate}.
That is, $\Exp(\secpar,\{\Pi_{\Vect{c}}\}_{\Vect{c}\in\bit^k},\rho')$ works as follows:

\begin{itemize}
    \item Choose $\Vect{c}\sample \bit^k$.
    \item Apply the measurement $\{\Pi_{\Vect{c}},I-\Pi_{\Vect{c}}\}$ on $\rho'$.
    \begin{itemize}
        \item If the state is projected onto $\Pi_{\Vect{c}}$, the experiment outputs $\Vect{c}$, the classical string $\val_\Com(\vcom)$, and the resulting state. 
        \item If the state is projected onto $I-\Pi_{\Vect{c}}$, the experiment outputs $\Vect{c}$, $\bot$, and the resulting state. 
    \end{itemize}
\end{itemize}
One can see that the state in the third output of 
$\Exp(\secpar,\rho')$ is similar to the final state of $C^*$ in the real execution except that $C^*$ applies 
the unitary $U_{\Vect{c}}$ instead of
the measurement $\{\Pi_{\Vect{c}},I-\Pi_{\Vect{c}}\}$ and measures $\regV$ and $\regD$.  
By noting that $\Pi_{\Vect{c}}^{\test}U_{\Vect{c}}=U_{\Vect{c}}\Pi_{\Vect{c}}$ and that measuring $\regV$ and $\regD$ is the same as first applying the measurement $\{\Pi_{\Vect{c}}^{\test},I-\Pi_{\Vect{c}}^{\test}\}$ and then measuring  $\regV$ and $\regD$, if we apply $U_{\Vect{c}}$ on the third output of $\Exp(\secpar,\{\Pi_{\Vect{c}}\}_{\Vect{c}\in\bit^k},\rho')$ and then measure $\regV$ and $\regD$, the state is exactly the same as the final state of $C^*$. 


Therefore, the following extractor $\SimExt_\weak$ works for the weak $\epsilon$-simulation extractability:

\smallskip
\noindent
$\SimExt_\weak^{C^*(\rho)}(1^\secpar,1^{\epsilon^{-1}}):$
\begin{enumerate}
    \item Run the commit stage of $\wExtCom$ between $C^*(\rho)$ and the honest receiver $R$ until $C^*$ sends $\vcom$ in \Cref{item:extcom:com}.
    Let $\rho'$ be the internal state of $C^*$ at this point.
    \item Run $(\Vect{c},m_\ext,\rho_\ext)\gets \widetilde{\SimExt}(1^\secpar,1^{\epsilon^{-1}},\{\Pi_{\Vect{c}}\}_{\Vect{c}\in\bit^k},\A,\rho')$ where $\A=(\A_0,\A_1)$ is as defined above.
    Remark that the definition of $\Pi_{\Vect{c}}$ depends on $\vcom$, and it uses $\vcom$ generated in the previous step.
    \item 
    Apply $U_{\Vect{c}}$ on $\rho_\ext$ to generate  $U_{\Vect{c}}\rho_\ext U_{\Vect{c}}^\dagger$ and measures registers $\regV$ and $\regD$ to get $\Vect{v}$ and $\vdecom$.
    Let $\rho_\final$ be the state after the measurement. 
    \item Output $(m_\ext,\rho_\final)$.
\end{enumerate}
\end{proof}
\para{On the Parallel Execution of $\algo{wExtCom}$.} We can prove that $\wExtCom$ satisfies a parallel version of the weak extractability with $\epsilon$-simulation in a similar way. 
In the following, we prove that $\wExtCom$ satisfies even a generalized version of that, which we call \emph{special parallel  weak extractability with $\epsilon$-simulation}.
Looking ahead, this will be used in the proof of the (parallel) $\epsilon$-simulation strong extractability of \Cref{protocol:e-strong-ExtCom} in \Cref{sec:handling-over-extraction}. 

Intuitively, it requires the following:
Suppose that a malicious committer $C^*$ interacts with $n$ copies of the honest receiver $R$ in parallel, and let $\com_j$ be the commitment generated in the $j$-th execution. 
Suppose that $\com_j$ is valid for all $j\in V$ for some subset $V\subseteq [n]$.
Let $F:\bit^{\ell}\cup\{\bot\}\rightarrow \bit^*$ be a function that is determined by $\{\val(\com_j)\}_{j\in V}$, i.e., 
$F(m_1,...,m_n)$ takes a unique value $m^*$ as long as $m_j=\val(\com_j)$ for all $j\in V$. 
Then, the extractor can extract $m^*$ while simulating the post-execution state of $C^*$. 
A formal definition is given below.

\begin{definition}[Special Parallel Weak Extractability with $\epsilon$-Simulation ]\label{def:epsilon-sim-ext-com:parallel_weak}
We say that a commitment scheme $\Prot$ satisfies the \emph{special parallel weak extractability with $\epsilon$-simulation} if the following is satisfied.
For any integer $n=\poly(\secpar)$ and an efficiently computable function $F:\{\bit^\ell\cup \{\bot\}\}^n \rightarrow \bit^*$,  
there exists $\SimExt_F$ that satisfies the following:
For commitments $\{\com_j\}_{j=1}^{n}$, we say that $\{\com_j\}_{j=1}^{n}$ is $F$-good if it satisfies the following:
\begin{enumerate}
    \item there exists $V\subseteq [n]$ such that $\com_j$ is valid (i.e., $\val_{\Prot}(\com_j)\neq \bot$) for all $j\in V$; {\bf and} 
    \item there exists a unique
    $m^*$ such that $F(m'_1,...,m'_n)=m^*$ for all $(m'_1,...,m'_n)$ such that $m'_j=\val_\Prot(\com_j)$ for all $j \in V$. 
\end{enumerate}

Then it holds that
\begin{align*}
&\left\{ \Gamma_{F,\{\com_j\}_{j=1}^{n}}\left(m_\ext,\ST_{C^*}\right): (\{\com_j\}_{j=1}^{n},m_\ext,\ST_{C^*})\gets\SimExt_F^{C^*(\rho)}(1^\secpar,1^{\epsilon^{-1}})\right\}_{\secpar}\\
\cind_\epsilon
&\left\{
\begin{array}{l}
\Gamma_{F,\{\com_j\}_{j=1}^{n}}(F\left(\val_\Prot(\com_1),...,\val_\Prot(\com_n)),\ST_{C^*}\right)\\ :(\{\com_j\}_{j=1}^{n},\ST_{C^*},\{b_{\mathrm{com},j}\}_{j=1}^{n})\gets\execution{C^*(\rho)}{R^n}(1^\SecPar)
\end{array}
\right\}_{\secpar },
\end{align*}
where 
$(\{\com_j\}_{j=1}^{n},\ST_{C^*},\{b_{\mathrm{com},j}\}_{j=1}^{n})\sample\execution{C^*(\rho)}{R^n}(1^\SecPar)$
means that $C^*(\rho)$ interacts with $n$ copies of the honest receiver $R$ in parallel and the execution results in transcripts  $\{\com_j\}_{j=1}^{n}$, the final state $\ST_{C^*}$, and outputs $\{b_{\mathrm{com},j}\}_{j=1}^{n}$ of each copy of $R$ and
$$\Gamma_{F,\{\com_j\}_{j=1}^{n}}\left(m, \ST_{C^*}\right) \coloneqq 
\begin{cases}
(m, \ST_{C^*}) & \text{if}~ \{\com_j\}_{j=1}^n \text{~is~}F\text{-good} \\
\bot & \text{otherwise}
\end{cases}.$$
\end{definition}
\begin{lemma}[Special Parallel Weak Extractability with $\epsilon$-Simulation]\label{lem:para_weak_ext}
$\wExtCom$ satisfies the special parallel weak extractability with $\epsilon$-simulation (as per \Cref{def:epsilon-sim-ext-com:parallel_weak}).
\end{lemma}
\begin{proof}
Since the proof is very similar to that of \Cref{lem:weak_ext}, we only highlight the differences from that. 
Similarly to the proof of \Cref{lem:weak_ext}, we assume that $\Com$ satisfies perfect binding for simplicity. 

In a parallel interaction between a non-uniform QPT malicious committer $C^*(\rho)$ and $n$ copies of $R$, 
let $\wExtCom.\com_j$ be the commitment generated in the $j$-th session and 
let $\vcom_j=\{\com_{j,i}^{0},\com_{j,i}^{1}\}_{i=1}^{k}$ be the part of $\wExtCom.\com_j$ that consists of the commitments of $\Com$ generated in \Cref{item:extcom:com}.  
Remark that the special parallel weak extractability with $\epsilon$-simulation only requires the extractor to correctly extract and simulate if $\{\wExtCom.\com_j\}_{j=1}^{n}$ is $F$-good.
In this case, $\{\vcom_j\}_{j=1}^{n}$ is also $F$-good in the following sense: 
\begin{itemize}
    \item there exists $V\subseteq [n]$ such that $\vcom_j$ is valid (as per \Cref{def:valid_vcom}) for all $j\in V$, and 
   \item there exists 
    $m^*$ such that $F(m'_1,...,m'_n)=m^*$ for all $(m'_1,...,m'_n)$ such that $m'_j=\val_\Com(\vcom_j)$ for all $j\in V$.
\end{itemize}
Therefore, it suffices to prove that the extractor works for any fixed $F$-good $\{\vcom_j\}_{j=1}^{n}$.

For $\{\Vect{c}_j\}_{j=1}^{n}\in(\bit^k)^n$, let $U_{\{\Vect{c}_j\}_{j=1}^{n}}$ be the unitary corresponding to the action of $C^*$ in \Cref{item:extcom:challenge} similarly to the proof of \Cref{lem:weak_ext}.
Remark that the unitary is indexed by $\{\Vect{c}_j\}_{j=1}^{n}$ since we are considering an $n$-parallell execution.
Similarly, we let $\Pi^{\test}_{\{\Vect{c}_j\}_{j=1}^{n}}$ be the projection that maps onto states that contain accepting openings of $\vcom_j$ w.r.t. $\Vect{c}_j$ for all $j\in[n]$ in the designated registers. 
Then, we define $\Pi_{\{\Vect{c}_j\}_{j=1}^{n}}\defeq U_{\{\Vect{c}_j\}_{j=1}^{n}}^{\dagger}
\Pi^{\test}_{\{\Vect{c}_j\}_{j=1}^{n}}
U_{\{\Vect{c}_j\}_{j=1}^{n}}$.

We apply \Cref{lem:extract_and_simulate} for $\{\Pi_{\{\Vect{c}_j\}_{j=1}^{n}}\}_{\{\Vect{c}_j\}_{j=1}^{n}\in(\bit^k)^n}$ with the following correspondence.
\begin{itemize}
    \item $\hil$ is the internal space of $C^*$.
    \item The initial state is $\rho'$.
    \item $C=(\bit^k)^n$.
    \item $S=\{(\{c_{j,1},...,c_{j,k})\}_{j=1}^{n},\{c'_{j,1},...,c'_{j,k}\}_{j=1}^{n}):\forall j\in[n]~
    \exists i\in[k]\text{~s.t.~}c_{j,i}\neq c'_{j,i}\}$
    \item $\A_0$ applies $U_{\{\Vect{c}_j\}_{j=1}^{n}}$ on its input, measures the designated registers to get the $\Vect{v}_j$ for each $j\in [n]$, applies $U_{\{\Vect{c}_j\}_{j=1}^{n}}^\dagger$, and outputs $\{\Vect{v}_j\}_{j=1}^{n}$.
    \item $\A_1$ is given as input $(\{\Vect{c}_j\}_{j=1}^{n},\{\Vect{c}'_j\}_{j=1}^{n})\in S$, $\Vect{v}_{\{\Vect{c}_j\}_{j=1}^{n}}=\{v_{j,1}^{c_{j,1}},...,v_{j,k}^{c_{j,k}}\}_{j=1}^{n}$, and $\Vect{v}_{\{\Vect{c}'_j\}_{j=1}^{n}}=\{v_{j,1}^{c'_{j,1}},...,v_{j,k}^{c'_{j,k}}\}_{j=1}^{n}$. $\A_1$ 
    computes 
    $m_j\defeq v_{j,i_j}^{c_{j,i_j}}\oplus v_{j,i_j}^{c'_{j,i_j}}$ for $j\in[n]$ where $i_j\in[k]$ is the smallest index such that $c_{j,i}\neq c'_{j,i}$
    and outputs $F(m_1,...,m_n)$. 
    Note that such $i_j$ exists for all $j\in[n]$ since we assume $(\{\Vect{c}_j\}_{j=1}^{n},\{\Vect{c}'_j\}_{j=1}^{n})\in S$.
\end{itemize}

If $\{\vcom_j\}_{j=1}^{n}$ is $F$-good, we can see that the assumptions for \Cref{lem:extract_and_simulate} are satisfied as follows.
\begin{enumerate}
    \item By the definition of $S$ and the union bound, it is easy to see that $\frac{|S|}{|C|^2}\geq 1-n2^{-k}=1-\negl(\secpar)$.
    \item 
    For any $\{\Vect{c}_j\}_{j=1}^{n}$, if $\A_0$ takes a state in the span of $\Pi_{\{\Vect{c}\}_{j=1}^{n}}$ as input, it outputs $s_{\{\Vect{c}_j\}_{j=1}^{n}}\defeq \{\val_\Com(\com_{j,1}^{c_{j,1}}),...,\val_\Com(\com_{j,k}^{c_{j,k}})\}_{j=1}^{n}$ with probability $1$ 
    by the definition of $\Pi_{\{\Vect{c}\}_{j=1}^{n}}$ and the perfect binding property of $\Com$. 
    \item 
    Let $V\subseteq [n]$ be a subset for which the conditions of the $F$-goodness are satisfied. 
    For any $(\{\Vect{c}_j\}_{j=1}^{n},\{\Vect{c}'_j\}_{j=1}^{n})\in S$, if $\A_1$ takes as input the $s_{\{\Vect{c}_j\}_{j=1}^{n}}$ and $s_{\{\Vect{c}'_j\}_{j=1}^{n}}$ defined as follows:
$$
\begin{cases}
s_{\{\Vect{c}_j\}_{j=1}^{n}}\defeq \{\val_\Com(\com_{j,1}^{c_{j,1}}),\allowbreak ...,\val_\Com(\com_{j,k}^{c_{j,k}})\}_{j=1}^{n}\\
s_{\{\Vect{c}'_j\}_{j=1}^{n}}\defeq \{\val_\Com(\com_{j,1}^{c'_{j,1}}),...,\val_\Com(\com_{j,k}^{c'_{j,k}})\}_{j=1}^{n}
\end{cases};
$$
   then, we have
    $m_j=\val_{\Com}(\vcom_j)$
    for all $j\in V$ since $\vcom_j$ is valid for all $j\in V$. 
    Note that this may not hold for $j\notin V$.
    However, by the second condition of the $F$-goodness, there is $m^*$ such that $F(m_1,...,m_n)=m^*$ regardless of the values of $\{m_j\}_{j\notin V}$.
    We can set $s^*\defeq m^*$.
\end{enumerate}
Then, the rest of the proof is identical to that of \Cref{lem:weak_ext}.
\end{proof}

\subsection{Strongly Extractable Commitment}
\label{sec:handling-over-extraction}
In this section, we construct {\em strongly} extractable commitment with $\epsilon$-simulation. 
The scheme is shown in \Cref{protocol:e-strong-ExtCom}.
It relies on the following building blocks:
\begin{enumerate}
\item
the $\epsilon$-simulatable {\em weakly} extractable commitment $\wExtCom$ given in \Cref{figure:wext-com}. We remark that the security of \Cref{protocol:e-strong-ExtCom} relies on the particular $\algo{wExtCom}$ presented in \Cref{figure:wext-com} because we also need the special parallel weak extractability with $\epsilon$-simulation (\Cref{def:epsilon-sim-ext-com:parallel_weak}); we do not know if \Cref{protocol:e-strong-ExtCom} can be based on any $\wExtCom$ satisfying the weak extractability with $\epsilon$-simulation as in \Cref{def:epsilon-sim-ext-com:weak}.
\item
a $(n+1,t)$-perfectly verifiable secret sharing scheme $\VSS = (\VSS_{\Share}, \VSS_{\Recon})$ (as per \Cref{def:VSS}). We require that $t$ is a constant fraction of $n$ such that  $t \le n/3$. There are known constructions (without any computational assumptions) satisfying these properties \cite{STOC:BenGolWig88,EC:CDDHR99}.
\end{enumerate}

\kaimin{minor inconsistency in step 7-(b)? \Cref{def:view-consistency} is for MPC-in-the-Head instead of VSS. }

\xiao{Technically, you are correct. But this is not a real issue as VSS is just a special MPC. I will make this more explict by give a remark after \Cref{def:view-consistency}}

\xiao{I added it as \Cref{rmk:VSS:view-consistency}.}

\begin{ProtocolBox}[label={protocol:e-strong-ExtCom}]{$\epsilon$-Simulatable Strongly Extractable Commitment $\ExtCom$}
Let $n(\SecPar)$ be a polynomial on $\SecPar$. Let $t$ be a constant fraction of $n$ such that $t \le n/3$.

\para{Input:}
both the (committer) $C$ and the receiver $R$ get security parameter $1^\SecPar$ as the common input; $C$ gets a string $m \in \bits^{\ell(\SecPar)}$ as his private input, where $\ell(\cdot)$ is a polynomial.

\para{Commit Stage:}
\begin{enumerate}[topsep=0pt,itemsep=0pt]
\item \label[Step]{item:strong-ExtCom:commit-stage:1}
$C$ emulates $n+1$ (virtual) players $\Set{P_i}_{i \in [n+1]}$ to execute the $\VSS_\Share$ protocol ``in his head'', where the input to $P_{n+1}$ (i.e., the Dealer) is $m$. Let $\Set{\view_i}_{i\in [n+1]}$  be the views of the $n + 1$ players describing the execution.
\item \label[Step]{item:strong-ExtCom:commit-stage:2}
$C$ and $R$ involve in $n$ executions of $\wExtCom$ in parallel, where in the $i$-th instance ($i\in[n]$), $C$ commits to $\view_i$.
\item \label[Step]{item:strong-ExtCom:commit-stage:3}
$R$ picks a random string $r_1$ and commits to it using $\wExtCom$.
\item \label[Step]{item:strong-ExtCom:commit-stage:4}
$C$ picks a random string $r_2$ and sends it to $R$.
\item \label[Step]{item:strong-ExtCom:commit-stage:5}
$R$ sends to $C$ the value $r_1$ together with the corresponding decommitment information w.r.t.\ the $\wExtCom$ in \Cref{item:strong-ExtCom:commit-stage:3}. Now, both parties learn a coin-tossing result $r = r_1 \xor r_2$, which specifies a size-$t$ random subset $T\subseteq [n]$.
\item \label[Step]{item:strong-ExtCom:commit-stage:6}
$C$ sends to $R$ in {\em one round} the following messages: $\Set{\view_i}_{i \in T}$ together with the corresponding decommitment information w.r.t.\ the $\wExtCom$ in \Cref{item:strong-ExtCom:commit-stage:2}.
\item
$R$ checks the following conditions:
\begin{enumerate}[topsep=0pt,itemsep=0pt]
\item
All the decommitments in \Cref{item:strong-ExtCom:commit-stage:6} are valid; {\bf and}
\item
for any $i,j\in T$, views $(\view_i, \view_j)$ are consistent (as per \Cref{def:view-consistency} and \Cref{rmk:VSS:view-consistency}) w.r.t.\ the $\VSS_\Share$ execution as described in \Cref{item:strong-ExtCom:commit-stage:1}. 
\end{enumerate}
If all the checks pass, $R$ accepts; otherwise, $R$ rejects.
\end{enumerate}
\para{Decommit Stage:}
\begin{enumerate}[topsep=0pt,itemsep=0pt]
\item \label[Step]{item:strong-ExtCom:decommit-stage:1}
$C$ sends $\Set{\view_i}_{i\in [n]}$ together with all the corresponding information w.r.t.\ the $\wExtCom$ in \Cref{item:strong-ExtCom:commit-stage:1} of the Commit Stage.
\item \label[Step]{item:strong-ExtCom:decommit-stage:2}
$R$ constructs $\Set{\view'_i}_{i\in [n]}$ as follows: in \Cref{item:strong-ExtCom:decommit-stage:1} of the Decommit Stage, if the $i$-th decommitment is valid, $R$ sets $\view'_i \coloneqq \view_i$; otherwise, $R$ sets $\view'_i \coloneqq \bot$.
\item  \label[Step]{item:strong-ExtCom:decommit-stage:3}
$R$ outputs $m' \coloneqq \VSS_\Recon(\view'_1, \ldots, \view'_n)$.
\end{enumerate}
\end{ProtocolBox}

\para{Proof of Security.} Correctness and statistically-binding property of $\ExtCom$ follows straightforwardly from that of $\wExtCom$.
In the following, we prove that $\ExtCom$ is computationally-hiding  and (parallel) strong extractable with $\epsilon$-simulation.

\begin{lemma}[Computational Hiding]\label{lem:ExtCom_comp_hiding}
$\ExtCom$ is computationally hiding.
\end{lemma}
\begin{proof}
We consider an interaction between the honest sender $C$ and a malicious non-uniform QPT receiver $R^*(\rho)$.
We consider the following two cases:

\para{Case 1:} If $C$ rejects the decommitment in \Cref{item:strong-ExtCom:commit-stage:5}, any commitment of $\wExtCom$ generated in \Cref{item:strong-ExtCom:commit-stage:2} is not decomitted. In this case, it is straightforward to reduce the computational hiding to that of $\wExtCom$.

\para{Case 2:}
If $C$ accepts the decommitment in \Cref{item:strong-ExtCom:commit-stage:5}, the commitment of $\wExtCom$ generated in \Cref{item:strong-ExtCom:commit-stage:3} is valid (as per \Cref{def:val}).
    We consider the following hybrid experiments, where $H_0$ denotes the real execution $\execution{C}{R^*(\rho)}(1^\secpar)$:
\begin{itemize}
\item {Hybrid $H_1$:} This hybrid works similarly to $H_0$  except that we run the $\epsilon$-simulation weak extractor to extract $r_1$ from the commitment in \Cref{item:strong-ExtCom:commit-stage:3}.
    Since the commitment is valid in this case as observed above, the final output of $R^*$ is $\epsilon$-close to that in $H_0$ for arbitrarily small noticeable $\epsilon$.
    
\item {Hybrid $H_2$:} In the next hybrid, we first randomly pick a size-$t$ random subset $T$ at the beginning, and define $r_2$ so that $r=r_1\oplus r_2$ specifies $T$ in \Cref{item:strong-ExtCom:commit-stage:4}. Since $r_2$ is uniformly distributed in either case, this is perfectly indistinguishable from $H_1$.

\item {Hybrid $H_3$:} This hybrid is identical to $H_2$, except that in \Cref{item:strong-ExtCom:commit-stage:1}, for all $i \in [n]\setminus T$, we set $\view_i$ to an all-0 string of proper length.

\underline{$H_2 \cind H_3:$} We now reduce the indistinguishability between $H_2$ and $H_3$ to the computationally-hiding property of (parallel executions of)\footnote{Note that the computationally-hiding property of parallel executions follows from that of the stand-alone execution by a standard hybrid argument.} $\algo{wExtCom}$. Consider an adversary $\Adv$ participating in the hiding game of $\algo{wExtCom}$. $\Adv$ finishes \Cref{item:strong-ExtCom:commit-stage:1} as in $H_2$, where the views of the $n$ parties are denoted as $\Set{\view_i}_{i \in [n]}$. It also prepare $(n-t)$ ``null views'' $\Set{\view'_i}_{i\in [n]\setminus T}$ where each of them is an all-0 string of proper length. $\Adv$ sends $\Set{\view_i}_{i\in [n]\setminus T}$ and $\Set{\view'_i}_{i\in [n]\setminus T}$ to the external challenger as his challenges for the parallel hiding game of $\wExtCom$. Then, $\Adv$ generates the \Cref{item:strong-ExtCom:commit-stage:2} commitment (in parallel) in the following manner:\begin{itemize}
\item
For $i \in T$: $\Adv$ commits to $\Set{\view_i}_{i\in T}$ to (the internal) $R^*$ using $\algo{wExtCom}$ himself;
\item
For $i \in [n]\setminus T$: $\Adv$ relays the external challenger's $\algo{wExtCom}$ commitments to the internal $R^*$.
\end{itemize} 
Finally, $\Adv$ finishes the remaining steps of the Commit Stage as in $H_3$ (or equivalently, $H_2$). We remark that in \Cref{item:strong-ExtCom:commit-stage:6}, $\Adv$ needs to decommit to the views corresponding to $T$, which he can because these commitments are generated by $\Adv$ himself.

Now, observe that if the external challenger commits to $\Set{\view_i}_{i\in [n]\setminus T}$, then $\Adv$ is identical to $H_2$; if the external challenger commits to $\Set{\view'_i}_{i\in [n]\setminus T}$, then $\Adv$ is identical to $H_3$. Therefore, $H_2 \cind H_3$ as otherwise $\Adv$ wins the parallel hiding game.

\item {Hybrid $H_4$:} This hybrid is identical to $H_3$, except that in \Cref{item:strong-ExtCom:commit-stage:1}, we run the simulator guaranteed by the Secrecy property (\Cref{item:def:VSS:secrecy}) of $\VSS$ to fake the views of parties in set $T$: $\Set{\view'_i}_{i \in [T]} \gets \Sim^\Adv(1^\secpar, T)$, and for each $j \in [n]\setminus T$, we set $\view'_j$ to all-0 strings of proper length. Note that this hybrid does not need to know $m$ anymore. It is easy to see that $H_3 \idind H_4$ due to the perfect secrecy of $\VSS$.
\end{itemize}   
    The above argument implies that for any $m_0$ and $m_1$, the outputs of $R^*$ when the message is $m_0$ or $m_1$ are indistinguishable with advantage at most $2\epsilon$.  
    Since $\epsilon$ is arbitrarily small noticeable function, the above implies that they are computationally indistinguishable in the standard sense.

Since we know that one of Case 1 and Case 2 must happen, the overall reduction algorithm can first guess which case occurs and run the reduction algorithm corresponding to the guessed case; when the guess turns out to be incorrect, it simply outputs a uniform bit. Since the guess is correct with probability $1/2$, this reduction works with a  security loss of the multiplicative factor $1/2$.\footnote{If we use Watrous' rewinding lemma (\Cref{lem:quantum_rewinding}), we may avoid the security loss; but this is not needed here.} 
\end{proof}

In the following, we prove the (parallel-)strong extractability with $\epsilon$-simulation.
Though we finally prove the parallel version, we first give a proof for the stand-alone version since that is simpler and the proof is readily extended to that of the parallel version. 

\begin{lemma}[Strong Extractability with $\epsilon$-Simulation]\label{lem:strong_ext}
$\ExtCom$ is strongly extractable with $\epsilon$-simulation (as per \Cref{def:epsilon-sim-ext-com:strong}).
\end{lemma}
\begin{proof}
Suppose that a non-uniform QPT committer  $C^*$ interacts with the honest receiver $R$ in the commit stage of $\ExtCom$.
We consider two cases where $R$ accepts or rejects, respectively.
By using Watrous' rewinding lemma (\Cref{lem:quantum_rewinding}) in a similar way to the proof of \Cref{lem:extract_and_simulate}, 
it suffices to construct a simulator that correctly extracts and simulates for each case separately. 
Moreover, when $R$ rejects, the commitment is invalid and thus the extractor does not need to extract anything. 
Thus, there is a trivial perfect simulation extractor for this case: it can simply run the interaction between $C^*(\rho)$ and $R$ by playing the role of $R$ and outputs the final state of $C^*$.
What is left is to construct an extractor that correctly extracts and simulates assuming that $R$ accepts in the committing stage. 
That is, it suffices to prove the following claim. 
\begin{myclaim}[Extraction and Simulation for Accepting Case]\label{lem:strong_ext_acc}
There exists a QPT algorithm $\SimExt_{\Acc}$ such that for any noticeable $\epsilon(\secpar)$ and any non-uniform QPT $C^*(\rho)$, it holds that 
\begin{align*}
&\big\{ \Gamma_{b_{\mathrm{com}}}(m_\ext, \ST_{C^*}):(m_\ext,\ST_{C^*},b_{\mathrm{com}}) \gets\SimExt_{\Acc}^{C^*(\rho)}(1^\secpar,1^{\epsilon^{-1}}) \big\}_\secpar\\
\cind_\epsilon
&\big\{\Gamma_{b_{\mathrm{com}}}(\val_\ExtCom(\com), \ST_{C^*}):(\com,\ST_{C^*},b_{\mathrm{com}}) \gets \langle C^*(\rho), R \rangle(1^\secpar)\big\}_\secpar, 
\end{align*}
where 
$\Gamma_{b_{\mathrm{com}}}(m, \ST_{C^*}) \coloneqq 
\begin{cases}
(m, \ST_{C^*}) & \text{if}~ b_{\mathrm{com}}= 1 \\
\bot & \text{otherwise}
\end{cases}$.
\end{myclaim}
\begin{remark}
One may think that the above claim is similar to the weak extractability with $\epsilon$-simulation (\Cref{def:epsilon-sim-ext-com:weak}). However, the crucial difference is that the extractor $\SimExt_{\Acc}$ should declare if the simulation has succeeded by outputting  $b_{\mathrm{com}}$ in the clear. 
On the other hand, in \Cref{def:epsilon-sim-ext-com:weak}, $\SimExt_{\weak}$ is only required to indirectly declare that depending on if $\com$ is valid, which may not be known by $\SimExt_{\weak}$. 
\end{remark}
\begin{proof}[Proof of \Cref{lem:strong_ext_acc}]
Let $\wExtCom.\com_i$ be the $i$-th commitment of $\wExtCom$ in \Cref{item:strong-ExtCom:commit-stage:2} in the commit stage.
In the execution of $(\com,\ST_{C^*},b_{\mathrm{com}}) \gets \langle C^*(\rho), R \rangle(1^\secpar)$,
let $\Good$ be the event  that $\{\wExtCom.\com_i\}_{i=1}^{n}$ is $\VSS_\Recon$-good in the sense of \Cref{def:epsilon-sim-ext-com:parallel_weak}, i.e.,
\begin{itemize}
    \item there exists $V\subseteq [n]$ such that   $\wExtCom.\com_i$ is valid (i.e., $\val_\wExtCom(\wExtCom.\com_i)\neq \bot$) for all $i\in V$, and
    \item there exists $m^*$ such that $\VSS_\Recon(\view'_1, \ldots, \view'_n)=m^*$  for all  $(\view'_1, \ldots, \view'_n)$ such that 
    $$\forall i\in V, ~\view'_i=\val_\wExtCom(\wExtCom.\com_i).$$
\end{itemize}
Let $\Bad$ be the complementary event of $\Good$.
We prove the following claim.
\begin{myclaim}\label{cla:bad_negl}
It holds that 
\begin{equation}\label{eq:cla:bad_negl}
\Pr[\Bad\land b_{\mathrm{com}}=1 :(\com,\ST_{C^*},b_{\mathrm{com}}) \gets \langle C^*(\rho), R \rangle(1^\secpar)]=\negl(\secpar).
\end{equation}
\end{myclaim}
Assuming \Cref{cla:bad_negl}, it is straightforward to finish the proof of \Cref{lem:strong_ext_acc} by using \Cref{lem:para_weak_ext}.
\Cref{cla:bad_negl} means that the $\Good$ occurs whenever $b_{\mathrm{com}}=1$ except for negligible probability.
Since $\SimExt_\Acc$ is only required to correctly extract and simulate when $b_{\mathrm{com}}=1$, it suffices to give an extractor that correctly extracts and simulates when $\{\wExtCom.\com_i\}_{i=1}^{n}$ satisfies the condition for $\Good$.
Since $\wExtCom$ satisfies the special parallel weak extractability with $\epsilon$-simulation as shown in \Cref{lem:para_weak_ext}, $\SimExt_{\VSS_\Recon}$ given in  \Cref{def:epsilon-sim-ext-com:parallel_weak} (where we set $F\defeq {\VSS_\Recon}$) directly gives $\SimExt_\Acc$. Specifically, $\SimExt_\Acc$ as described below suffices for \Cref{lem:strong_ext_acc}.

\smallskip
\noindent
$\SimExt_\Acc^{C^*(\rho)}(1^\secpar,1^{\epsilon^{-1}})$:
\begin{enumerate}
    \item Run $(\{\wExtCom.\com_i\}_{i=1}^{n},m_\Ext,\ST_{C^*_2})\gets \SimExt_{\VSS_\Recon}^{C^*_{2}(\rho)}(1^\secpar,1^{\epsilon^{-1}})$
    where $C^*_2$ denotes the action of $C^*$ until \Cref{item:strong-ExtCom:commit-stage:2} in the commit stage where it outputs $\{\wExtCom.\com_i\}_{i=1}^{n}$. 
    \item Simulate the interaction between $C^*$ and $R$ from \Cref{item:strong-ExtCom:commit-stage:3} where the state of $C^*$ is initialized to be $\ST_{C^*_2}$. 
    Let 
    $b_{\mathrm{com}}$ be $R$'s decision (i.e., $b_{\mathrm{com}}=1$ if and only if $R$ accepts) and
    $\ST_{C^*}$ be the post-execution state of $S$
    \item Output $(m_\Ext,\ST_{C^*},b_{\mathrm{com}})$.
\end{enumerate}
Now, the only thing left is to prove \Cref{cla:bad_negl}.
\begin{proof}[Proof of \Cref{cla:bad_negl}]
This proof follows from a similar argument which has been used to establish the soundness of the \cite{STOC:IKOS07} commit-and-prove protocol \cite{STOC:IKOS07,FOCS:GLOV12,STOC:GOSV14,C:LiaPan21}. 

Let $\view^*_i \coloneqq \algo{val_{wExtCom}}(\algo{wExtCom.com}_i)$ for all $i \in [n]$. We now define an object called {\em inconsistency graph}. This is an undirected graph $G$ with $n$ vertices, where the $i$-th vertex corresponds to $\view^*_i$; there is an edge between vertices $i$ and $j$ in $G$ if only only if $\view^*_i$ and $\view^*_j$ are {\em inconsistent} (as per \Cref{def:view-consistency}) w.r.t.\ the $\VSS_\Share$ execution. Let $B_G$ denote that set of vertices that form a {\em minimum vertex cover}\footnote{Recall that a {\em vertex cover} of a graph is a set of vertices that includes at least one endpoint of every edge of the graph.} of $G$ (When $B_G$ is not unique, pick one arbitrarily). Next, we prove \Cref{eq:cla:bad_negl} by considering the following two possibilities based on the size of $B_G$:

\subpara{If $|B_G|\le t$:} we argue that except with negligible probability, either the even $\Good$ will happen, or we must have $b_\mathrm{com} = 0$.

To see that, consider an execution of $\VSS_\Share$ where an adversary
corrupts the set of players in $B_G$, and behaves in a way that the views of any player $P_j$, for $j \notin B_G$, is $\view^*_j$. Such an execution can be obtained by choosing all the messages from $P_j \in B_G$ to $P_j \notin B_G$ as in the view $\view^*_j$; since $B_G$ is a vertex cover, every pair of views $(\view^*_i, \view^*_j)$ with $i,j \in \overline{B_G}$  are not connected in the graph G and hence consistent. Finally, by the $(n+1, t)$-perfect verifiable-committing property of $\VSS$ (see \Cref{item:def:VSS:vc}), such a corruption should not influence the output of the honest players in the $\VSS_\Share$ stage. That is, one of the following two cases (corresponding to the two possibilities listed in \Cref{item:def:VSS:vc}) must happen:
\begin{enumerate}
\item
 During the Sharing Phase, all honest players (i.e., those in $\overline{B_G}$) disqualify the dealer. In this case, for each $i \in \overline{B_G}$, $\view^*_i$ contains a special symbol $\bot$ indicating the failure of the Sharing Phase. Recall that $R$ checks a size-$t$ random subset (determined by the coin-flipping in \Cref{item:strong-ExtCom:commit-stage:3,item:strong-ExtCom:commit-stage:4,item:strong-ExtCom:commit-stage:5}) of $\Set{v^*_i}_{i\in[n]}$. Since $|\overline{B_G}|>n-t$ and $t$ is a constant fraction of $n$, $R$ will pick at least one $v^*i$ for $i \in \overline{B_G}$ with overwhelming probability; in this case, $R$ learns the failure of the Sharing Phase and thus rejects the Commit Phase (i.e., $b_\mathrm{com} = 0$). 
 \item
 During the Sharing Phase, honest players do not disqualify the dealer. Therefore such a phase determines a unique value $m^*$ such that $\VSS_\Recon(\view^*_1, \ldots, \view^*_n) = m^*$, which implies that $\{\wExtCom.\com_i\}_{i=1}^{n}$ is $\VSS_\Recon$-good in the sense of \Cref{def:epsilon-sim-ext-com:parallel_weak} (with $\overline{B_G}$ playing the role of the set $V$ in \Cref{def:epsilon-sim-ext-com:parallel_weak}). Put in other words, the even $\Good$ is happening now.
\end{enumerate}

\begin{remark} Notice that in the above argument, it is essential that the verifiable-committing property of $\VSS$ is {\em perfect} (see \Cref{def:VSS}), because it implies that that following types of corruption do not hurt the security of $\VSS$: (1) semi-honest corruption; (2) semi-honest corruption with maliciously chosen randomness. Therefore, an ``effective corruption'' must create inconsistency with at least one honest party. Indeed, if this property is only statistical, extra efforts are needed to finish this proof. See \cite{STOC:IKOS07} for details.
\end{remark} 

\subpara{If $|B_G|> t$:} we argue that $b_{\mathrm{com}} = 1$ (i.e., $R$ accepts at the end of the Commit Stage) with at most negligible probability. Recall that $R$ checks the consistency of a size-$t$ random subset of all the views $\Set{\view^*_i}_{i\in[n]}$ (i.e., vertices in $G$). We only need to argue that such a checking will hit an edge in $G$ with overwhelming probability. For this, we use the well-known connection between the size of a minimum vertex cover to the size of a {\em maximum matching}. Concretely, the graph $G$ must have a {\em matching}\footnote{Recall that a matching is a set of edges without common vertices.} $\mathcal{M}$ of size at least $t/2$. (Otherwise, if the maximum matching contains less than $t/2$ edges, then the vertices of this matching form a vertex cover set $B$ with $|B|<t$.) Recall that if $R$ hits any edge of $G$, he will reject. The probability that the $t$ vertices (views) that $R$ picks miss all the edges of $G$ is smaller than the probability that he misses all edges
of the matching, which is again at most $2^{-\Omega(t)} = 2^{-\Omega(\secpar)}$. To see that, suppose that the first $t/2$ vertices picked by $R$ do not hit an edge of the matching. Denote this set of vertices as $S_{t/2}$. It follows from Serfling's Inequality (see \Cref{lem:Serfling}) that with overwhelming probability over $\secpar$, $S_{t/2}$ contains $\Omega(t)$ vertices that are the vertices of the edges $\mathcal{M}$. Then, their $\Omega(t)$ matching neighbors will have $\Omega(t/n) = \Omega(1)$ probability of being hit by each subsequent vertex picked by $R$. Since $R$ will pick $t/2$ more vertices, the probability that $R$ misses all the $\Omega(t)$ matching neighbors  with probability at most $2^{-\Omega(t)} = 2^{-\Omega(\secpar)}$.

This finishes the proof of \Cref{cla:bad_negl}.
\end{proof}
This finishes the proof of \Cref{lem:strong_ext_acc}.
\end{proof}
This eventually concludes the proof of \Cref{lem:strong_ext}.
\end{proof}

\begin{lemma}[Parallel-Strong Extractability with $\epsilon$-Simulation]\label{lem:para_strong_ext}
$\ExtCom$ is parallel-strongly extractable with $\epsilon$-simulation (as per \Cref{def:epsilon-sim-ext-com:parallel_strong}).
\end{lemma}
\begin{proof}
Since this lemma can be proven similarly to \Cref{lem:strong_ext}, we only highlight the differences from that.
Similarly to the proof of \Cref{lem:strong_ext}, by using Watrous' rewinding lemma \Cref{lem:quantum_rewinding}, we only have to construct an extractor that correctly extracts and simulates when $R$ accepts in all the parallel sessions. That is, it suffices to prove the following lemma.

\begin{myclaim}[Extraction and Simulation for Accepting Case]\label{lem:para_strong_ext_acc}
For any integer $N=\poly(\secpar)$, 
there exists a QPT algorithm $\SimExt_{\sfpar,\Acc}$ such that for any noticeable $\epsilon(\secpar)$ and any non-uniform QPT $C^*(\rho)$, 
\begin{align*}
&\big\{(\Gamma_{\{b_{\mathrm{com},j}\}_{j=1}^{N}}(\{m_{\Ext,j}\}_{j=1}^{N},\ST_{C^*}):
(\{m_{\Ext,j}\}_{j=1}^{N},\ST_{C^*},\{b_{\mathrm{com},j}\}_{j=1}^{N})
\gets \SimExt_{\sfpar,\Acc}^{C^*(\rho)}(1^\secpar,1^{\epsilon^{-1}}) \big\}_\secpar\\
\cind_\epsilon
&\big\{(\Gamma_{\{b_{\mathrm{com},j}\}_{j=1}^{N}}(\{\val_\ExtCom(\com_j)\}_{j=1}^{N},\ST_{C^*})):(\{\com_j\}_{j=1}^{N},\ST_{C^*},\{b_{\mathrm{com},j}\}_{j=1}^{N})\gets\execution{C^*(\rho)}{R^N}(1^\SecPar)\big\}_\secpar
\end{align*}
where 
$(\{\com_j\}_{j=1}^{N},\ST_{C^*},\{b_{\mathrm{com},j}\}_{j=1}^{N})\sample\execution{C^*(\rho)}{R^N}(1^\SecPar)$
means that $C^*(\rho)$ interacts with $N$ copies of the honest receiver $R$ in parallel and the execution results in transcripts  $\{\com_j\}_{j=1}^{N}$, the final state $\ST_{C^*}$, and outputs $\{b_{\mathrm{com},j}\}_{j=1}^{N}$ of each copy of $R$ and 
$$\Gamma_{\{b_{\mathrm{com},j}\}_{j=1}^{N}}(\{m_j\}_{j=1}^{N},\ST_{C^*}) \coloneqq 
\begin{cases}
(\{m_j\}_{j=1}^{N},\ST_{C^*}) & \text{if}~\forall~j\in[N]~ b_{\mathrm{com},j} = 1 \\
\bot & \text{otherwise}
\end{cases}.
$$
\end{myclaim}
\begin{remark}
One may think that the above claim is similar to the parallel-strong extractability with $\epsilon$-simulation (\Cref{def:epsilon-sim-ext-com:parallel_strong}). However, the crucial difference is that the above claim does not require the extractor to simulate $\ST_{C^*}$ when $b_{\mathrm{com},j} = 0$ for some $j$.
\end{remark}
Below, we prove \Cref{lem:para_strong_ext_acc} using 
the special parallel weak extractability with $\epsilon$-simulation of $\wExtCom$ (\Cref{lem:para_weak_ext}) similarly to 
the proof of \Cref{lem:strong_ext_acc}.

In an $N$-parallel execution $\execution{C^*(\rho)}{R^N}(1^\SecPar)$, 
Let $\wExtCom.\com_{j,i}$ be the $i$-th commitment of $\wExtCom$ in \Cref{item:strong-ExtCom:commit-stage:2} in the commit stage in the $j$-th session of $\ExtCom$ for $i\in[n]$ and $j\in[N]$.
In the execution of $(\{\com_j\}_{j=1}^{N},\ST_{C^*},\{b_{\mathrm{com},j}\}_{j=1}^{N}) \gets \langle C^*(\rho), R^N \rangle(1^\secpar)$,
let $\Good_j$ be the event that $\{\wExtCom.\com_{j,i}\}_{i=1}^{n}$ is $\VSS_\Recon$-good in the sense of \Cref{def:epsilon-sim-ext-com:parallel_weak}, i.e.,
\begin{itemize}
    \item there exists $V_j\subseteq [n]$ such that   $\wExtCom.\com_{j,i}$ is valid (i.e., $\val_\wExtCom(\wExtCom.\com_{j,i})\neq \bot$) for all $i\in V_j$, and
    \item 
    there exists $m^*_j\neq \bot$ such that $\VSS_\Recon(\view'_1, \ldots, \view'_n)=m^*$  for all  $(\view'_1, \ldots, \view'_n)$ such that 
    $$
    \forall i\in V_j, ~ 
    \view'_i=\val_\wExtCom(\wExtCom.\com_{j,i}).
    $$
\end{itemize}
Let $\Bad_j$ be the complementary event of $\Good_j$.
We prove the following claim.
\begin{myclaim}\label{cla:bad_negl_para}
For all $j\in[N]$, It holds that 
\begin{align*}
\Pr[\Bad_j\land  b_{\mathrm{com},j}=1 :(\{\com_j\}_{j=1}^{N},\ST_{C^*},\{b_{\mathrm{com},j}\}_{j=1}^{N}) \gets \langle C^*(\rho), R^N \rangle(1^\secpar)]=\negl(\secpar).
\end{align*}
\end{myclaim}
We can prove \Cref{cla:bad_negl_para} in exactly the same way as the proof of \Cref{cla:bad_negl} by focusing on one session while ignoring all the other sessions.

By \Cref{cla:bad_negl_para} and the union bound, $\Good_j$ occurs for all $j\in[N]$ simultaneously whenever $b_{\mathrm{com},j}=1$ for all $j\in[N]$ except for negligible probability.
Since $\SimExt_{\sfpar,\Acc}$ is only required to correctly extract and simulate when $b_{\mathrm{com},j}=1$ for all $j\in[N]$, it suffices to give an extractor that correctly extracts and simulates when $\Good_j$ occurs, i.e.,  $\{\wExtCom.\com_{j,i}\}_{i=1}^{n}$ is $\VSS_\Recon$-good for all $j\in[N]$.
In this case, it is easy to see that $(\{\wExtCom.\com_{1,i}\}_{i=1}^{n},...,\{\wExtCom.\com_{N,i}\}_{i=1}^{n})$ is $F$-good if we define
$$
F((\view_{1,1},...,\view_{1,n}),...,(\view_{N,1},...,\view_{N,n}))\defeq (\VSS_\Recon(\view_{1,1},...,\view_{1,n}),...,\VSS_\Recon(\view_{N,1},...,\view_{N,n}))
$$
with the corresponding subset  
$$
V\defeq V_1\times V_2 ...\times V_N.
$$
Thus, we can prove \Cref{lem:para_strong_ext_acc} by using the special parallel weak extractability with $\epsilon$-simulation of $\wExtCom$ (\Cref{lem:para_weak_ext}) similarly to the proof of \Cref{lem:strong_ext_acc}.
\end{proof}

\section{Black-Box $\epsilon$-Simulatable ExtCom-and-Prove in Constant Rounds}
\label{sec:e-extcom-n-prove}

\subsection{Definition}
The following definition is taken from \cite{ICALP:ChaLiaPan20,C:LiaPan21} with modifications to admit an $\epsilon$-simulation-extractable Commit Stage and an $\epsilon$-ZK Prove Stage.

\begin{definition}[$\epsilon$-Simulatable ExtCom-and-Prove]
\label{def:com-n-prove}
An $\epsilon$-Simulatable ExtCom-and-Prove scheme consists of a pair of protocols $\Prot_{\textsc{ECnP}} = (\algo{ExtCom}, \algo{Prove})$ executed between a pair of \PPT machines $P$ and $V$. Let $m\in \bits^{\ell(\secpar)}$ (where $\ell(\cdot)$ is some polynomial) is a message that $P$ wants to commit to. The protocol consists of the following stages (we omit the input $1^\secpar$ to $P$ and $V$): 
\begin{itemize}
\item 
{\bf Commit Stage:} 
$P(m)$ and $V$ execute $\algo{ExtCom}$, which generates a transcript (commitment) $\com$, 
$P$'s state $\ST_P$, and  
$V$'s decision bit $b\in \bit$ indicating acceptance (i.e., $b=1$) or rejection (i.e., $b=0$). We denote this execution as $(\com, \ST_P, b) \gets \langle P(m),V \rangle_\mathsf{EC}$. 
A malicious verifier is allowed to output any quantum state, which we denote by $\ST_{V^*}$ instead of $b$, and to keep the state for the prove stage. 

\item 
{\bf Decommit Stage:}\footnote{This stage is rarely executed in applications.} $P(\ST_P)$ generates a decommitment $\decom$ and sends it to $V$ along with a message $m$. $V$ accepts or rejects. 

\item 
{\bf Prove Stage:} 
Let $\phi$ be any predicate. $P(\ST_P, \phi)$ and $V(\com,\phi)$ execute $\algo{Prove}$, after which $V$ outputs 1 (accept) or 0 (reject). We denote the execution of this stage as $b' \gets \langle P(\ST_P), V(\com) \rangle^\phi_\mathsf{Pr}$, where $b' \in \bits$ is $V$'s output. 
A malicious verifier is allowed to output an arbitrary quantum state, which we denote by $\OUT_{V^*}$ instead of $b'$. 
\end{itemize}
The following requirements are satisfied:
\begin{enumerate}
\item \label[Property]{item:com-n-prove:condition:SimExt}
{\bf Security as $\epsilon$-Simulation Extractable Commitment.} The Commit Stage and Decommit Stage constitute a post-quantum commitment scheme (as per \Cref{def:commitment} where $P$ and $V$ play the roles of $C$ and $R$, respectively) that is 
computationally hiding (as per \Cref{def:comp_hiding}), 
statistically binding (as per \Cref{def:stat-binding}),   
and strongly extractable with $\epsilon$-simulation (as per \Cref{def:epsilon-sim-ext-com:strong}).

\item 
{\bf Completeness.} For any $m \in \bits^{\ell(\secpar)}$ and any polynomial-time computable predicate $\phi$ s.t.\ $\phi(m) = 1$, it holds that
\begin{equation}
\Pr[b=1 ~\wedge~ b' =1 : 
\begin{array}{l}
(\com,\ST_P,b) \gets \langle P(m),V \rangle_\mathsf{EC} \\
b' \gets \langle P(\ST_P), V(\com) \rangle^\phi_\mathsf{Pr}
\end{array}
] = 1.
\end{equation}

\item {\bf Soundness.} \label[Property]{item:com-n-prove:condition:soundness}
For any predicate $\phi$ and any non-uniform QPT prover $P^*(\rho)$, 
\begin{equation}
\Pr[
\begin{array}{l}
b=1 ~\wedge~ b' =1  \\ 
\wedge~ \phi(\val_{\ExtCom}(\com))=0 
\end{array}: 
\begin{array}{l}
(\com, \ST_{P^*}, b) \gets \langle P^*(\rho),V \rangle_\mathsf{EC} \\
b' \gets \langle P^*(\ST_{P^*}), V(\com) \rangle^\phi_\mathsf{Pr}
\end{array}
] = \negl(\secpar),
\end{equation}
where $\val_{\ExtCom}(\com)$ is as defined in \Cref{def:val} 
and we stipulate that $\phi(\bot) = 0$.


\item {\bf $\epsilon$-Zero-Knowledge.} \label[Property]{item:com-n-prove:condition:e-zk}
There exists a pair of QPT simulators $(\Sim_\algo{EC}, \Sim_\algo{Pr})$ such that for any $m \in \bits^{\ell(\secpar)}$, polynomial-time computable predicate $\phi$  s.t.\ $\phi(m) = 1$, any non-uniform QPT verifier $V^*(\rho)$, and any noticeable function $\epsilon(\secpar)$, the following conditions hold:
\begin{align}
& \big\{\tilde{\ST}_{V^*} : (\tilde{\ST}_{V^*}, \ST_\algo{EC})\gets\Sim_\algo{EC}^{V^*(\rho)}\big\}_{\secpar} \cind \big\{\ST_{V^*} : (\com, \ST_P, \ST_{V^*}) \gets \langle P(m),V^*(\rho) \rangle_\mathsf{EC}\big\}_\secpar \label{item:com-n-prove:ZK:condition:1}\\
&\bigg\{\tilde{\OUT}_{V^*} : 
\begin{array}{l}
(\tilde{\ST}_{V^*}, \ST_\algo{EC})\gets\Sim_\algo{EC}^{V^*(\rho)}\\
\tilde{\OUT}_{V^*}  \gets \Sim^{V^*}_\algo{Pr}(1^{\epsilon^{-1}},\tilde{\ST}_{V^*}, \ST_\algo{EC},\phi)
\end{array}
\bigg\}_{\secpar} \cind_\epsilon 
\bigg\{ \OUT_{V^*} : 
\begin{array}{l}
(\com, \ST_P, \ST_{V^*}) \gets \langle P(m),V^*(\rho) \rangle_\mathsf{EC} \\
\OUT_{V^*} \gets \langle P(\ST_P), V^*(\ST_{V^*}) \rangle^\phi_\mathsf{Pr}
\end{array}
\bigg\}_\secpar. \label{item:com-n-prove:ZK:condition:2} 
\end{align}

We refer to $\Sim_\algo{EC}$ (resp.\ $\Sim_\algo{Pr}$) as the Commit-Stage (resp.\ Prove-Stage) simulator.
\end{enumerate}
\end{definition}
\begin{remark}[On the ZK Conditions]\label{remark:com-n-prove:ZK-conditions}
\Cref{item:com-n-prove:ZK:condition:1} is optional. We include it because our construction achieves it. The $\epsilon$-zero-knowledge property defined by \Cref{item:com-n-prove:ZK:condition:2} alone should suffice for most applications. 
\end{remark}
In \Cref{lem:e-extcom-n-prove}, we show the existence of a construction satisfying \Cref{def:com-n-prove}, only assuming black-box access to post-quantum secure OWFs.
\begin{lemma}\label{lem:e-extcom-n-prove}
Assume the existence of post-quantum secure OWFs. Then, there exists a constant-round construction of $\Prot_\textsc{ECnP}$ satisfying \Cref{def:com-n-prove}. Moreover, this construction makes only black-box use of the assumed OWF.
\end{lemma}
We will prove \Cref{lem:e-extcom-n-prove} by presenting the construction (and security proof) in \Cref{sec:extcom-n-prove:construction}. Before that, we will first present three applications of such an $\epsilon$-simulatable ExtCom-and-Prove protocol in \Cref{sec:extcom-n-prove:application:EqCom:Coin-Tossing,sec:extcom-n-prove:application:e-sim-ZKAoK,sec:extcom-n-prove:application:e-ZK:QMA}.

\xiao{Sections for applications, i.e., \Cref{sec:extcom-n-prove:application:EqCom:Coin-Tossing,sec:extcom-n-prove:application:e-sim-ZKAoK,sec:extcom-n-prove:application:e-ZK:QMA} can go the the appendix.}

\subsection{Application I: $\epsilon$-Simulatable Coin-Flipping Protocols}
\label{sec:extcom-n-prove:application:EqCom:Coin-Tossing}

Let us first recall a canonical construction of two-party coin-flipping protocol from any extractable-and-equivocal string-commitment $\algo{EqExtCom}$ in the classical setting (\cite{C:CanFis01,TCC:PasWee09}). To toss $\ell(\secpar)$ coins, this construction works in 3 steps: (1) $P_1$ commits to a random string $r_1 \in \bits^{\ell(\secpar)}$ using $\algo{EqExtCom}$; (2) $P_2$ sends to $P_1$ a random string $r_2 \in \bits^{\ell(\secpar)}$; (3) $P_1$ then decommits to the $r_1$ he committed to in step (1). Both parties set $r \coloneqq r_1 \xor r_2$ as the coin-flipping result.

This protocol flips $\ell(\secpar)$ coins securely as per the simulation-based definition: To simulate for a malicious $P^*_1$, the simulator $\Sim$ extracts from step (1) the value $r^*_1$ committed by $P^*_1$ relying on the extractability of $\algo{EqExtCom}$. $\Sim$ then sends $r_2 = r^*_1 \xor \tilde{r}$ as the simulated step-(2) message, where $\tilde{r}$ is the random string $\Sim$ received from the ideal functionality in the ideal world (i.e., $\Sim$ needs to ``bias'' the coin-tossing result to $\tilde{r}$). To simulate for a malicious $P^*_2$, $\Sim$ first commit to an arbitrary string of length $\ell(\secpar)$ using $\algo{EqExtCom}$, as the simulated step-(1) message; then in step (3), relying on the equivocality of $\algo{EqExtCom}$, $\Sim$ ``decommits'' the step-(1) value to $r^*_2 \xor \tilde{r}$, where $r^*_2$ is the step-(3) message received from $P^*_2$ and $\tilde{r}$ again is the random string $\Sim$ obtained from the ideal functionality.

Our post-quantum $\epsilon$-simulatable coin-flipping protocol follows the above classical protocol but with the following modification. We observe that the $\epsilon$-simulatable ExtCom-and-Prove protocol $\Prot_\textsc{ECnP}$ can be used to achieve the same effect as a extractable-and-equivocal commitment (albeit with $\epsilon$-simulation). To see that, we just need to ``interpret'' $\Prot_\textsc{ECnP}$ in the following way:
\begin{itemize}
\item
{\bf Commit:} $C(m)$ and $R$ simply run the Commit Stage of $\Prot_\textsc{ECnP}$, where $C$ commits to its message $m$.
\item
{\bf Decommit:} To decommit to a message $m$, $C$ sends to $R$ the message $m$ only ({\em without decommitment}); then, they execute the Prove Stage of $\Prot_\textsc{ECnP}$ where $C$ proves a special predicate the $\phi_{m}(\cdot)$, which equals to $1$ if and only if the input equals to (the hard-wired) $m$.
\end{itemize}  
Such a commitment is extractable with $\epsilon$-simulation as the Commit Stage of $\Prot_\textsc{ECnP}$ is so. Also, it is equivocal with $\epsilon$-simulation because $\Prot_\textsc{ECnP}$ is  $\epsilon$-zero-knowledge. That is, a equivocator $\mathcal{E}$ can be constructed by running the $\epsilon$-ZK simulator guaranteed by \Cref{item:com-n-prove:condition:e-zk}. We present in \Cref{prot:e-sim-coin-flipping} our coin-flipping protocol constructed in black-box from $\Prot_\textsc{ECnP}$.
\begin{ProtocolBox}[label={prot:e-sim-coin-flipping}]{$\epsilon$-Simultable Coin-Flipping}
Let $\ell(\secpar)$ be a polynomial specifying the length of the desired coin-flipping result. Both parties only take the security parameter $1^\secpar$ as their input.
\begin{enumerate}
\item \label[Step]{item:e-sim-coin-flipping:step:1}
$P_1$ and $P_2$ execute the Commit Stage of $\Prot_\textsc{ECnP}$, where $P_1$ commits to a random string $r_1 \in \bits^{\ell(\secpar)}$;
\item \label[Step]{item:e-sim-coin-flipping:step:2}
$P_2$ samples a random string $r_2 \pick \bits^{\ell(\secpar)}$ and sends it to $P_1$;
\item \label[Step]{item:e-sim-coin-flipping:step:3}
$P_1$ sends $r_1$ to $P_2$, and then uses the Prove Stage of $\Prot_\textsc{ECnP}$ to prove to $P_2$ the predicate $\phi_{r_1}(\cdot)$, which equals $1$ if and only if the input equals $r_1$.
\end{enumerate}
Both parties set $r \coloneqq r_1\xor r_2$ as the coin-flipping result.
\end{ProtocolBox}
\para{On Security.} The security of \Cref{prot:e-sim-coin-flipping} can be proved following a similar argument as the above one in the classical setting: If $P^*_1$ is corrupted, we build the simulator $\Sim$ by extracting $r^*_1$ from \Cref{item:e-sim-coin-flipping:step:1} using the $\epsilon$-simulation extractor for the Commit Stage of $\Prot_\textsc{ECnP}$ (of course, setting $r_2 = r^*_1\xor \tilde{r}$ as in the above classical setting). Since such an extractor ensures that the state of $P^*_1$ after extraction is at most $\epsilon$-far from the real execution, $\Sim$ works as expected. If $P^*_2$ is corrupted, $\Sim$ will simulate the \Cref{item:e-sim-coin-flipping:step:1} by committing to an arbitrary string of length $\ell(\secpar)$, and then ``equivocate'' $r_1$ to $r_2\xor \tilde{r}$ in \Cref{item:e-sim-coin-flipping:step:3} using the simulator $\Sim_\textsc{ECnP}$ of $\Prot_\textsc{ECnP}$ (i.e., it simulates a proof for the correctness of $\phi_{r_1}$). It then follows from the $\epsilon$-ZK property of $\Prot_\textsc{ECnP}$ (more accurately, \Cref{item:com-n-prove:ZK:condition:2}) that $\Sim$ is a proper $\epsilon$-simulator for the corrupted $P^*_2$. We suppress further details as this argument is standard. In summary, we have the following corollary of \Cref{lem:e-extcom-n-prove}.
\begin{corollary}\label{cor:coin-flipping}
Assume the existence of post-quantum secure OWFs. Then, there exists a constant-round construction of $\epsilon$-simulatable coin-flipping. Moreover, this construction makes only black-box use of the assumed OWF.
\end{corollary}

\subsection{Application II: ZKAoK with $\epsilon$-Simulatable Knowledge Extractor}
\label{sec:extcom-n-prove:application:e-sim-ZKAoK}
\para{Definitions.}
We define post-quantum argument for $\NP$ and its quantum $\epsilon$-zero-knowledge property and argument of knowledge with $\epsilon$-simulation extractor property.
For a language $\Lang\in \NP$, let $\Relation_\Lang$ be the corresponding relation function, i.e., $\Relation_\Lang(x,w)=1$ if and only if $w$ is a valid witness of $x$.
We write $\Relation_\Lang(x)$ to mean the set of all valid witnesses for the statement $x$. 
. 

\begin{definition}[Post-Quantum Argument Systems for NP]\label{def:IP:NP}
A classical protocol $(P,V)$ 
with an honest PPT prover $P$ and an honest PPT verifier $V$ for a language $\Lang \in \NP$ is said to be \emph{post-quantum argument} if it  satisfies the following requirements where $\algo{OUT}_V$ denotes the final output of $V$. 
\begin{enumerate} \item {\bf Completeness.} For any $x \in \Lang$ and any $w \in \Relation_\Lang(x)$,
$$\Pr[\algo{OUT}_V\langle P(w), V\rangle(x)=1] \ge 1 - \negl(\secpar).$$ 
\item
{\bf  Computational Soundness:} For any quantum polynomial-size prover $P^*=\Set{P^*_\secpar, \rho_\secpar}_{\secpar \in \Naturals}$ and any $x \in \bits^\secpar\setminus \Lang$,
$$\Pr[\algo{OUT}_V\langle P^*_\secpar(\rho_\secpar), V\rangle(x)=1	] \le \negl(\secpar).$$

\end{enumerate}
\end{definition}

\begin{definition}[Quantum $\epsilon$-ZK for
$\NP$]\label{def:e-zk:NP}
Let $(P,V)$ be a post-quantum argument for a language $\Lang \in \NP$ as in \Cref{def:IP:NP}. 
The protocol is quantum $\epsilon$-zero-knowledge if it satisfies:
\begin{enumerate} 
\item
{\bf Quantum $\epsilon$-Zero-Knowledge.}  There exists an oracle-aided QPT simulator $\Sim$, such that for any
quantum polynomial-size verifier $V^* = \Set{V^*_\secpar, \rho_\secpar}_{\secpar \in \Naturals}$ and any noticeable function $\epsilon(\secpar)$, 
$$\big\{\algo{OUT}_{V^*_\secpar}\langle P(w), V^*_\secpar(\rho_\secpar)\rangle(x) \big\}_{\secpar, x,w} \cind_\epsilon \big\{\Sim^{V^*_\secpar(\rho_\secpar)}(1^{\epsilon^{-1}}, x) \big\}_{\secpar, x, w}, $$
where $\secpar \in \Naturals$, $x \in \Lang \cap \bits^\secpar$,  $w \in \Relation_\Lang(x)$, and $\algo{OUT}_{V^*_\secpar}$ denotes $V^*_\secpar$'s final output.
\end{enumerate} 
\end{definition}
\begin{definition}[Argument of Knowledge with $\epsilon$-Simulation Extractor]\label{def:e-aok}
A post-quantum argument system $(P,V)$ for an \NP language $\Lang$ is an argument of knowledge with $\epsilon$-simulation extractor if there exists a QPT machine $\SimExt$ such that the following holds: for any non-uniform QPT machine $P^*(\rho)$, any $x\in \bit^\secpar$, and any noticeable function $\epsilon(\cdot)$, the following condition is satisfied:
$$\big\{ (\tilde{\ST}_{P^*}, \Relation_\Lang(x, \tilde{w})): (\tilde{\ST}_{P^*}, \tilde{w}) \gets \SimExt^{P^*}(1^{\epsilon^{-1}},x,\rho)\big\}_\secpar \cind_\epsilon \big\{ (\ST_{P^*}, b): (\ST_{P^*}, b) \gets\langle P^*(x,\rho), V(x) \rangle\big\}_\secpar.$$
\end{definition}

\para{Our Construction.} To give an $\epsilon$-ZK protocol also satisfying \Cref{def:e-aok}, we first recall a canonical (constant-round but non-black-box) construction in the classical setting: the prover commits to the witness $w$ using an extractable commitment, and then proves using a zero-knowledge protocol to the verifier that the committed value is a valid witness for the concerned statement $x$. 

Our construction is obtained by observing that in the above protocol,  what the prove does is essentially an ExtCom-and-Prove. I.e., the prover's initial commitment can be interpreted as the Commit Stage of the ExtCom-and-Prove, committing to the witness $w$. The subsequent zero-knowledge protocol can be viewed as executing the Prove Stage of the ExtCom-and-Prove where the prover prove a special predicate $\phi_x(\cdot)$, for which $\phi_x(w)$ if and only if its input $w$ is a valid witness $w$ (i.e., $\Relation_\Lang(x, w) = 1$). We present the construction in \Cref{prot:e-sim-ZKAoK}.
\begin{ProtocolBox}[label={prot:e-sim-ZKAoK}]{Zero-Knowledge Argument of Knowledge with $\epsilon$-Simulation Extractor}
{\bf Input:} both parties have the statement $x$ and the security parameter $1^\secpar$ as the common input; $P$ additionally obtains the witness $w \in \Relation_\Lang(x)$ as its private input.
\begin{enumerate}
\item \label[Step]{item:e-sim-ZKAoK:1}
$P$ and $V$ execute the Commit Stage of $\Prot_\textsc{ECnP}$, where $P$ commits to $w$.
\item \label[Step]{item:e-sim-ZKAoK:2}
$P$ and $V$ execute the Prove Stage of $\Prot_\textsc{ECnP}$, where $P$ proves the special predicate $\phi_x(\cdot)$, for which $\phi_x(w)=1$ if and only if $\Relation_\Lang(x, w) = 1$.
\end{enumerate}
{\bf Verifier's Decision.} $V$ accepts if and only if both the Commit and Prove Stage are convincing.
\end{ProtocolBox}

\para{On Security.} It is easy to see that soundness of \Cref{prot:e-sim-ZKAoK} follows from the soundness of $\Prot_\textsc{ECnP}$ (\Cref{item:com-n-prove:condition:soundness}), and that $\epsilon$-ZK property of \Cref{prot:e-sim-ZKAoK} follows from that of $\Prot_\textsc{ECnP}$ (in particular, \Cref{item:com-n-prove:ZK:condition:2} in \Cref{item:com-n-prove:condition:e-zk}). To show that \Cref{prot:e-sim-ZKAoK} is a argument of knowledge with $\epsilon$-simulation extractor (as per \Cref{def:e-aok}), we rely on the $\epsilon$-simulatable extractability (\Cref{item:com-n-prove:condition:SimExt}) of the Commit Stage of $\Prot_\textsc{ECnP}$. In more detail, the $\epsilon$-simulation knowledge extractor $\SimExt$ can be construct as follows: $\SimExt$ extracts the value committed by the malicious prover $P^*$ in \Cref{item:e-sim-ZKAoK:1}, while also performing a $\epsilon$-close simulation for $P^*$'s internal state after the extraction. We emphasize that the Commit Stage of $\Prot_\textsc{ECnP}$ is a $\epsilon$-simulatable {\em strongly} extractable commitment (as per \Cref{def:epsilon-sim-ext-com:strong}); that is, $\SimExt$ (using the $\epsilon$-simulation extractor ensured by \Cref{def:epsilon-sim-ext-com:strong}) can always extract the value committed in \Cref{item:e-sim-ZKAoK:1} while performing a $\epsilon$-simulation for $P^*$'s state, {\em even if the \Cref{item:e-sim-ZKAoK:1} commitment is invalid (in which case, the committed value is define as $\bot$).} Thus, \Cref{def:e-aok} will be satisfied. Since the above arguments for security are standard, we omit the details. In summary, we obtain the following corollary of \Cref{lem:e-extcom-n-prove}.
\begin{corollary}\label{cor:e-ZKAoK}
Assume the existence of post-quantum secure OWFs. Then, there exists a constant-round construction of $\epsilon$-zero-knowledge argument of knowledge with an $\epsilon$-simulation knowledge extractor for $\NP$. Moreover, this construction makes only black-box use of the assumed OWF.
\end{corollary}

\begin{xiaoenv}{Compare with \cite{ananth2021concurrent}}
Compare with \cite{ananth2021concurrent}:
\begin{itemize}
\item
\cite{ananth2021concurrent} is a proof of knowledge; ours is an argument of knowledge,
\item
\cite{ananth2021concurrent} achieves $\negl(\secpar)$-simulation; ours achieve only $\epsilon$-simulation.
\item
\cite{ananth2021concurrent} works in the bounded-concurrent setting; ours only works in the stand-alone setting.
\item
\cite{ananth2021concurrent} has polynomial rounds; ours is constant-round.
\item
\cite{ananth2021concurrent} assumes post-quantum OTs (IND-based definition) where receiver's privacy is statistical, which they build from QLWE; ours assumes only QOWF. By replacing the assumption to statistically-hiding, collapse-binding commitment (also known from QLEW), we our result can also be Proof of knowledge while reserving all other properties (i.e., black-box, constant-round, $\epsilon$-simulatable knowledge extractor).
\item
\cite{ananth2021concurrent} is non-black-box; ours is black-box.
\item
\cite{ananth2021concurrent} say nothing about simulation when $P^*$ does not convince the verifier (this is unclear because the definitino of $\tilde{\rho}$ in \cite{ananth2021concurrent}  is ambiguous); our definition requires simulation regardless of whether $P^*$ convinces $V$ or not. We remark that this stronger version is important for many applications. For example, when the ZKAoK protocol is used as a building black for some larger 2PC protocol to achieve simulation-based definition, simulating $P^*$'s state is necessary even if $V$ rejects.
\end{itemize}
\end{xiaoenv}
\kaimin{For the last point, I checked \cite{ananth2021concurrent}. $\tilde{\rho}$ refers to the final state of the prover without conditioning on accept, so it also simulates $P^*$ when $V$ rejects as well.}
\takashi{Xiao and I discussed about the definition, and we found that the simulation without conditioning is too weak for some applications where the joint distribution of the state and the decision is visible to an adversary. That's why we introduce a slightly different (stronger) definition.} \kaimin{If the decision is known to $P^*$, then it's also encoded in $\tilde{\rho}$} \xiao{only true for publicly verifiable ZK??}
\xiao{Btw, we won't keep the above ``comparison with [ACP21]'', as takashi already included some comparison/discussion in the intro.} \kaimin{I see, so it's not in a rush to figure this out :)}

\subsection{Application III: Black-Box $\epsilon$-ZK for QMA}
\label{sec:extcom-n-prove:application:e-ZK:QMA}
\para{Definitions.} We first present the definition of $\QMA$. Note that we formalize $\QMA$ problems as promise problems. This is because the most ZK-friendly $\QMA$-complete problem (known currently) is the Consistency of Local Density Matrices (CLDM) problem, which is in the form of a promise problem. We refer interested readers to \cite{FOCS:BroGri20} for details.
\begin{definition}[$\QMA$]\label{def:QMA}
We say that a promise problem $\Lang=(\Lang_\yes,\Lang_\no)$ is in $\QMA$ if there is 
a polynomial $\ell$ and a QPT algorithm $V$ such that the following is satisfied: 
\begin{itemize}
    \item For any $x \in \Lang_\yes $, there exists a quantum state $w$ of $\ell(|x|)$-qubit (called a witness) such that we have $\Pr[V(x,w)=1]\geq 2/3$.
    \item For any $x\in \Lang_\no$ and any quantum state $w$ of $\ell(|x|)$-qubit, we have $\Pr[V(x,w)=1]\leq 1/3$.
\end{itemize}
For any $x \in \Lang_\yes$, we denote by $R_\Lang(x)$ to mean the (possibly infinite) set of all quantum states 
$w$ such that $\Pr[V(x,w)=1]\geq 2/3$.
\end{definition}
Next, we define
quantum $\epsilon$-ZK for $\QMA$. This definition is taken from
\cite{STOC:BitShm20} with modifications to accommodate the
$\epsilon$-simulation. 
\begin{definition}[Quantum Proof and Argument Systems for QMA]\label{def:IP:QMA}
A quantum protocol $(P,V)$ 
with an honest QPT prover $P$ and an honest QPT verifier $V$ for a promise problem $\Lang=(\Lang_\yes,\Lang_\no) \in \QMA$ is said to be a quantum proof or argument system if it satisfies the following requirements where $\algo{OUT}_V$ denotes $V$'s final output.
\begin{enumerate} \item {\bf Completeness.} There is a polynomial $k(\cdot)$
s.t.\ for any $x \in \Lang_\yes$ and any $w \in \Relation_{\Lang}(x)$,
$$\Pr[\algo{OUT}_V\langle P(w^{\otimes k(\secpar)}), V\rangle(x)=1] \ge 1 - \negl(\secpar).$$ 
\item
{\bf Soundness:} the protocol satisfies one of the following
\begin{itemize}
\item
{\bf  Computational Soundness:} For any quantum polynomial-size prover $P^*=\Set{P^*_\secpar, \rho_\secpar}_{\secpar \in \Naturals}$ and any $x \in \Lang_\no \cap \bit^\secpar$,
$$\Pr[\algo{OUT}_V\langle P^*_\secpar(\rho_\secpar), V\rangle(x)=1	] \le \negl(\secpar).$$
A protocol with computational soundness is called an argument.
\item
{\bf  Statistical Soundness:} For any (potentially unbounded) $P^*$ and any $x \in \Lang_\no\cap \bit^\secpar$,
$$\Pr[\algo{OUT}_V\langle P^*, V\rangle(x)=1	] \le \negl(\secpar).$$
A protocol with statistical soundness is called a proof.
\end{itemize}

\end{enumerate}
\end{definition}

\begin{definition}[Quantum $\epsilon$-ZK for
$\QMA$]\label{def:e-zk:QMA}
 Let (P,V) be a quantum protocol (argument or
proof) for a language $\Lang \in \QMA$ as in \Cref{def:IP:QMA}, where the prover uses $k(\secpar)$ copies of a witness.
The protocol is quantum $\epsilon$-zero-knowledge if it satisfies:
\begin{enumerate} 
\item
{\bf Quantum $\epsilon$-Zero-Knowledge.}  There exists an oracle-aided QPT simulator $\Sim$, such that for any
quantum polynomial-size verifier $V^* = \Set{V^*_\secpar, \rho_\secpar}_{\secpar \in \Naturals}$ and any noticeable function $\epsilon(\secpar)$, 
$$\big\{\algo{OUT}_{V^*_\secpar}\langle P(w^{\otimes k(\secpar)}), V^*_\secpar(\rho_\secpar)\rangle(x) \big\}_{\secpar, x,w} \cind_\epsilon \big\{\Sim^{V^*_\secpar(\rho_\secpar)}(1^{\epsilon^{-1}}, x) \big\}_{\secpar, x, w}, $$
where $\secpar \in \Naturals$, $x \in \Lang_\yes \cap \bits^\secpar$,  $w \in \Relation_\Lang(x)$, and $\algo{OUT}_{V^*_\secpar}$ denotes the $V^*_\secpar$'s final output. 
\end{enumerate} 
\end{definition}

We present in \Cref{def:sigm:QMA} the definition of quantum sigma protocols for $\QMA$, which will be used as a building block for our construction. This definition is again take from \cite{STOC:BitShm20}. We emphasize that the verifier's message $\beta$ must be a classical string. As observed in \cite{STOC:BitShm20},  the parallel version of \cite{FOCS:BroGri20} satisfies \Cref{def:sigm:QMA}.
\begin{definition}[Quantum Sigma Protocol for QMA]\label{def:sigm:QMA}
 A quantum sigma protocol for $\Lang=(\Lang_\yes,\Lang_\no) \in \QMA$ is a
quantum proof system $(\Xi.P, \Xi.V)$ (as per \Cref{def:IP:QMA}) with 3 messages and the following syntax.
\begin{itemize}
\item
$(\alpha, \tau)\gets\Xi.P_1(x, w^{\otimes k(\secpar)}):$ Given an $x \in \Lang_\yes \cap \bits^\secpar$ and $k(\secpar)$ witnesses $w \in \Relation_\Lang(x)$ (for a polynomial $k(\cdot)$),  the first prover execution outputs a public message $\alpha$ for $\Xi.V$ and a private
inner state $\tau$.
\item
$\beta \gets \Xi.V(x):$  The verifier simply outputs a {\em classical string} of $\poly(|x|)$ random bits.
\item
$\gamma \gets \Xi.P_3(\beta, \tau):$ Given the verifier's string $\beta$ and the private state $\tau$, the prover outputs a response $\gamma$.
\end{itemize}
The protocol satisfies the following:
\begin{enumerate}
\item \label[Property]{item:sigma:QMA:special-zk}
{\bf Special Zero-Knowledge:} There exists a QPT simulator $\Xi.\Sim$ such that,
$$\big\{(\alpha, \gamma): (\alpha, \tau)\gets\Xi.P_1(x, w^{\otimes k(\secpar)}); \gamma \gets \Xi.P_3(\beta, \tau) \big\}_{\secpar, x, w, \beta} \cind \big\{ (\alpha, \gamma): (\alpha, \gamma) \gets \Xi.\Sim(x, \beta)\big\}_{\secpar, x, w, \beta},$$
where $\secpar \in \Naturals$, $x\in \Lang_\yes\cap \bit^\secpar$, $w \in \Relation_\Lang(x)$, and $\beta \in \bits^{\poly(\secpar)}$.
\end{enumerate}
\end{definition}
\para{The \cite{STOC:BitShm20} Protocol.} Since our construction is essentially a black-box version of the protocol from \cite[Section 5]{STOC:BitShm20}, it is helpful to recall the \cite{STOC:BitShm20} protocol and its security proof. We will show how to make it black-box (but $\epsilon$-ZK) using our black-box $\epsilon$-simulatable ExtCom-and-Prove.

In the \cite{STOC:BitShm20} construction, the verifier first executes the $\beta\gets \Xi.V(x)$ algorithm of the quantum sigma protocol defined in \Cref{def:sigm:QMA}, and commits to $\beta$ using a post-quantum simulation-extractable commitment. Then, the prover and the verifier will execute the quantum sigma protocol but with one modification: in the 2nd round of the quantum sigma protocol (where the verifier is supposed to send his message $\beta$), the verifier will send the value $\beta$ that he committed at the beginning {\em without decommitment}, and then give a post-quantum ZK argument to convince the prover that the $\beta$ he sends is indeed the one he committed earlier; Other parts of the quantum sigma protocol are executed as they should be. 

To see why this construction is sound, consider a hybrid where the verifier commits at the beginning an arbitrary value (say, an all-0 string) of proper length. Then, in the 2nd round of the quantum sigma protocol, the verifier generates $\beta$ and sends it to the (malicious) prover; the verifier uses the zero-knowledge simulator to ``fake'' the ZK argument for the consistency between $\beta$ and his initial commitment. Because of the hiding property of the verifier's commitment and the zero-knowledge property of the ZK argument, this hybrid is indistinguishable with the real execution from the prover's point of view. However, notice in this hybrid that we successfully delayed the sampling of $\beta$ to the 2nd round of the quantum sigma protocol. Thus, the soundness of this protocol can be reduced to that of the quantum sigma protocol in a straightforward manner. 

To show that the construction is zero-knowledge, a simulator will extract the $\beta$ from the (malicious) verifier's initial commitment; this can be done without being noticed by the malicious verifier because of the simulatable-extractability of the verifier's commitment. That is, the simulator learns $\beta$ even before the quantum sigma protocol started; moreover, such an extraction only disturbs the state of the malicious verifier negligibly. Therefore, the ZK property can be reduced to the special-ZK property (\Cref{item:sigma:QMA:special-zk}) of the quantum sigma protocol using a standard argument. 

\para{Our Construction.} The \cite{STOC:BitShm20} protocol is not black-box because: (1) the verifier runs a zero-knowledge argument on a {\em cryptographic} statement (i.e., $\beta$ is the value committed in the beginning); (2) the constant-round post-quantum zero-knowledge protocol constructed in \cite{STOC:BitShm20} is not black-box; indeed, they use non-black-box techniques both for the construction and simulation. 

Our protocol is obtained by observing that what the verifier does in the \cite{STOC:BitShm20} protocol is exactly a fully-simulatable ExtCom-and-Prove. I.e., the verifier's initial commitment can be viewed as a simulation-extractable commitment to $\beta$, constituting the Commit Stage; later when he sends $\beta$ and proofs the consistency, he is essentially giving a post-quantum ZK proving the value in his initial commitment satisfies a special predicate $\phi_\beta(\cdot)$, for which $\phi_\beta(x)=1$ if and only if $x = \beta$.

Thus, we can simply replace the verifier's commitment and zero-knowledge argument with our black-box $\epsilon$-simulatable ExtCom-and-Prove protocol $\Prot_\textsc{ECnP}$. Notice that the Prove Stage of $\Prot_\textsc{ECnP}$ is black-box on the commitment of its Commit Stage; also, the construction of the Prove Stage itself is black-box (albeit $\epsilon$-ZK, instead of fully-simulatable ZK). This bypasses the aforementioned two sources of non-black-boxness in the \cite{STOC:BitShm20} protocol.

It is also worth noting that our construction does not change the structure of the \cite{STOC:BitShm20} protocol. We simply replace the non-black-box components with a black-box counterpart. Therefore, the security proof of the \cite{STOC:BitShm20}  protocol extends smoothly to our construction (with slight modifications to accommodate the $\epsilon$-simulation in proving the ZK property). Therefore, we suppress further details for the security proof of our construction. For completeness, we present our construction in \Cref{prot:e-zk:QMA}, which makes black-box use of a quantum sigma protocol $(\Xi.P, \Xi.V)$ as per \Cref{def:sigm:QMA}, and our $\epsilon$-simulatable ExtCom-and-Prove $\Prot_\textsc{ECnP}$ as per \Cref{def:com-n-prove}.
\begin{ProtocolBox}[label={prot:e-zk:QMA}]{Quantum $\epsilon$-Zero-Knowledge Argument for \bf{QMA}}
{\bf Common Input:} An instance $x \in \Lang_\yes \cap \bits^\secpar$, for security parameter $\secpar\in \Naturals$.

\para{Prover's private input:} Polynomially many identical witnesses for $x$: $w^{\otimes k(\secpar)}$ s.t.\ $w \in \Relation_\Lang(x)$;
\begin{enumerate}
\item
$V$ computes $\beta \gets \Xi.V(x)$. $V$ and $P$ then execute the Commit Stage of $\Prot_\textsc{ECnP}$, where $V$ commits to $\beta$;
\item
$P$ computes $(\alpha,\tau) \gets \Xi.P_1(x, w^{\otimes k(\secpar)})$ and sends $\alpha$ to $V$;
\item
$V$ sends $\beta$ to $P$. Now, both parties agree on a predicate $\phi_\beta(\cdot)$, for which $\phi_\beta(x) = 1$ if and only if $x = \beta$;
\item
$V$ and $P$ execute the Prove Stage of $\Prot_\textsc{ECnP}$, where $V$ proves the predicate $\phi_\beta$ defined in the last step.  If
the argument was not convincing; $P$ terminates communication outputting $\bot$; otherwise, the protocol continues.
\item
 $P$ computes $\gamma \gets \Xi.P_3(\beta, \tau)$ and sends $\gamma$;
\end{enumerate}
{\bf Verifier's Decision:} $V$ accepts if and only if $1 = \Xi.V(\alpha, \beta, \gamma)$.
\end{ProtocolBox} 
\subsection{Our Construction of ExtCom-and-Prove (Proof of \Cref{lem:e-extcom-n-prove})}
\label{sec:extcom-n-prove:construction}
The construction is shown in \Cref{protocol:e-ExtCom-n-Prove}. It makes black-box use of the following building blocks:
\begin{enumerate}
\item
The $\epsilon$-simulatable, {\em parallel-strong} extractable (as per \Cref{def:epsilon-sim-ext-com:parallel_strong}) commitment $\ExtCom$ constructed in \Cref{sec:handling-over-extraction}, which in turn makes black-box use of any post-quantum secure OWFs. 

\item
	A statistically-binding, computationally-hiding (against QPT adversaries) commitment $\Com$. This is also known assuming only black-box access to post-quantum secure OWFs.
\item
	 A $(n+1,t)$-perfectly secure verifiable secret sharing scheme $\VSS = (\VSS_{\Share}, \VSS_{\Recon})$ (see \Cref{add-prelim:VSS});
\item
	A $(n,t)$-perfectly secure MPC protocol $\Prot_\textsc{mpc}$ (see \Cref{add-prelim:mpc_in_the_head});
\end{enumerate}
For the VSS and MPC protocols, we require that $t$ is a constant fraction of $n$ such that  $t \le n/3$. There are information-theoretical constructions satisfying these properties \cite{STOC:BenGolWig88,EC:CDDHR99}. 

\begin{ProtocolBox}[label={protocol:e-ExtCom-n-Prove}]{$\epsilon$-Simulatable ExtCom-and-Prove}
{\bf Parameter Setting:} Let $n(\SecPar)$ be a polynomial on $\SecPar$. Let $t$ be a constant fraction of $n$ such that $t \le n/3$.

\para{Input:}
Both $P$ and the receiver $V$ get $1^\SecPar$ as the common input; $P$ gets a string $m \in \bits^{\ell(\SecPar)}$ as his private input, where $\ell(\cdot)$ is a polynomial.

\para{Commit Stage:}
\begin{enumerate}[topsep=0pt,itemsep=0pt]
\item \label[Step]{item:e-ExtCom-n-Prove:commit-stage:1}
$P$ emulates $n+1$ (virtual) players $\Set{P_i}_{i \in [n+1]}$ to execute the $\VSS_\Share$ protocol ``in his head'', where the input to $P_{n+1}$ (i.e., the Dealer) is $m$. Let $\Set{\view_i}_{i\in [n+1]}$  be the views of the $n + 1$ players describing the execution.
\item \label[Step]{item:e-ExtCom-n-Prove:commit-stage:2}
$P$ and $V$ involve in $n$ executions of $\ExtCom$ in parallel, where in the $i$-th instance ($i\in[n]$), $P$ commits to $\view_i$.
\end{enumerate}
\para{Decommit Stage:}
\begin{enumerate}[topsep=0pt,itemsep=0pt]
\item  \label[Step]{item:e-ExtCom-n-Prove:decommit-stage:1}
$P$ sends $\Set{\view_i}_{i\in [n]}$ together with the corresponding decommitment information w.r.t.\ the $\ExtCom$ in \Cref{item:e-ExtCom-n-Prove:commit-stage:2} of the Commit Stage.
\item
$V$ checks that all the decommitments in \Cref{item:e-ExtCom-n-Prove:decommit-stage:1} of the Decommit Stage are valid. If so, $V$ outputs $\VSS_\Recon(\view_1, \ldots, \view_n)$ and then halts; otherwise, $V$ outputs $\bot$ and then halts.
\end{enumerate}

\para{Prove Stage:} both parties learn a polynomial-time computable predicate $\phi$.
\begin{enumerate}[topsep=0pt,itemsep=0pt]
\item \label[Step]{item:e-ExtCom-n-Prove:prove-stage:1}
$P$ emulates ``in his head'' $n$ (virtual) players $\Set{P_i}_{i \in [n]}$, where $P_i$'s input is $\view_i$ (from \Cref{item:e-ExtCom-n-Prove:commit-stage:1} of the Commit Stage). These $n$ parties execute $\Prot_\textsc{mpc}$ for the following functionality: the functionality reconstructs $m' \coloneqq \VSS_\Recon(\view_1, \ldots, \view_n)$ and sends the value $\phi(m')$ to all the parties as their output. For $i \in [n]$, let $\view'_i$ be the view of party $P_i$ during $\Prot_\textsc{mpc}$. 
\item \label[Step]{item:e-ExtCom-n-Prove:prove-stage:2}
$P$ and $V$ involve in $n$ executions of $\Com$ in parallel, where in the $i$-th instance ($i\in[n]$), $P$ commits to $\view'_i$. 
\item \label[Step]{item:e-ExtCom-n-Prove:prove-stage:3}
$V$ picks a random string $r_1$ and commits to it using $\ExtCom$.
\item \label[Step]{item:e-ExtCom-n-Prove:prove-stage:4}
$P$ picks a random string $r_2$ and sends it to $V$.
\item \label[Step]{item:e-ExtCom-n-Prove:prove-stage:5}
$V$ sends to $P$ the value $r_1$ together with the corresponding decommitment information w.r.t.\ the $\ExtCom$ in \Cref{item:e-ExtCom-n-Prove:prove-stage:3}. Now, both parties learn a coin-tossing result $r = r_1 \xor r_2$, which specifies a size-$t$ random subset $T\subseteq [n]$.
\item \label[Step]{item:e-ExtCom-n-Prove:prove-stage:6}
$P$ sends to $V$ in {\em one round} the following messages:
\begin{enumerate}[topsep=0pt,itemsep=0pt]
\item \label[Step]{item:e-ExtCom-n-Prove:prove-stage:6:a}
$\Set{\view_i}_{i \in T}$ together with the corresponding decommitment information w.r.t.\ the $\ExtCom$ in \Cref{item:e-ExtCom-n-Prove:commit-stage:2} of the Commit Stage; {\bf and}
\item \label[Step]{item:e-ExtCom-n-Prove:prove-stage:6:b}
$\Set{\view'_i}_{i \in T}$ together with the corresponding decommitment information w.r.t.\ the $\Com$ in \Cref{item:e-ExtCom-n-Prove:prove-stage:2} of the Prove Stage.
\end{enumerate}
\item 
$V$ checks the following conditions:
\begin{enumerate}[topsep=0pt,itemsep=0pt]
\item \label[Step]{item:e-ExtCom-n-Prove:prove-stage:7:a}
All the decommitments in \Cref{item:e-ExtCom-n-Prove:prove-stage:6:a,item:e-ExtCom-n-Prove:prove-stage:6:b} are valid; {\bf and}
\item \label[Step]{item:e-ExtCom-n-Prove:prove-stage:7:b}
for any $i\in T$, $\view_i$ is the prefix of $\view'_i$ ; {\bf and}
\item \label[Step]{item:e-ExtCom-n-Prove:prove-stage:7:c}
for any $i,j\in T$, views $(\view'_i, \view'_j)$ are consistent (as per \Cref{def:view-consistency} and \Cref{rmk:VSS:view-consistency}) w.r.t.\ the 
$\VSS_\Share$ execution in \Cref{item:e-ExtCom-n-Prove:commit-stage:1} of the Commit Stage
and the $\Prot_\textsc{mpc}$ execution as described in \Cref{item:e-ExtCom-n-Prove:prove-stage:1} of the Prove Stage.
\end{enumerate}
If all the checks pass, $V$ accepts; otherwise, $V$ rejects.
\end{enumerate}
\end{ProtocolBox}
\para{Proof of Security.} We now prove that \Cref{protocol:e-ExtCom-n-Prove} satisfies \Cref{def:com-n-prove}. It is straightforward to see that \Cref{protocol:e-ExtCom-n-Prove} is constant-round and makes only black-box access to OWFs. Completeness follows from that of $\VSS$, $\ExtCom$, $\Com$, and $\Prot_\textsc{mpc}$. In the following, we show $\epsilon$-simulatable extractability (in \Cref{lem:e-ExtCom-n-Prove:e-sim-ext}), soundness (in \Cref{lemma:e-ExtCom-n-Prove:soundness}), and $\epsilon$-zero-knowledge (in \Cref{lemma:e-ExtCom-n-Prove:e-zk}).


\begin{lemma}[$\epsilon$-Simulation Extractability]\label{lem:e-ExtCom-n-Prove:e-sim-ext}
Assume $\algo{ExtCom}$ is parallel-strongly extractable with $\epsilon$-simulation (as per \Cref{def:epsilon-sim-ext-com:parallel_strong}). Then,  \Cref{protocol:e-ExtCom-n-Prove} satisfies security as $\epsilon$-simulation extractable commitment defined in \Cref{item:com-n-prove:condition:SimExt} in \Cref{def:com-n-prove}.
\end{lemma}

\begin{proof}
First, notice that the statistically-binding property and computationally-hiding property follows straightforwardly from those of $\algo{ExtCom}$ via standard argument. In the following, we prove  strong extractability with $\epsilon$-simulation.

A commitment $\com$ generated in the commit stage of \Cref{protocol:e-ExtCom-n-Prove} consists of $n$ commitments of $\ExtCom$, which we denote by $\{\ExtCom.\com_i\}_{i=1}^{n}$. 
By the statistical binding property of $\ExtCom$, $\ExtCom.\com_i$ can be opened to only $\view_i\defeq \val_{\ExtCom}(\ExtCom.\com_i)$ for all $i\in[n]$ except for negligible probability. 
Then, by the definition of the decommit stage of \Cref{protocol:e-ExtCom-n-Prove}, we can see that $\val_{\text{\Cref{protocol:e-ExtCom-n-Prove}}}(\com)=\VSS_\Recon(\view_1,...,\view_n)$ where we define $\VSS_\Recon(\view_1,...,\view_n)$ to be $\bot$ when one of $\view_i$'s is $\bot$. 
Moreover, remark that the verifier (which plays the role of a receiver as an extractable commitment scheme) of \Cref{protocol:e-ExtCom-n-Prove} accepts in the commit stage if and only if the prover passes verification of the commit stage of $\ExtCom$ in all the sessions and thus
$\val_{\text{\Cref{protocol:e-ExtCom-n-Prove}}}(\com)=\bot$ when the verifier rejects in any of the sessions.  
Thus, the extractor for the parallel strong extractability with $\epsilon$-simulation of $\ExtCom$ can be directly used as an extractor for the strong extractability with $\epsilon$-simulation of  \Cref{protocol:e-ExtCom-n-Prove}.
\end{proof}

\begin{lemma}[Soundness]\label{lemma:e-ExtCom-n-Prove:soundness}
Assume $\ExtCom$ and $\Com$ are statistically binding, $\ExtCom$ is computationally-hiding, $\VSS$ is $(n+1,t)$-perfectly verifiable-committing (see \Cref{def:VSS}) and $\Prot_\textsc{mpc}$ is $(n,t)$-perfectly robust (see \Cref{def:t-robustness}). Then, \Cref{protocol:e-ExtCom-n-Prove} satisfies the soundness   defined in \Cref{item:com-n-prove:condition:soundness} in \Cref{def:com-n-prove}.
\end{lemma}
\begin{proof}
This follows from a similar argument from previous black-box commit-and-prove literature \cite{STOC:IKOS07,FOCS:GLOV12,STOC:GOSV14,C:LiaPan21}. We provide here a self-contained proof.

We want to show that no non-uniform QPT $P^*$ can commit to a $m_\algo{com}$ for which $\phi(m_\algo{com}) = 0$ and make $V$ accept with non-negligible probability. There are two cases where $\phi(m_\algo{com}) = 0$:
\begin{enumerate}
\item \label[Case]{item:lemma:e-ExtCom-n-Prove:soundness:case:1}
$m_\algo{com} = \bot$; {\bf or}
\item \label[Case]{item:lemma:e-ExtCom-n-Prove:soundness:case:2}
$m_\algo{com}$ is in $\bits^{\ell(\secpar)}$, but $\phi(m_\algo{com}) = 0$.
\end{enumerate}
(Note that \Cref{item:e-ExtCom-n-Prove:prove-stage:3,item:e-ExtCom-n-Prove:prove-stage:4,item:e-ExtCom-n-Prove:prove-stage:5} constitute a coin-flipping protocol such that the resulting $T$ must be a size-$t$ pseudo-random subset of $[n]$ even if $P^*$ is malicious. We henceforth assume w.l.o.g.\ that $T$ is a size-$t$ random subset of $[n]$.)

\subpara{For \Cref{item:lemma:e-ExtCom-n-Prove:soundness:case:1}:}
If this case happens, we know that the $\Set{\view_i}_{i\in[n]}$ statistically-bounded in \Cref{item:e-ExtCom-n-Prove:commit-stage:2} are such that $\VSS_{\Recon}(\view_1, \ldots, \view_n) = \bot$. 
We now define an undirected graph $G$ with $n$ vertices corresponding to the $n$ views $\Set{\view_i}_i$. Assign an edge between vertices $i$ and $j$ in $G$ if $\view_i$ and $\view_j$ are {\em inconsistent} w.r.t.\ $\VSS_\Share$ execution. We first argue that the {\em minimum
vertex cover} set $B$ of $G$ must have size $\ge t$. To see this, consider an execution of $\VSS_\Share$ where the adversary
corrupts the set of players in $B$ with $|B|<t$, and behaves in a way that the views of any player $P_j$, for $j \notin B$, is $\view_j$. Such an execution is obtained by choosing all the messages from $P_j \in B$ to $P_j \notin B$ as in the view $\view_j$; since $B$ is a vertex cover, every pair of views $(\view_i, \view_j)$ with $i,j \in \bar{B}$  are not connected in the graph G and hence consistent. Finally, by the $(n+1, t)$-perfect verifiable-committing of $\VSS$,  such a corruption should not influence the output of the honest players in the $\VSS_\Share$ stage, which must be $\bot$ (otherwise, $\VSS_{\Recon}(\view_1, \ldots, \view_n) \ne \bot$). This means there will be at least $(n-t)$ views in $\Set{\view_i}_{i\in [n]}$ indicating that the $\VSS_\Share$ execution fails. Since $V$ checks $t$ out of them randomly, $V$ will learn this information and rejects in \Cref{item:e-ExtCom-n-Prove:prove-stage:7:c} of the Prove Stage {\em except} with probability $\le (t/n)^t$, which is negligible in $\secpar$ due to our parameter setting. That is, $|B|\ge t$ with overwhelming probability.

	
Now, recall that $V$ checks the consistency of a size-$t$ random subset of all the views $\Set{\view_i}_{i\in[n]}$ (i.e., vertices in $G$). We only need to argue that such a checking will hit an edge in $G$ with overwhelming probability. For this, we use the well-known connection between the size of a minimum vertex cover to the size of a {\em maximum matching}. Concretely, the graph $G$ must have a {\em matching}\footnote{Recall that a matching is a set of edges without common vertices.} $\mathcal{M}$ of size at least $t/2$. (Otherwise, if the maximum matching contains less than $t/2$ edges, then the vertices of this matching form a vertex cover set $B$ with $|B|<t$.) Recall that if $V$ hits any edge of $G$, he will reject. The probability that the $t$ vertices (views) that $V$ picks miss all the edges of $G$ is smaller than the probability that he misses all edges
of the matching, which is again at most $2^{-\Omega(t)} = 2^{-\Omega(\secpar)}$. To see that, suppose that the first $t/2$ vertices picked by $V$ do not hit an edge of the matching. Denote this set of vertices as $S_{t/2}$. It follows from Serfling's Inequality (see \Cref{lem:Serfling}) that with overwhelming probability over $\secpar$, $S_{t/2}$ contains $\Omega(t)$ vertices that are the vertices of the edges $\mathcal{M}$. Then, their $\Omega(t)$ matching neighbors will have $\Omega(t/n) = \Omega(1)$ probability of being hit by each subsequent vertex picked by $V$.  Since $V$ will pick $t/2$ more vertices, the probability that $V$ misses all the $\Omega(t)$ matching neighbors with probability at most $2^{-\Omega(t)} = 2^{-\Omega(\secpar)}$.



\subpara{For \Cref{item:lemma:e-ExtCom-n-Prove:soundness:case:2}:} In this case, we know that $m_\algo{com}$ does not satisfy $\phi$. However, the $\Set{\view'_i}_{i\in[n]}$ committed by $\Com$ in \Cref{item:e-ExtCom-n-Prove:prove-stage:2} of the Prove Stage are supposed to be the views of $n$ parties executing $\Prot_\textsc{mpc}$. Then, we can use the same argument as for \Cref{item:lemma:e-ExtCom-n-Prove:soundness:case:1} to show that $V$ must rejects except with negligible probability. That is, we define the ``inconsistency graph'' $G$ corresponding to $\Set{\view'_i}_{i\in[n]}$, and argue either that there are too many inconsistent views (such that $V$ will catch by checking $t$ of them), or that most parties are honest and report $\phi(m_\algo{com}) = 0$ (such that $V$ will learn this with overwhelming probability). The only change is, we now rely on the the $(n, t)$-perfect robustness of $\Prot_\textsc{MPC}$, instead of the perfectly verifiable-committing of $\VSS$.  One caveat is that we need to ensure that $P$ use $\Set{\view_i}_{i\in[n]}$ from the Commit Stage to execute $\Prot_\textsc{mpc}$ in \Cref{item:e-ExtCom-n-Prove:prove-stage:1} of the Prove Stage. By checking \Cref{item:e-ExtCom-n-Prove:prove-stage:7:b}, $V$ is convinced with overwhelming probability that this is indeed the case for at least $(n-t)$ views out of $n$. This suffices to use the above inconsistency-graph argument. Since this argument is almost identical to \Cref{item:lemma:e-ExtCom-n-Prove:soundness:case:1}, we omit the details.
\end{proof}

\begin{lemma}[$\epsilon$-Zero-Knowledge]\label{lemma:e-ExtCom-n-Prove:e-zk}
Assume $\ExtCom$ and $\Com$ are computationally-hiding, $\ExtCom$ is weakly extractable with $\epsilon$-simulation, $\VSS$ is $(n+1,t)$-secret (see \Cref{def:VSS}), and $\Prot_\textsc{mpc}$ is $(n,t)$-semi-honest computationally private (see \Cref{def:t-privacy}). Then, \Cref{protocol:e-ExtCom-n-Prove} satisfies the $\epsilon$-zero-knowledge property  defined in \Cref{item:com-n-prove:condition:e-zk} in \Cref{def:com-n-prove}.
\end{lemma}
\begin{proof}
This proof is also standard in existing black-box commit-and-prove literature. We provide it for completeness.

The Commit Stage simulator $\Sim_\algo{EC}$ behaves identically as the honest prover, except that he executes the $\VSS_\Share$ of \Cref{item:e-ExtCom-n-Prove:commit-stage:1} (in his head) to secret-share an arbitrary value, say $0^{\ell(\secpar)}$. Note that $\Sim_\algo{EC}$ is straight-line and simulates the honest prover with only $\negl(\secpar)$ error (i.e., satisfying \Cref{item:com-n-prove:ZK:condition:1}), thanks to the computationally-hiding property of $\ExtCom$. 

To define the Prove Stage simulator, first notice that \Cref{item:e-ExtCom-n-Prove:prove-stage:3,item:e-ExtCom-n-Prove:prove-stage:4,item:e-ExtCom-n-Prove:prove-stage:5} of the Prove Stage constitute a $\epsilon$-simulatable coin-flipping protocol against the malicious non-uniform QPT $V^*(\rho)$, because $\ExtCom$ is a $\epsilon$-simulatable weakly-extractable commitment. The Prove Stage simulator $\Sim_\algo{Pr}$ works as follows:
\begin{enumerate}
\item \label[Step]{item:e-ExtCom-n-Prove:e-zk:Prove-Stage-Sim:1}
 Sample at random a size-$t$ subset $\tilde{T} \subseteq[n]$;
 \item \label[Step]{item:e-ExtCom-n-Prove:e-zk:Prove-Stage-Sim:2}
 Invoke the simulator for $\Prot_\textsc{mpc}$ on parties $\Set{P_i}_{i\in[T]}$ to obtain the simulated view $\Set{\tilde{\view}'_i}_{i \in [T]}$; set $\tilde{\view}'_j$ to all-0 strings of proper length for all $j \in [n]\setminus \tilde{T}$;
 \item \label[Step]{item:e-ExtCom-n-Prove:e-zk:Prove-Stage-Sim:3}
 Commit to $\Set{\tilde{\view}'_i}_{i \in [n]}$ using $\Com$ in parallel as \Cref{item:e-ExtCom-n-Prove:prove-stage:2};
 \item \label[Step]{item:e-ExtCom-n-Prove:e-zk:Prove-Stage-Sim:4}
For  \Cref{item:e-ExtCom-n-Prove:prove-stage:3,item:e-ExtCom-n-Prove:prove-stage:4,item:e-ExtCom-n-Prove:prove-stage:5}, invoke the coin-flipping simulator to force the resulting $r$ such that it will determine the set $\tilde{T}$ he sampled in the 1st step;
\item \label[Step]{item:e-ExtCom-n-Prove:e-zk:Prove-Stage-Sim:5}
Finish the remaining steps as the honest prover.
\end{enumerate}
Because of the computationally-hiding property of $\Com$, $\Sim_\algo{Pr}$ is computationally-indistinguishable from the honest prover until \Cref{item:e-ExtCom-n-Prove:prove-stage:2}. Then, because of the security of $\epsilon$-simulatable coin-flipping, at the end of \Cref{item:e-ExtCom-n-Prove:prove-stage:5}, $\Sim_\algo{Pr}$ is at most $\epsilon$-far from the honest prover and will force the result to be $\tilde{T}$ successfully. Finally, the remaining steps are again computationally-indistinguishable from the honest prover, because of the $(n+1, t)$-secret of $\VSS$ and the $(n, t)$-semi-honest computational privacy of $\Prot_\textsc{mpc}$. In total, $\Sim_\algo{Pr}$ is at most $\epsilon$-computationally distinguishable from the honest prover (i.e., satisfying \Cref{item:com-n-prove:ZK:condition:2}).
\end{proof}

\section{Black-Box $\epsilon$-Simulatable PQ-2PC in Constant Rounds}
\label{sec:e-2pc}

\subsection{Definition and Notation}
We first present the formal definition for $\epsilon$-simulatable two-party computation. It is identical to the standard 2PC definition in the classical setting except that:
\begin{enumerate}
\item
The malicious party can be QPT;
\item
The indistinguishability between the real-world execution and the simulated one is parameterized by a noticeable function $\epsilon(\secpar)$. 
\end{enumerate}
Consider two parties $P_1$ and $P_2$ with inputs $x_1$ and $x_2$ that wish to interact in a protocol $\Prot$ to evaluate a 2-party
functionality $f$ on their joint inputs. The ideal and real executions follow the standard description as in, e.g., \cite{Goldreich04}.

In the real world, a QPT adversary $\Adv_\secpar$ with a quantum auxiliary input $\rho_\secpar$ corrupting $P_i(x_i)$ ($i\in \bits$) interacts with $P_{1-i}(x_{1-i})$. Let $\vb{x} = (x_1, x_2)$ denote the inputs to the two parties. Let $\REAL_{\Prot, \Adv, i}(\secpar, \vb{x}, \rho_\secpar)$ denote the random variable consisting of the output of the adversary (which may be an arbitrary function of its view and in particular may be a quantum state) and the outputs of the honest party $P_{1-i}$. 

In the ideal world, a QPT machine $\Sim$ controls the same party $P_i$ as $\Adv_\secpar$. It gets $x_i$ and $\rho_\secpar$ as input. Similar as in the $\epsilon$-ZK definition \cite{C:ChiChuYam21}, $\Sim$ additionally takes as input a ``slackness parameter''\footnote{Actually, \cite{C:ChiChuYam21} refers to it as the ``accuracy parameter''. But we think ``slackness'' is a better name as $\epsilon$ measures how {\em far} two ensembles are.} $\epsilon(\secpar)$, which is a noticeable function on $\secpar$. Henceforth, we always require that $\Sim$'s running time is a polynomial on both $\secpar$ and $\epsilon^{-1}$. Let $\IDEAL_{f,\Sim,i}(\secpar, \epsilon, \vb{x}, \rho_\secpar)$ denote the outputs of $\Sim$ (with slackness $\epsilon$) and the uncorrupted party $P_{1-i}$ from the ideal-world execution.

 We remark that, throughout this paper, we only focus on {\em static} adversaries and {\em security with abortion} (i.e., the ideal-world adversary (aka the simulator) learns the its output first, and then can instruct the ideal functionality to deliver the output to the honest party or not). This is standard in 2PC literature as security without abortion (aka {\em fairness}) is impossible for {\em general-purpose} two-party protocols \cite{STOC:Cleve86,TCC:ABMO15}.

\begin{definition}[Post-Quantum $\epsilon$-Simulatable 2PC] \label{def:e-2pc}
 Let $f$ be a classical 2-party functionality, and \Prot be a classical 2-party protocol. We say that $\Prot$ is a {\em $\epsilon$-simulatable} protocol for $f$ if there exists a QPT simulator $\Sim$ such that for any non-uniform QPT adversary $\Adv = \Set{\Adv_\secpar, \rho_\secpar}_{\secpar\in\Naturals}$, any $i \in \bits$, any $\vb{x}\in (\bits^*)^2$, and any noticeable function $\epsilon(\secpar)$, it holds that:
$$\Set{\REAL_{\Prot, \Adv, i}(\secpar, \vb{x}, \rho_\secpar)}_{\secpar \in \Naturals} ~\cind_\epsilon~ \Set{\IDEAL_{f,\Sim,i}(\secpar, \epsilon, \vb{x}, \rho_\secpar)}_{\secpar \in \Naturals}.$$
\end{definition}

\subsection{Non-Concurrent Composition of Post-Quantum $\epsilon$-Simulatable Protocols} \label{sec:e-non-concurrent-composition}
We now prove a lemma that will allow us to securely compose different $\epsilon$-simulatable protocols in the post-quantum setting, as long as the composition happens in a ``non-concurrent'' manner (explained later). It will be used later since our $\epsilon$-simulatable 2PC construction in \Cref{sec:PQ-2PC:blueprint} relies on such composition. This lemma is a straightforward extension of the non-concurrent composition lemma from \cite{JC:Canetti00} to the post-quantum $\epsilon$-simulatable 2PC protocols in \Cref{def:e-2pc}.

\para{The Hybrid Model.} We start by specifying the model for evaluating a 2-party function $g$ with the assistance of a trusted party computing some 2-party functionalities $(f_1, \ldots, f_m)$. The trusted party is invoked at special rounds, determined by the protocol. In each such round, a function $f$ (out of $f_1, \ldots, f_m$) is specified. At this point, both parties pause the execution of $\pi$, store their current state, and then start to make the call to the ideal functionality $f$. Upon receiving the output back from the trusted party, the protocol $\pi$ continues.

The protocol $\pi$ is such that $f_{i+1}$ can be called only if the invocation of $f_{i+1}$ is completely finished. It is possible that $f_i = f_j$ for some $i \ne j$, representing that the same ideal functionality is invoked twice at different time point. We emphasize that, during the invocation of some $f_i$, the honest party does not send/respond any other messages until it finishes the execution with $f_i$. This is called ``non-concurrent requirement'' in \cite{JC:Canetti00}. For an execution of $\pi$ in the $(f_1, \ldots, f_m)$-hybrid model, we use $\EXEC^{f_1, \ldots, f_m}_{\pi, \Adv, i}(\secpar, \vb{x}, \rho_\secpar)$ to denote the joint outputs of the adversary corrupting $P_i(x_i)$ and the honest $P_{1-i}(x_{1-i})$. 

Let $(\rho_1, \ldots, \rho_m)$ be ``subroutine'' protocols that are supposed to compute $(f_1, \ldots, f_m)$ respectively. We use $\pi^{\rho_1, \ldots, \rho_m}$ to denote the (plain-model) protocol obtained by replacing the ideal calls in the $(f_1, \ldots, f_m)$-hybrid protocol $\pi$ with the corresponding ``subroutine'' protocols in $(\rho_1, \ldots, \rho_m)$.

With these notations, we present the composition lemma in \Cref{lem:e-composition}. 
\begin{lemma}[Non-Concurrent Composition of Post-Quantum $\epsilon$-Simulatable Protocols]\label{lem:e-composition}Let $m \in \Naturals$ be a constant. For $i \in [m]$,
let $\rho_i$ be a post-quantum $\epsilon$-simulatable protocol for a 2-party functionality $f_i$. Let $\pi$ be a $\epsilon$-simulatable protocol for a 2-party functionality $g$  in the $(f_1, \ldots, f_m)$-hybrid model {\em where no more than one ideal evaluation call is made at each round}. Then, $\pi^{\rho_1, \ldots, \rho_m}$ is a $\epsilon$-simulatable protocol for $g$.
\end{lemma}
\begin{proof}
\xiao{This proof can go to the appendix, as it is a straightforward generalization of known techniques.} We will show the proof for the case $m=1$. The proof for $m > 1$ follows straightforwardly by sequentially repeating the following argument for $m = 1$ to replace $(f_1, \ldots, f_m)$ one-by-one, because the calls to each $f_i$ happen sequentially  (see also \cite[Theorem 5 and Corollary 7]{JC:Canetti00} for more details).

Since $\pi$ is a $\epsilon$-simulatable protocol for $g$ in the $f$-hybrid model, there is a QPT $\Sim'$ such that for any non-uniform QPT $\Adv' = \Set{\Adv'_\secpar, \alpha_\secpar}_{\secpar\in\Naturals}$, any $i \in \bits$, any $\vb{x}\in (\bits^*)^2$, and any noticeable $\epsilon(\secpar)/2$, 
\begin{equation}\label{eq:e-composition:ideal'-hybrid:original}
\Set{\IDEAL_{g, \Sim', i}(\secpar, \frac{\epsilon}{2}, \vb{x}, \alpha_\secpar)}_{\secpar \in \Naturals} ~\cind_{\frac{\epsilon}{2}}~ \Set{\EXEC^f_{\pi, \Adv', i}(\secpar, \vb{x}, \alpha_\secpar)}_{\secpar \in \Naturals}.
\end{equation}
Now, for the LHS execution of \Cref{eq:e-composition:ideal'-hybrid:original}, consider a new $\Sim$ that on input $\epsilon$, it runs $\Sim'$ with slackness parameter $\epsilon/2$. Note that $\Sim$'s running time is also a polynomial on $\secpar$ and $\epsilon^{-1}$. Indeed, $\Sim$ is identical to $\Sim'$ with only syntactical changes. Thus, we have
\begin{equation}\label{eq:e-composition:ideal'-hybrid}
\Set{\IDEAL_{g, \Sim, i}(\secpar, \epsilon, \vb{x}, \alpha_\secpar)}_{\secpar \in \Naturals} ~\idind~ \Set{\IDEAL_{g, \Sim', i}(\secpar, \frac{\epsilon}{2}, \vb{x}, \alpha_\secpar)}_{\secpar \in \Naturals} ~\cind_{\frac{\epsilon}{2}}~ \Set{\EXEC^f_{\pi, \Adv', i}(\secpar, \vb{x}, \alpha_\secpar)}_{\secpar \in \Naturals}.
\end{equation}
Therefore, \Cref{lem:e-composition}, follows from the following \Cref{claim:e-composition:hybrid-real}.
\end{proof}

\begin{myclaim}\label{claim:e-composition:hybrid-real}
Let $\rho$, $f$ and $\pi$ be the same as in \Cref{lem:e-composition}.
There exists a quantum adversary $\Adv'$ in the $f$-hybrid execution of $\pi$ such that for any non-uniform QPT $\Adv = \Set{\Adv_\secpar, \alpha_\secpar}_{\secpar\in\Naturals}$ in the real execution of $\pi^\rho$, any $i \in \bits$, any $\vb{x}\in (\bits^*)^2$, and any noticeable $\epsilon(\secpar)$, 
\begin{equation}\label{eq:e-composition:hybrid-real}
\Set{\EXEC^f_{\pi, \Adv', i}(\secpar, \vb{x}, \alpha_\secpar)}_{\secpar \in \Naturals} ~\cind_{\frac{\epsilon}{2}}~ \Set{\REAL_{\pi^\rho, \Adv, i}(\secpar, \vb{x}, \alpha_\secpar)}_{\secpar \in \Naturals},
\end{equation}
where $\Adv'$'s running time is a polynomial on $\secpar$ and $\epsilon^{-1}$.
\end{myclaim}
\begin{proof}
This proof proceeds as follows:
\begin{enumerate}
\item
We construct out of $\Adv = \Set{\Adv_\secpar, \alpha_\secpar}_{\secpar\in\Naturals}$ a QPT real-world adversary $\Adv^\rho = \Set{\Adv^\rho_{\secpar}, \alpha^\rho_\secpar}_{\secpar\in\Naturals}$ that operates against protocol $\rho$ as a stand-alone protocol. The security of $\rho$ guarantees that $\Adv^\rho$ has a QPT simulator $\Sim^\rho$ such that for any $i$, $\vb{x}$, and $\epsilon/2$, 
\begin{equation}\label{eq:ideal-real:rho}
\Set{\IDEAL_{f, \Sim^\rho, i}(\secpar, \frac{\epsilon}{2}, \vb{x}, \alpha^\rho_\secpar)}_{\secpar \in \Naturals} ~\cind_{\frac{\epsilon}{2}}~ \Set{\REAL_{\rho, \Adv^\rho, i}(\secpar, \vb{x}, \alpha^\rho_\secpar)}_{\secpar \in \Naturals}.
\end{equation}
\item
Out of $\Adv$ and $\Sim^\rho$, we construct an adversary $\Adv'$ that operates against protocol $\pi$ in the $f$-hybrid model. We then finish the proof by showing that $\Adv'$ satisfies \Cref{eq:e-composition:hybrid-real} and the running time requirement.
\end{enumerate}

\subpara{The Simulator $\Sim^\rho$ for Protocol $\rho$.} We provide more details. Intuitively, $\Adv^\rho$ represents the ``segment'' of $\Adv$ that is involved in the execution of $\rho$. That is, $\Adv^\rho$ takes a non-uniform quantum auxiliary input $\Set{\alpha^\rho_\secpar}_{\secpar\in\Naturals}$. This auxiliary input contains the internal state of $\Adv$, controlling party $P_i(x_i)$, and executing the protocol $\pi^\rho$ with the honest party $P_{1-i}(x_{1-i})$ up to round $\ell_\rho$, where $\rho$ is invoked. We note that $\Set{\alpha^\rho_\secpar}_{\secpar\in\Naturals}$ can be constructed from $\Set{\alpha_\secpar}_{\secpar\in\Naturals}$ (i.e., $\Adv$'s auxiliary input). At the end of its execution of $\rho$ with $P_{1-i}(x_{1-i})$, adversary $A^\rho$ outputs the current state of the simulated $\Adv$. 

By assumption, $\rho$ is a $\epsilon$-simulatable protocol for $f$. Thus, there exists a QPT $\Sim^\rho$ satisfying \Cref{eq:ideal-real:rho}.

\subpara{$\Adv'$ in the $f$-Hybrid Model.} Adversary $\Adv'$ represents the ``segment'' of $\Adv$ that executes $\pi$ in the $f$-hybrid model, where $\Adv$'s execution of $\rho$ is handled by $\Sim^\rho$. Recall that $\Adv$ expects to execute the protocol $\pi_\rho$, not $\pi$ in the $f$-hybrid model. Therefore, we will need $\Sim^\rho$ to simulate the $\rho$ part for $\Adv$.  Formally, $\Adv'$ starts by invoking $\Adv_\secpar$ on auxiliary input $\alpha_\secpar$, and follows $\Adv$'s instructions up to round $\ell_\rho$.
At this point, $\Adv$ expects to execute $\rho$ with $P_i$, whereas $\Adv'$ invokes the ideal functionality $f$ (of the $f$-hybrid model). To continue the execution of $\Adv$, adversary $\Adv'$ runs $\Sim^\rho$. For this purpose, $\Sim^\rho$ is given the auxiliary input, denoted as $\alpha^\rho_\secpar$, that describes the current state of $\Adv$ at round $\ell_\rho$. The information from $\Sim^\rho$'s trusted party is emulated by $\Adv'$, using $\Adv'$'s own ideal functionality $f$. Recall that the output of $\Sim^\rho$ is a (simulated) internal state of $\Adv$ at the completion of protocol $\rho$. Once protocol $\rho$ completes its execution and the parties return to running $\pi$, adversary $\Adv'$ returns to running $\Adv$ (starting from the state in $\Sim^\rho$'s output) and follows the instructions of $\Adv$. When $\Adv$ terminates, $\Adv'$ outputs whatever $\Adv$ outputs.

With $\Adv'$ defined above, the LHS and RHS executions of \Cref{eq:e-composition:hybrid-real} can be divided into the following 3 stages:
\begin{enumerate}
\item \label[Stage]{stage:e-composition:hybrid-real-1}
Both the LHS and RHS executions are identical up to the starting of round $\ell_\rho$
\item \label[Stage]{stage:e-composition:hybrid-real-2}
At round $\ell_\rho$, the RHS $\Adv$ starts executing $\rho$ with the honesty $P_{1-i}$, whereas the LHS $\Adv$ (controlled by $\Adv'$) talks with $\Sim^\rho$. At the end of this stage, the internal state of $\Adv$ in the RHS is $\epsilon/2$-far from the ($\Sim^\rho$-simulated) internal state of $\Adv$ (controlled by $\Adv'$) in the LHS. 
\item \label[Stage]{stage:e-composition:hybrid-real-3}
Then, both the LHS and RHS again proceed identically until the end. 
\end{enumerate} 
Since both \Cref{stage:e-composition:hybrid-real-1,stage:e-composition:hybrid-real-3} are polynomial time, \Cref{eq:e-composition:hybrid-real} follows from a straightforward reduction to the $\epsilon$-simulatability of $\Sim^\rho$ in \Cref{stage:e-composition:hybrid-real-2} (by setting the slackness parameter to $\epsilon/2$).

Finally, we need to argue that $\Adv'$'s running time is a polynomial on $\secpar$ and $\epsilon'$. This is true as $\Adv'$ does nothing more than running $\Adv$ and $\Sim^\rho$, and both of them is QPT on $\secpar$ and $\epsilon$. (Though $\Sim^\rho$ is invoked with slackness $\epsilon/2$, $(\epsilon/2)^{-1}$ is also a polynomial on $\epsilon^{-1}$.) 
\end{proof}

\subsection{The Classical Compiler and Parallel Commitments and OTs}
\label{sec:e-2pc:classical-recipe}
 We start by recalling the black-box constant-round compiler from semi-honest OT to secure (stand-alone) 2PC in the classical setting. Such a compiler works in 2 steps:
\begin{enumerate}
\item \label[Step]{item:e-2pc:road-mad:step:1}
Given black-box access to any semi-honest bit-OT, it constructs a malicious-secure string-OT (via \cite{DBLP:journals/siamcomp/HaitnerIKLP11,TCC:CDMW09}). This step incurs only constant overhead on round complexity.
\item \label[Step]{item:e-2pc:road-mad:step:2}
Given black-box access to any maliciously-secure string-OT, it constructs a maliciously-secure 2PC protocol (via \cite{C:IshPraSah08}). This step incurs only constant overhead on round complexity, {\em assuming the given OT is parallelly-secure} (see the discussion below).
\end{enumerate}
Although the above steps are widely known to the community, there are subtleties {\em regarding the stand-alone scenario} that we feel obliged to address. This is because the main focus of \cite{TCC:CDMW09} and \cite{C:IshPraSah08} is the UC setting, and thus their implications for {\em constant-round constructions in the stand-alone setting} has not been thoroughly discussed. Since this is crucial to the current paper, we provide further clarification in the following.
(While all the aforementioned works are for the more general {\em multi-party} computation, our discuss here will only focus on the 2-party case.)

\para{Two Flavors of the Hybrid Model.} We first need to distinguish between two flavors of the $\Func$-hybrid model. This model is a helpful tool enabling modular design and composition of 2PC (and in general, MPC) protocols. For example, to design a 2PC protocol implementing some ideal functionality $\mathcal{G}$ in the $\Func$-hybrid model, one can assume that both parties have access to the ideal functionality $\Func$. The resulting protocol will be denoted as $\pi^\Func$, and it will lead to a secure implementation of $\mathcal{G}$ in the plain model if the parties replace  the calls to the ideal functionality $\Func$ with a protocol $\phi$ which securely implements $\Func$, as long as they do not interleave other parts of $\pi^{(\cdot)}$ with the execution of $\phi$. 

The meaning of $\Func$-hybrid model may vary depending on whether the concerned security is in the stand-alone model or the UC model. In the stand-alone model (e.g., \cite{JC:Canetti00}), the $\Func$-hybrid model only allows the parties to make a single call of the ideal $\Func$, because stand-alone protocols may not be concurrently (or even parallelly) composable. Traditionally, if $n$ calls to the idea functionality is needed, people denote this hybrid model as the $(\Func_1, \ldots, \Func_n)$-hybrid model, and the protocol in this model as $\pi^{\Func_1, \ldots, \Func_n}$, even if all the $\Func_i$'s are actually the same functionality. Importantly, no calls to $\Func_i$ (or other parts of $\pi^{(\cdot)}$) are allowed when some $\Func_j$ is running. Such composition is called ``non-concurrent composition'' in \cite{JC:Canetti00}. To distinguish with the one that we will discuss next, we refer to this model as the {\em stand-alone} $\Func$-hybrid model.

In the UC model (\cite{FOCS:Canetti01}), the ideal functionality is extended by including a session ID, and multiple calls to the same functionality will be distinguished by different session IDs. Moreover, the strong composability of the UC security allows us to schedule the calls to $\Func$ arbitrarily when designing $\pi^\Func$ in the $\Func$-hybrid model. In particular, parties executing $\pi^\Func$ can make multiple calls to the ideal $\Func$ in parallel (or interleaved arbitrarily), and these calls can also be interleaved with other parts of $\pi^{(\cdot)}$; moreover, parties executing a protocol $\pi^{\Func_1, \Func_2}$ can even interleave their calls to the two distinct ideal functionalities if necessary. We refer to this hybrid model as the {\em UC} $\Func$-hybrid model. It is worth noting that if we want  to replace the ideal $\Func$ call(s) in the UC-secure protocol $\pi^\Func$ (in the UC $\Func$-hybrid model), we must use a protocol $\phi$ that {\em UC-securely realizes} $\Func$. In particular, if $\phi$ only securely implements $\Func$ in the stand-alone sense, no security (not even stand-alone security) is guaranteed for the resulting protocol.

\para{For \Cref{item:e-2pc:road-mad:step:1}.} \cite{DBLP:journals/siamcomp/HaitnerIKLP11}\footnote{This paper is the journal version merging two previous works \cite{STOC:IKLP06,TCC:Haitner08}.} presents the first constant-round black-box compiler from semi-honest bit-OT to maliciously-secure {\em bit}-OT in the stand-alone setting. This compiler requires a coin-tossing protocol (satisfying the simulation-based security), which can be built from any protocol that implements the ideal commitment functionality $\Func_\textsc{com}$ in the stand-alone setting. It is also worth noting that the security proof in \cite{DBLP:journals/siamcomp/HaitnerIKLP11} relies heavily on rewindings, {\em even when being analyzed in the stand-alone $\Func_\textsc{com}$-hybrid model}. Also, at that time, it was unclear if the resulting protocol is parallelly composable (or if it leads to a maliciously-secure {\em string}-OT). 
 
Later, \cite{TCC:CDMW09} shows that \cite{DBLP:journals/siamcomp/HaitnerIKLP11} with slight modification yields a black-box constant-round compiler from semi-honest bit-OTs to maliciously-secure {\em string}-OTs. Moreover, \cite{TCC:CDMW09} also simplifies the security proof such that no rewinds are needed in the $\Func_\textsc{com}$-hybrid model. One caveat here is that the ``$\Func_\textsc{com}$-hybrid model'' in \cite{TCC:CDMW09} is different from that in \cite{DBLP:journals/siamcomp/HaitnerIKLP11}. The $\Func_\textsc{com}$ in \cite{DBLP:journals/siamcomp/HaitnerIKLP11} is just the stand-alone string-commitment functionality. Such an $\Func_\textsc{com}$ suffices for the application of coin-flipping as required by the \cite{DBLP:journals/siamcomp/HaitnerIKLP11} compiler. In contrast, the $\Func_\textsc{com}$ employed by the \cite{TCC:CDMW09} needs to be a commitment that captures {\em bounded-parallel security with selectively-opening}. That is, it can be used by a committer to commit to an a-priori bounded number, say a polynomial $t(\secpar)$, of strings by a single invocation; later, the receiver can specify an arbitrary subset $T \subseteq [t]$ of positions, and the committer will decommit to the $i$-th commitment for all $i\in T$. More accurately, we denote this ideal functionality as $\Func^t_\textsc{so-com}$ and present it in \Cref{figure:functionality:so-com}.\footnote{We note that this $\Func^t_\textsc{so-com}$ has recently been formalized in \cite{EC:GLSV21}.} 
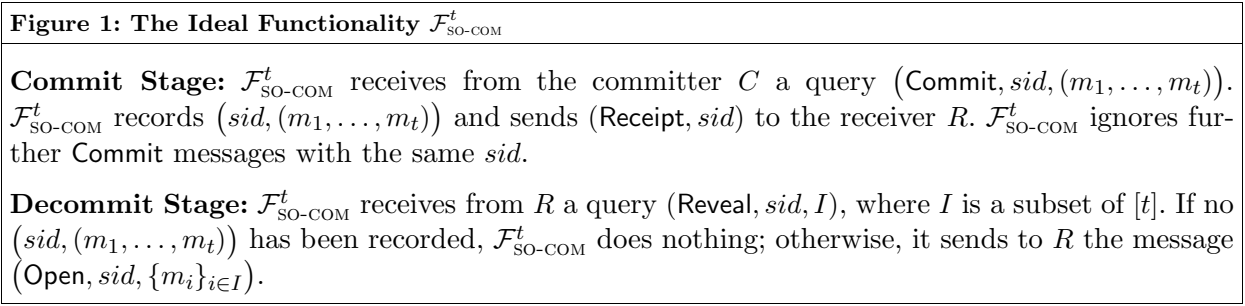
\begin{FigureBox}[label={figure:functionality:so-com}]{The Ideal Functionality \textnormal{$\Func^t_\textsc{so-com}$}}
\para{Commit Stage:} $\Func^t_\textsc{so-com}$ receives from the committer $C$ a query $\big(\algo{Commit}, sid, (m_1, \ldots, m_t)\big)$. $\Func^t_\textsc{so-com}$ records $\big(sid, (m_1, \ldots, m_t)\big)$ and sends $(\algo{Receipt}, sid)$ to the receiver $R$. $\Func^t_\textsc{so-com}$ ignores further $\algo{Commit}$ messages with the same $sid$.

\para{Decommit Stage:} $\Func^t_\textsc{so-com}$ receives from $R$ a query $(\algo{Reveal}, sid, I)$, where $I$ is a subset of $[t]$. If no $\big(sid, (m_1, \ldots, m_t)\big)$ has been recorded, $\Func^t_\textsc{so-com}$ does nothing; otherwise, it sends to $R$ the message $\big(\algo{Open}, sid, \Set{m_i}_{i\in I} \big)$. 
\end{FigureBox}
The reason why \cite{TCC:CDMW09} needs such an $\Func^t_\textsc{so-com}$ is that their compiler relies on the ``cut-and-choose'' technique where a party first commits to polynomially-many random strings {\em in parallel}, and later opens a subset of them (determined by a coin-tossing) for the other party to check. We emphasize that these commitments must be done in parallel; otherwise (i.e., when done sequentially), the resulting protocol will not be in constant rounds.
In summary, \cite{TCC:CDMW09} actually proved the following lemma.
\begin{lemma}[{\cite[Proposition 1]{TCC:CDMW09}}]\label{lem:compiler:Com-to-OT}
There is a polynomial $t(\secpar)$ such that there exists a black-box construction of a string-OT protocol
secure against static, malicious adversaries in the stand-alone $\Func^t_\textsc{so-com}$-hybrid model (or alternatively, the UC $\Func_\textsc{com}$-hybrid model), starting from any bit-OT protocol secure against  static, semi-honest adversaries. Moreover, the construction achieves a constant multiplicative blow up in the number of rounds, and has a strictly polynomial-time and straight-line
simulator.
\end{lemma}

\para{For \Cref{item:e-2pc:road-mad:step:2}.} We now discuss the \cite{C:IshPraSah08} compiler from OTs to 2PC. As mentioned above, \cite{C:IshPraSah08} mainly focuses on the UC setting. It proves that in the UC $\Func_\textsc{ot}$-hybrid model, there is a constant-round
black-box protocol $\Prot^{\Func_\textsc{ot}}_\textsc{2pc}$ of general-purpose UC-secure 2PC. As mentioned earlier, it is {\em in general} unclear what would happen if the $\Func_\textsc{ot}$ in $\Prot^{\Func_\textsc{ot}}_\textsc{2pc}$ is replaced by a stand-alone secure OT protocol. However, for the special case of \cite{C:IshPraSah08}, it is known that the two participants of the $\Prot^{\Func_\textsc{ot}}_\textsc{2pc}$ protocol only make parallel calls to $\Func_\textsc{ot}$ {\em for an a-priori bounded number $t$},\footnote{This has been observed and employed in earlier works (e.g., \cite{TCC:PasWee09,FOCS:Wee10,STOC:Goyal11}). Thus, we suppress further explanation and refer the reader to \cite{C:IshPraSah08}.} which is a polynomial on $\secpar$. Therefore, similar as the above $\Func^t_\textsc{so-com}$, we can also define a bounded-parallel OT functionality $\Func^t_\textsc{ot}$ (in \Cref{figure:functionality:para-ot}) in the stand-alone setting, and interpret the \cite{C:IshPraSah08} compiler in the stand-alone setting.
This leads to following special case of the \cite{C:IshPraSah08} result (see also \Cref{rmk:parall-vs-sequential:IPS}).
\begin{lemma}[Special Case of {\cite[Theorem 3]{C:IshPraSah08}}]\label{lem:compiler:OT-to-2PC}
There exists an a-priori known polynomial $t(\secpar)$ such that in the stand-alone $\Func^t_\textsc{ot}$-hybrid model (or alternatively, the UC $\Func_\textsc{ot}$-hybrid model), there is a  general-purpose 2PC protocol that achieves stand-alone security against static, malicious adversaries. Moreover, the protocol is constant-round and has a strictly polynomial-time and straight-line simulator.
\end{lemma}

\begin{remark}\label{rmk:parall-vs-sequential:IPS}
Alternatively, one can also replace the $t$ parallel-OT instances with $t$ sequential stand-alone secure OT instances. While this will indeed lead to a secure 2PC protocol in the stand-alone setting, the resulting protocol will not be constant-round as $t=\Omega(\secpar)$ for the \cite{C:IshPraSah08} construction. Another caveat here is that in a larger protocol where parallel OTs are used, replacing these parallel OTs with sequential OTs may jeopardize security. Nevertheless, there exist standard techniques to achieve the same effect of $t$ parallel OTs using $t$ sequential executions of random OTs.
\end{remark}
\begin{FigureBox}[label={figure:functionality:para-ot}]{The Ideal Functionality \textnormal{$\Func^t_\textsc{ot}$}}
\para{Sender's Message:}
$\Func^t_\textsc{ot}$ receives from the sender $S$ a query $\big(\algo{Send}, sid, \Set{(x^i_0, x^i_1)}_{i \in [t]}\big)$. $\Func^t_\textsc{ot}$ records $\big(sid, \Set{(x^i_0, x^i_1)}_{i \in [t]}\big)$. $\Func^t_\textsc{ot}$ ignores further $\algo{Send}$ messages with the same $sid$.

\para{Receiver's Message:}
$\Func^t_\textsc{ot}$ receives from the receiver $R$ a query $\big(\algo{Receive}, sid, c \in \bits^t\big)$. If no $\big(sid, \Set{(x^i_0, x^i_1)}_{i \in [t]}\big)$ has been recorded, $\Func^t_\textsc{ot}$ does nothing; otherwise, it sends to $R$ the message $\big(\algo{Open}, sid, \Set{x^i_{c_i}}_{i\in [t]} \big)$, where $c_i$ is the $i$-th bit of $c$.
\end{FigureBox}

\para{The Final Compiler (Classical).} Let us summarize the semi-honest bit-OTs to 2PC compiler in the classical setting. First, because of \Cref{lem:compiler:OT-to-2PC}, it suffices to build a protocol that realizes the $\Func^t_\textsc{ot}$ functionality w.r.t.\ stand-alone security, where $t$ is some a-priori known polynomial on $\secpar$. To do that, one may want to use \Cref{lem:compiler:Com-to-OT}. However, \Cref{lem:compiler:Com-to-OT} only securely implements the {\em stand-alone} $\Func_\textsc{ot}$, instead of $\Func^t_\textsc{ot}$ (i.e., the $t$-parallel version of $\Func_\textsc{ot}$) as required by \Cref{lem:compiler:OT-to-2PC}. To address this issue, we need the following observation regarding parallel composability in the (stand-alone) hybrid model.
\begin{lemma}[{\cite[Theorem 3.3]{EC:GLSV21}}]\label{lem:stand-alone-parallel-composition}
Assume we have a protocol $\pi$ that securely realizes an ideal functionality $\mathcal{G}$ in the stand-alone $\Func$-hybrid model and {\em with a straight-line simulator}. Then, a parallel repetition of $\pi$ (denoted as $\pi^{||}$) implements $\mathcal{G}^{||}$ in the stand-alone $\Func^{||}$-hybrid model, where $\mathcal{G}^{||}$ and $\Func^{||}$ are the parallel version of $\mathcal{G}$ and $\Func$ respectively.
\end{lemma}
\Cref{lem:stand-alone-parallel-composition} had been a folklore and was recently formally proven in \cite{EC:GLSV21}. The intuition behind it is that since the simulator for $\pi$ is straight-line, the advantage of the simulator (i.e., the ability to extract malicious parties' ``secrets'') must come from the fact that it emulates the ideal $\Func$ for the malicious parties (recall that we are in the $\Func$-hybrid model). Therefore, when $\pi$ is executed in parallel in the $\Func^{||}$-hybrid model, the simulator can extract the ``secrets'' for all the sessions by emulating $\Func^{||}$, which allows it to finish the simulation as other parts of the simulation are straight-line. We emphasize that \cite{EC:GLSV21} proved \Cref{lem:stand-alone-parallel-composition} {\em in the post-quantum setting} (i.e., it considers classical protocols but requires security against QPT adversaries).

We can combine \Cref{lem:compiler:Com-to-OT} and \Cref{lem:stand-alone-parallel-composition} to obtain a secure implementation of $\Func^t_\textsc{ot}$---Observe that the simulator in \Cref{lem:compiler:Com-to-OT} is straight-line; therefore, according to \Cref{lem:stand-alone-parallel-composition}, $t$-parallel repetition of the \Cref{lem:compiler:Com-to-OT} compiler will lead to a secure implementation of the desired $\Func^t_\textsc{OT}$ in the stand-alone $(\Func^{t}_\textsc{so-com})^{||_t}$-hybrid model, where $(\Func^{t}_\textsc{so-com})^{||_t}$ is the {\em $t$-parallel version of} the functionality $\Func^{t}_\textsc{so-com}$. Moreover, notice that $(\Func^{t}_\textsc{so-com})^{||_t}$ can be recast as $\Func^{t'}_\textsc{so-com}$ with a different $t'$, which is also an a-priori known polynomial on $\secpar$. 

In summary, we obtain the following \Cref{thm:OT-to-2PC:final-compiler:classical} by combining \Cref{lem:compiler:Com-to-OT,lem:stand-alone-parallel-composition,lem:compiler:OT-to-2PC}. It is worth noting that previous works claiming constant-round black-box {\em stand-alone} secure 2PC/MPC from semi-honest OTs follows (albeit implicitly sometimes) this recipe \cite{TCC:PasWee09,FOCS:Wee10,TCC:CDMW09,STOC:Goyal11}.
\begin{theorem}\label{thm:OT-to-2PC:final-compiler:classical}
There is a polynomial $t'(\secpar)$ such that there exists a black-box construction of 2PC protocol
secure against static, malicious adversaries in the stand-alone $\Func^{t'}_\textsc{so-com}$-hybrid model, starting from any bit-OT protocol secure against static, semi-honest adversaries. Moreover, the construction achieves a constant multiplicative blow up in the number of rounds, and has a strictly polynomial-time and straight-line
simulator.
\end{theorem}

\subsection{Our Construction of $\epsilon$-Simulatable Post-Quantum 2PC}\label{sec:PQ-2PC:blueprint}
Now we are ready to establish the following theorem:
\begin{theorem}\label{thm:e-2pc:main}
Assuming the existence of a constant-round semi-honest bit-OT secure against QPT adversaries, there exists a black-box, constant-round construction of $\epsilon$-simulatable 2PC protocol secure against QPT adversaries. 
\end{theorem}
To prove \Cref{thm:e-2pc:main}, we follows the approach shown in \Cref{sec:e-2pc:classical-recipe}. There are two key differences that require special attention:
\begin{enumerate}
\item \label{item:e-2pc:classical-vs-quantum:1}
The recipe from \Cref{sec:e-2pc:classical-recipe} gives 2PC secure against (classical) PPT adversaries; but we want to to achieve security against QPT adversaries.
\item \label{item:e-2pc:classical-vs-quantum:2}
We only require $\epsilon$-simulatable security. That is, for any noticeable function $\epsilon(\secpar)$, the simulator will generate a view for the corrupted party that is at most $\epsilon$-far from that party's view in the real-world execution; the running time of the simulator should  be a polynomial on both the security parameter and $1/\epsilon$ (\Cref{def:e-2pc}). We remark that $\epsilon$-simulatable security is effectively the best one can hope for, because constant-round post-quantum 2PC satisfying the standard $\negl(\secpar)$-close simulation cannot exist, unless either {\em non-black-box simulation} is used to prove security or $\NP \subseteq \BQP$ \cite{FOCS:CCLY21}.
\end{enumerate}

To deal with these issues, we first note that the classical recipe from \Cref{sec:e-2pc:classical-recipe} actually extends to the post-quantum setting (i.e., when the adversaries are QPT), because the simulators in both \Cref{lem:compiler:Com-to-OT,lem:compiler:OT-to-2PC} are straight-line and non-cloning (this has also been observed in previous works, e.g., \cite{EC:Unruh10,DBLP:conf/eurocrypt/AgarwalBGKM21}); and as mentioned earlier, \Cref{lem:stand-alone-parallel-composition} was originally proven in the post-quantum setting directly. Therefore, it seems that we immediately obtain a post-quantum version of \Cref{thm:OT-to-2PC:final-compiler:classical}. But there is one more caveat---\Cref{thm:OT-to-2PC:final-compiler:classical} also relies on the aforementioned non-concurrent composition lemma. In particular, the $\Func^t_\textsc{ot}$ functionality assumed in \Cref{lem:compiler:OT-to-2PC} is realized by the real-world protocol (albeit in the $F^t_\textsc{so-com}$-hybrid model) induced by \Cref{lem:compiler:Com-to-OT} (together with \Cref{lem:stand-alone-parallel-composition}). Therefore, we need to make sure that such a non-concurrent composition is also applicable in the post-quantum setting. Fortunately, this follows from \Cref{lem:e-composition} which we proved in \Cref{sec:e-non-concurrent-composition}.
The above discussion leads to the following post-quantum and $\epsilon$-simulatable version of \Cref{thm:OT-to-2PC:final-compiler:classical}: 
\begin{theorem}\label{thm:OT-to-2PC:final-compiler:post-quantum}
There is a polynomial $t(\secpar)$ such that there exists a black-box construction of $\epsilon$-simulatable 2PC protocol
secure against static, malicious QPT adversaries in the stand-alone $\Func^{t}_\textsc{so-com}$-hybrid model, starting from any bit-OT protocol secure against static, semi-honest QPT adversaries. Moreover, the construction achieves a constant multiplicative blow up in the number of rounds.
\end{theorem}
Now, to finish the proof of \Cref{thm:e-2pc:main}, we only need to show the following \Cref{lem:OWF-to-paracom:post-quantum}, and then invoke the non-concurrent composition lemma (\Cref{lem:e-composition}) again, to replace the $\Func^t_\textsc{so-com}$ in \Cref{thm:OT-to-2PC:final-compiler:post-quantum} with the semi-honest OT based construction from \Cref{lem:OWF-to-paracom:post-quantum}.
\begin{lemma}\label{lem:OWF-to-paracom:post-quantum}
For any polynomial $t(\secpar)$, there exists a constant-round protocol that implements $\Func^t_\textsc{so-com}$ w.r.t.\ $\epsilon$-simulatable security against QPT adversaries. This construction makes only black-box access to any OWFs secure against QPT adversaries\footnote{Note that such OWFs can be constructed from any post-quantum semi-honest OT in black-box.}.
\end{lemma}
\begin{proof}
We show a $\epsilon$-simulatable protocol implementing $\Func^t_\textsc{so-com}$, assuming only black-box access to OWFs secure against QPT adversaries. 
We will again rely on the black-box ExtCom-and-Prove as per \Cref{def:com-n-prove}. The committer will commit to the $t$ messages $(m_1, \ldots, m_t)$ using the Commit Stage of the ExtCom-and-Prove. To decommit to the messages determined by the receiver's choice of $I \subseteq [n]$, the committer first sends $\Set{m_i}_{i\in I}$ to the receiver; then, both parties execute the Prove Stage, where the committer proves the following predicate:
\begin{equation}
\phi_{I, \Set{m_i}_{i\in I}}(x) = 
\begin{cases}
1 & \text{if}~ (x = m'_1 \| \ldots\| m'_t) \wedge (\forall i\in I, m'_i = m_i)\\
0 & \text{otherwise}
\end{cases}.
\end{equation} 
That is, $\phi_{I, \Set{m_i}_{i\in I}}(\cdot)$ has $I$ and $|I|$-many messages hard-wired; it evaluates to $1$ if and only if its input can be parsed as $(m'_1 \| \ldots, \|m'_t)$ of the correct length, and $m'_i$ agrees with $m_i$ for all $i$'s specified by $I$.

This construction is constant-round and based on black-box access to post-quantum secure OWFs, because the ExtCom-and-Prove scheme is so. To prove security, if the committer is corrupted, simulation can be done via the $\epsilon$-simulatable extractability of the Commit Stage (\Cref{item:com-n-prove:condition:SimExt}); if the receiver is corrupted, simulation can be done via the $\epsilon$-ZK property (\Cref{item:com-n-prove:condition:e-zk}) of the ExtCom-and-Prove. 
\end{proof}

\section{Black-Box Constant-Round $\epsilon$-2PC using Quantum Communication}

\xiao{This section can go to the appendix.}

If quantum communication is allowed, we can obtain a constant-round $\epsilon$-simulatable PQ-2PC assuming only black-box access to post-quantum secure OWFs (instead of post-quantum secure semi-honest OTs). Recently, \cite{EC:GLSV21} and \cite{C:BCKM21b} (based on earlier works  \cite{C:CreKil88,C:BBCS91,C:DFLSS09,C:BouFeh10}) constructed fully simulatable (in contrast to $\epsilon$-simulatable) post-quantum 2PC (actually, they obtained MPC) assuming only post-quantum secure OWFs and quantum communication. The protocol from \cite{C:BCKM21b} makes only black-box use of the assumed OWF. But neither of these constructions is in constant rounds (without trusted assumptions like common reference strings).

Our construction follows the same path shown in \Cref{sec:PQ-2PC:blueprint}. That is, we break the task of constructing $\epsilon$-simulatable PQ-2PC in to the following steps:
\begin{enumerate}
\item \label[Step]{classical-PQ-2PC:blueprint:1}
Construct a constant-round $\epsilon$-simulatable protocol implementing $\Func^{t'}_\textsc{so-com}$;
\item \label[Step]{classical-PQ-2PC:blueprint:2}
In the $\Func^{t'}_\textsc{so-com}$-hybrid model, construct a constant-round $\epsilon$-simulatable protocol implementing $\Func^t_\textsc{ot}$;
\item \label[Step]{classical-PQ-2PC:blueprint:3}
In the $\Func^t_\textsc{ot}$-hybrid model, construct a constant-round $\epsilon$-simulatable PQ-2PC protocol.
\end{enumerate} 
Notice that in \Cref{sec:e-2pc} where only classical communication is allowed, the only place where the post-quantum semi-honest OT is used is \Cref{classical-PQ-2PC:blueprint:2}. (\Cref{classical-PQ-2PC:blueprint:1,classical-PQ-2PC:blueprint:3} only need to make black-box access to a post-quantum secure OWF.) Therefore, we will obtain the desired construction if we can make \Cref{classical-PQ-2PC:blueprint:2} work without relying on post-quantum secure semi-honest OTs.

We observe that \cite{EC:GLSV21} already did this relying on the \cite{C:BBCS91} OT construction, which makes use of quantum communication. We refer the read to \cite{EC:GLSV21} for further details. Here, we simply import the related lemma from  \cite{EC:GLSV21}.
\begin{lemma}[{\cite[Theorem 3.2, Corollary 3.4]{EC:GLSV21}}]\label{lem:quantum-2PC:p-ot-to-so-com}
For any polynomial $t(\secpar)$, there exists another polynomial $t'(\secpar)$ such that there is a constant-round protocol implementing $\Func^t_\textsc{ot}$ w.r.t.\ $\epsilon$-simulatable security against QPT adversaries in the $\Func^{t'}_\textsc{so-com}$-hybrid model. This protocol makes use of quantum communication.
\end{lemma}
Replacing \Cref{classical-PQ-2PC:blueprint:2} with the construction promised by \Cref{lem:quantum-2PC:p-ot-to-so-com} yields our final theorem. (Recall that we already finished \Cref{classical-PQ-2PC:blueprint:1,classical-PQ-2PC:blueprint:3} in \Cref{sec:e-2pc}.)
\begin{theorem}
Assuming the existence of OWFs secure against QPT adversaries, there exists a black-box, constant-round construction of $\epsilon$-simulatable 2PC protocol secure against QPT adversaries. This protocol makes use of quantum communication. 
\end{theorem}

\section*{Acknowledgments}\addcontentsline{toc}{section}{Acknowledgment}
We thank Susumu Kiyoshima for answering questions regarding the strongly extractable commitment in \cite{C:Kiyoshima14}. We also thank Omkant Pandey for helpful discussions about the black-box constant-round compiler from semi-honest OTs to 2PC.
We thank a reviewer of STOC 2022 for suggesting a simplified proof of \Cref{lem:state-close} with a better bound. 

\bibliographystyle{alpha}
\bibliography{main.bbl}
\addcontentsline{toc}{section}{References}

\newpage
\appendix
\section*{Appendix}

\section{From Extractable Commitment to ZK Argument}\label{sec:extcom_to_zk}
We give a proof sketch for that constant-round post-quantum extractable commitment with black-box extraction implies constant-round post-quautum ZK argument for $\NP$. 
The construction also relies on $\Sigma$-protocol (say, Blum's Hamiltonicity protocol \cite{Blum86}). 
Then the ZK argument works as follows:
\begin{enumerate}
    \item The prover sends the first message of the  $\Sigma$-protocol. 
    \item The prover and verifier jointly run one-sided simulation coin-flipping based on the extractable commitment to determine the second message of $\Sigma$-protocol. That is, they do the following:
    \begin{enumerate}
        \item The verifier commits to a uniformly random string $r_1$ of the same length as the second message of $\Sigma$-protocol by using the extractable commitment. 
        \item The prover sends a uniformly random string $r_2$ of the same length in the clear.
        \item The verifier opens $r_1$, and the second message of the $\Sigma$-protocol is set to be $r_1\oplus r_2$.
    \end{enumerate}
    \item The prover sends the third message of the $\Sigma$-protocol.
    \item The verifier runs the verification algorithm of the $\Sigma$-protocol.
\end{enumerate}
It is clear that the above protocol is constant-round if the extractable commitment is constant-round. 
For soundness, we show that no QPT cheating prover can bias $r_1\oplus r_2$ based on the computational hiding of the extractable commitment. Then, the soundness of the above protocol can be easily reduced to that of the underlying $\Sigma$-protocol. 
For ZK, recall that $\Sigma$-protocol satisfies a special honest-verifier ZK, which enables one to simulate the transcript if the second message is known in advance. Based on that, we construct a simulator for the above protocol as follows.
It first randomly chooses the second message $\beta$ of the $\Sigma$-protocol. 
Then it simulates the transcript $(\alpha,\beta,\gamma)$ of the $\Sigma$-protocol and sends $\alpha$ to the malicious verifier as the first message. 
Then it extracts $r_1$ from the malicious verifier while simulating its state by running the extractor and sets $r_2\defeq r_1\oplus \beta$ and sends $r_2$ to the malicious verifier. 
Finally, it sends $\gamma$ to the malicious verifier as the final message. 
It is straightforward to show that this simulator satisfies the requirement for ZK. Moreover, this simulator is black-box as long as the extractor is black-box.

\section{Postponed Proofs in \Cref{sec:extract_and_simulate}}
\subsection{Proof of \Cref{lem:amplification}}\label{sec:proof_lem_amplification}
\begin{proof}[Proof of \Cref{lem:amplification}]
\revise{Since the proof is very similar to that of \cite[Lemma 3.3]{C:ChiChuYam21}, we only explain the differences. 
  We define projections $\Pi_0$ and $\Pi_1$ over $\hil=\hil_\regX\times \hil_\regY$ as
\begin{align*}
\Pi_0:=I_{\regX} \ot (\ket{0}\bra{0})_{\regY}, \Pi_1:=\Pi,
\end{align*} 
and apply Jordan's lemma similarly to in the proof of  \cite[Lemma 3.2]{C:ChiChuYam21}. 
That is, $S_{\ge \delta}$ (resp. $S_{< \delta}$) is defined to be the subspace spanned by eigenvectors of $\Pi_0\Pi_1\Pi_0$ with eigenvalues $\ge \delta$ (resp. $<\delta$). 
Then \Cref{item:amplification_invariance_projection,item:amplification_success_probability} directly follow from Jordan's lemma. For proving \Cref{item:amplification,item:amplification_efficiency}, we slightly modify the definitions of $U_{\amp,T}$ and $\Amp$ from those in  \cite[Lemma 3.2]{C:ChiChuYam21}. 
We first consider an algorithm $\widetilde{\Amp}$  described as follows:
\begin{description}
\item $\widetilde{\Amp}(1^T,\ket{\psi}_{\regX,\regY})$: 
This algorithm takes a repetition parameter $T$ and a quantum state $\ket{\psi}_{\regX,\regY}\in \hil$ as input and works as follows:\footnote{Strictly speaking, we need to consider descriptions of quantum circuits to perform measurements $\{\Pi_0, I_{\regX,\regY}-\Pi_0\}$  and $\{\Pi_1, I_{\regX,\regY}-\Pi_1\}$ as part of its input so that we can make the description of $\widetilde{\Amp}$ independent on them. (Looking ahead, this is needed for showing Item \ref{item:amplification_efficiency} in Lemma \ref{lem:amplification} where $\Amp$ is required to be a uniform QPT machine.) We omit to explicitly write them in the input of $\widetilde{\Amp}$ for notational simplicity.}
\begin{enumerate}
\item Repeat the following for $i=1,...,T$:
\begin{enumerate}
\item Perform a measurement $\{\Pi_0, I_{\regX,\regY}-\Pi_0\}$. If the outcome is $1$, i.e., if $\Pi_0$ is applied, then set $A_i:=1$. 
\item Perform a measurement $\{\Pi_1, I_{\regX,\regY}-\Pi_1\}$ If the outcome is $1$, i.e., if $\Pi_1$ is applied, then set $B_i:=1$.  
\item If $A_i=B_i=1$, output the state in the registers $(\regX,\regY)$ and a classical bit $b=1$ indicating a success and immediately halt. 
\end{enumerate}
\item Output the state in the registers $(\regX,\regY)$ and a classical bit $b=0$ indicating a failure.
\end{enumerate}
\end{description}
\begin{remark}
    In the proof of \cite[Lemma 3.2]{C:ChiChuYam21}, $\widetilde{\Amp}$ outputs the state of $(\regX,\regY)$ as soon as observing $B_i=1$ for some $i$, but here, it outputs the state only after observing $A_i=1$ and $B_i=1$ for some $i$. This modification is needed to ensure that the output state is in the span of $\Pi_1 \Pi_0$ (rather than in the span of $\Pi_1$ as in \cite[Lemma 3.2]{C:ChiChuYam21}). 
\end{remark}
We define an algorithm $\Amp$ as a purified version of $\widetilde{\Amp}$.
That is, $\Amp$ works similarly to $\widetilde{\Amp}$ except that intermediate measurement results are stored in designated registers in  $\reganc$ without being measured and the output $b$ is stored in register $\regB$. 
Let $U_{\amp,T}$ be the unitary part of $\Amp(1^T,\cdot)$. 
Then \Cref{item:amplification_map_to_pi,item:amplification_efficiency} follows from the definition and \Cref{item:amplification_invariance} follows from the fact that $S_{\ge \delta}$ and $S_{< \delta}$ are invariant under $\Pi_0$ and $\Pi_1$. 
For proving \Cref{item:amplification_amplification}, it suffices to consider the case where  $\ket{\phi}_{\regX}\ket{0}_{\regY}$ is an eigenvector of $\Pi_0\Pi_1\Pi_0$ with an eigenvalue $t\ge \delta$ by Jordan's lemma as argued in the proof of \cite[Lemma 3.2]{C:ChiChuYam21}.
In this case, $\{A_i\}$ and $\{B_i\}$ follow the following distribution:
\begin{itemize}
\item $\Pr[A_1=1]=1$ and for all $i\ge 2$, the distribution of $A_i$ only depends on $B_{i-1}$ and $\Pr[A_i= B_{i-1}]=t$.
\item For $i\ge 1$, the distribution of $B_i$ only depends on $A_{i}$ and $\Pr[B_i= A_{i}]=t$.
\end{itemize}
It suffices to show that for any noticeable $t=t(\secpar)$ and $\nu=\nu(\secpar)$, there is $T=\poly(\secpar)$ such that 
\begin{align}\label{eq:goal_amplification}
\Pr[\exists i\in[T]~A_i=B_i=1]\ge 1-\nu.
\end{align}
First, note that 
\begin{align*}
    \Pr[A_1=B_1=1]=t.
\end{align*}
Thus, if $t\geq 1-\nu$, then \Cref{eq:goal_amplification} is satisfied for $T=1$. 
Below, we consider the case where $t< 1-\nu$. 

For any $i\geq 1$, we have
\begin{align*}
\Pr[A_{i+1}=B_{i+1}=1 \mid B_i=0]= t(1-t)
\end{align*}
and 
\begin{align*}
\Pr[A_{i+1}=B_{i+1}=1 \mid B_i=1]= t^2.
\end{align*}
Thus, for any positive integer $T$, we have 
\begin{align*}
    \Pr[\exists i\in[T]~A_i=B_i=1]\ge t+(1-t)(1-(1-\min\{t(1-t),t^2\})^{T-1}). 
\end{align*}
Since $\min\{t(1-t),t^2\}$ is noticeable, we can take $T=\poly(\secpar)$ in such a way that $t+(1-t)(1-(1-\min\{t(1-t),t^2\})^{T-1})\ge 1-\nu$. This completes the proof.} 
\end{proof}

\subsection{Proof of \Cref{lem:state-close}}\label{sec:proof_state-close}
\begin{proof}[Proof of \Cref{lem:state-close}]

If $(12\gamma^{1/2}+2\delta)^{1/2}>1$, then the desired inequality trivially holds. Thus, we assume $(12\gamma^{1/2}+2\delta)^{1/2}\le 1$ in the rest of the proof.  

$F$ can be represented by a projector $\Pi$ over the input register and auxiliary input register with which $F$ works as follows.
$F$ appends the ancillary register initialized to $\ket{0^n}$, performs the measurement $\{\Pi,I-\Pi\}$ on the input and ancillary registers, and outputs the state in a designated output register tracing out all other registers if the state is projected onto $\Pi$, and otherwise outputs $\fail$.

We consider an additional one-qubit register and define 
\begin{align*}
\ket{\psi_b}\defeq \sqrt{1-p_b}\ket{0}\ket{0^m}\ket{0^n} + \ket{1}\Pi\ket{\phi_b}\ket{0^n}
\end{align*}
for $b\in \bit$ where 
$m$ is the number of qubits in the register for $\ket{\phi_b}$ and 
\begin{align*}
p_b \defeq \|\Pi\ket{\phi_b}\ket{0^n}\|^2.
\end{align*}
Without loss of generality, we assume $p_0\geq p_1$. 
It suffices to prove 
\begin{align}
\label{eq:conclusion_trace_distance_pure}
   \|\ket{\psi_0}\bra{\psi_0}-\ket{\psi_1}\bra{\psi_1}\|_{tr}\leq  (12\gamma^{1/2}+2\delta)^{1/2}
\end{align}
because a distinguisher that distinguishes $F(\ket{\phi_0}\bra{\phi_0})$ and $F(\ket{\phi_1}\bra{\phi_1})$ can be easily converted into a distinguisher that distinguishes 
$\ket{\psi_0}$ and $\ket{\psi_1}$ with the same advantage.

We show the following claim.
\begin{myclaim}\label{cla:useful_properties_new}
The following holds:
\begin{enumerate}
    \item \label{item:norm_of_projected_state_new} 
    $\|\Pi\ket{\phi_{b,0}}\ket{0^n}\|^2\leq \gamma$ for $b\in \bit$.
    \item \label{item:bound_probability_new}
    $\|\Pi\ket{\phi_{b,1}}\ket{0^n}\|^2\geq p_b-3\gamma^{1/2}$
    for $b\in \bit$.
\end{enumerate}
\end{myclaim}
\begin{proof}[Proof of \Cref{cla:useful_properties_new}]
\Cref{item:norm_of_projected_state_new} follows from the definition of $\Pi$ and the assumption that $\Pr[F\left(\frac{\ket{\phi_{b,0}}\bra{\phi_{b,0}}}{\|\ket{\phi_{b,0}}\|^2}\right)=\fail]\geq 1-\gamma$. 
\Cref{item:bound_probability_new} can be shown as follows:
\begin{align*}
    p_b&=\|\Pi\ket{\phi_b}\ket{0^n}\|^2\\
    &=\|\Pi\ket{\phi_{b,0}}\ket{0^n}+\Pi\ket{\phi_{b,1}}\ket{0^n}\|^2\\
    &\leq \|\Pi\ket{\phi_{b,0}}\ket{0^n}\|^2+\|\Pi\ket{\phi_{b,1}}\ket{0^n}\|^2 
    + 2\|\Pi\ket{\phi_{b,0}}\|\cdot \|\Pi\ket{\phi_{b,1}}\|\\
    &\leq \|\Pi\ket{\phi_{b,1}}\ket{0^n}\|^2+ 3\gamma^{1/2}
\end{align*}
where the last inequality follows from $\|\Pi\ket{\phi_{b,0}}\ket{0^n}\|^2\leq \|\Pi\ket{\phi_{b,0}}\ket{0^n}\|\leq \gamma^{1/2}$ by \Cref{item:norm_of_projected_state_new} and
$\|\Pi\ket{\phi_{b,1}}\|\leq 1$. 
\end{proof}

We give a lower bound for $|\braket{\psi_0}{\psi_1}|$.
By the definition of $\ket{\psi_b}$, 
\revise{\begin{align*}
|\braket{\psi_0}{\psi_1}|
&=|\sqrt{(1-p_0)(1-p_1)}+ \bra{\phi_0}\bra{0^n}\Pi\ket{\phi_1}\ket{0^n}| \notag\\
&=
\left|
\begin{array}{ccc}
\sqrt{(1-p_0)(1-p_1)}&+\bra{\phi_{0,0}}\bra{0^n}\Pi\ket{\phi_{1,0}}\ket{0^n}&+
\bra{\phi_{0,0}}\bra{0^n}\Pi\ket{\phi_{1,1}}\ket{0^n}  \\
&+\bra{\phi_{0,1}}\bra{0^n}\Pi\ket{\phi_{1,0}}\ket{0^n}
&+\bra{\phi_{0,1}}\bra{0^n}\Pi\ket{\phi_{1,1}}\ket{0^n}
\end{array}
\right| \notag \\
&=
\left|
\begin{array}{lll}
\sqrt{(1-p_0)(1-p_1)}&+\bra{\phi_{0,0}}\bra{0^n}\Pi\ket{\phi_{1,0}}\ket{0^n}&+
\bra{\phi_{0,0}}\bra{0^n}\Pi\ket{\phi_{1,1}}\ket{0^n}  \\
&+\bra{\phi_{0,1}}\bra{0^n}\Pi\ket{\phi_{1,0}}\ket{0^n}
&+\bra{\phi_{0,1}}\bra{0^n}\Pi\ket{\phi_{0,1}}\ket{0^n}\\
&+\bra{\phi_{0,1}}\bra{0^n}\Pi(\ket{\phi_{1,1}}-\ket{\phi_{0,1}})\ket{0^n}
\end{array}
\right| \notag \\
&\geq (1-p_0)+\|\Pi\ket{\phi_{0,1}}\ket{0^n}\|^2- \sum_{(c,d)\in \{(0,0),(0,1),(1,0)\}}\left\|\Pi\ket{\phi_{0,c}}\ket{0^n}\right\|\cdot \|\Pi\ket{\phi_{1,d}}\ket{0^n}\|
-\|\ket{\phi_{1,1}}-\ket{\phi_{0,1}}\|
\\
&\geq 
(1-p_0)+(p_0-3\gamma^{1/2})-3\gamma^{1/2}-\delta\\
&=1-6\gamma^{1/2}-\delta 
\end{align*}
where we used the assumption that $p_0\geq p_1$ in the first inequality 
and 
\Cref{cla:useful_properties_new} and the assumption that $\|\ket{\phi_{1,1}}-\ket{\phi_{0,1}}\|\le \delta$  
in the second inequality.
We note that $1-6\gamma^{1/2}-\delta>0$ since we assume $(12\gamma^{1/2}+2\delta)^{1/2}\le 1$.

Then, we have 
\begin{align*}
 \|\ket{\psi_0}\bra{\psi_0}-\ket{\psi_1}\bra{\psi_1}\|_{tr}
 &=\sqrt{1-|\braket{\psi_0}{\psi_1}|^2}\\
 &\leq  \sqrt{12\gamma^{1/2}+2\delta}
\end{align*}
This completes the proof of \Cref{lem:state-close}. }
\end{proof}

\end{document}